\documentclass[longauth]{aa}

\usepackage{graphicx}
\usepackage{txfonts}
\usepackage{placeins}
\begin{document} 

\def\ms{m\,s$^{-1}$}
\def\kms{km\,s$^{-1}$}
\def\teff{$T_{\rm eff}$}
\def\logg{$\log{g}$}
\def\vmic{$\upsilon_t$}
\def\loggf{$\log{gf}$}

   \title{A new and homogeneous metallicity scale \\ for Galactic classical Cepheids}
   
   \subtitle{II. The abundance of iron and $\alpha$ elements
   \thanks{
   Partly based on observations made with ESO Telescopes at the La Silla/Paranal Observatories under program IDs: 072.D-0419, 073.D-0136, and 190.D-0237 for HARPS spectra; 084.B-0029, 087.A-9013, 074.D-0008, 075.D-0676, and 60.A-9120 for FEROS spectra; 081.D-0928, 082.D-0901, 089.D-0767, and 093.D-0816 for UVES spectra.}$^,$
   \thanks{
   Partly based on data obtained with the STELLA robotic telescopes in Tenerife, a facility of The Leibniz Institute for Astrophysics Potsdam (AIP) jointly operated by the AIP and by the Instituto de Astrofísica de Canarias (IAC).}$^,$
   \thanks{
   Tables~\ref{table:linelist_iron}, \ref{table:linelist_alpha}, \ref{table:atm_params_rv_spectra}, \ref{table:alpha_spectra}, and \ref{table:alpha_stars} are only available in electronic form at the CDS via anonymous ftp to cdsarc.u-strasbg.fr (130.79.128.5) or via http://cdsweb.u-strasbg.fr/cgi-bin/qcat?J/A+A/.}
   }

   \author{R. da Silva \inst{1,2}
          \and
          J. Crestani \inst{3,1}
          \and
          G. Bono \inst{3,1}
          \and
          V.F. Braga \inst{1,2}
          \and
          V. D'Orazi \inst{4}
          \and
          B. Lemasle \inst{5}
          \and
          M. Bergemann \inst{6}
          \and
          M. Dall'Ora \inst{7}
          \and \\
          G. Fiorentino \inst{1}
          \and
          P. François \inst{8,9}
          \and
          M.A.T. Groenewegen \inst{10}
          \and
          L. Inno \inst{11}
          \and
          V. Kovtyukh \inst{12}
          \and
          R.-P. Kudritzki \inst{13,14}
          \and \\
          N. Matsunaga \inst{15,16}
          \and
          M. Monelli \inst{17,18}
          \and
          A. Pietrinferni \inst{19}
          \and
          L. Porcelli \inst{20}
          \and
          J. Storm \inst{21}
          \and
          M. Tantalo \inst{3,1}
          \and
          F. Thévénin \inst{22}
          }

   \institute{
   INAF - Osservatorio Astronomico di Roma, via Frascati 33, 00078 Monte Porzio Catone, Italy
   \and
   Agenzia Spaziale Italiana, Space Science Data Center, via del Politecnico snc, 00133 Rome, Italy
   \and
   Dipartimento di Fisica, Università di Roma Tor Vergata, via della Ricerca Scientifica 1, 00133 Roma, Italy
   \and
   INAF - Osservatorio Astronomico di Padova, Vicolo dell’Osservatorio 5, 35122 Padova, Italy
   \and
   Astronomisches Rechen-Institut, Zentrum für Astronomie der Universität Heidelberg, Mönchhofstr. 12-14, 69120, Heidelberg, Germany
   \and
   Max Planck Institute for Astronomy, D-69117 Heidelberg, Germany; Niels Bohr International Academy, Niels Bohr Institute, Blegdamsvej 17, DK-2100 Copenhagen Ø, Denmark
   \and
   INAF - Osservatorio Astronomico di Capodimonte, Napoli, Italy
   \and
   GEPI, Observatoire de Paris, PSL Research University, CNRS, 61 Avenue de l'Observatoire, 75014, Paris, France
   \and
   UPJV, Université de Picardie Jules Verne, 33 rue St Leu, 80080, Amiens, France
   \and
   Koninklijke Sterrenwacht van België, Ringlaan 3, 1180, Brussels, Belgium
   \and
   Science and Technology Department, Parthenope University of Naples, Naples, Italy
   \and
   Astronomical Observatory, Odessa National University, Shevchenko Park, 65014, Odessa, Ukraine
   \and
   LMU München, Universitätssternwarte, Scheinerstr. 1, D-81679 München, Germany
   \and
   Institute for Astronomy, University of Hawaii at Manoa, 2680 Woodlawn Drive, Honolulu, HI 96822, USA
   \and
   Department of Astronomy, School of Science, The University of Tokyo, 7-3-1, Hongo, Bunkyo-ku, Tokyo 113-0033, Japan
   \and
   Laboratory of Infrared High-Resolution spectroscopy (LiH), Koyama Astronomical Observatory, Kyoto Sangyo University, Motoyama, Kamigamo, Kita-ku, Kyoto 603-8555, Japan
   \and
   Instituto de Astrofísica de Canarias (IAC), La Laguna, E-38205, Spain
   \and
   Departamento de Astrofísica, Universidad de La Laguna (ULL), E-38200, La Laguna, Spain
   \and
   INAF-Osservatorio Astronomico d'Abruzzo, Via M. Maggini, s/n, I-64100, Teramo, Italy
   \and
   Istituto Nazionale di Fisica Nucleare, Laboratori Nazionali di Frascati (INFN-LNF), Frascati, Italy
   \and
   Leibniz-Institut für Astrophysik Potsdam (AIP), An der Sternwarte 16, 14482, Potsdam, Germany
   \and
   Université Côte d’Azur, Observatoire de la Côte d’Azur, CNRS, Lagrange UMR 7293, CS 34229, 06304, Nice Cedex 4, France
   }

   \date{Received ...; accepted ...}

  \abstract
   {Classical Cepheids are the most popular distance indicators and tracers of young stellar populations. The key advantage is that they are bright and they can be easily identified in Local Group and Local Volume galaxies. Their evolutionary and pulsation properties depend on their chemical abundances.}
   {The main aim of this investigation is to perform a new and accurate abundance analysis of two tens of calibrating Galactic Cepheids using high spectral resolution (R$\sim$40,000--115,000) and high S/N spectra ($\sim$400) covering the entire pulsation cycle.}
   {We focus our attention on possible systematics aﬀecting the estimate of atmospheric parameters and elemental abundances along the pulsation cycle. 
   We cleaned the line list by using atomic transition parameters based on laboratory measurements and by removing lines that are either blended or display abundance variations along the pulsation cycle.}
   {The spectroscopic approach that we developed brings forward small dispersions in the variation of the atmospheric parameters ($\sigma$(\teff)$\sim$50~K, $\sigma$(\logg)$\sim$0.2~dex, and $\sigma$($\xi$)$\sim$0.2~\kms) and in the abundance of both iron ($\lesssim$ 0.05~dex) and $\alpha$ elements ($\lesssim$ 0.10~dex) over the entire pulsation cycle. We also provide new and accurate eﬀective temperature templates by splitting the calibrating Cepheids into four diﬀerent period bins, ranging from short to long periods. For each period bin, we performed an analytical ﬁt with Fourier series providing $\theta = 5040~/{T_{\rm eff}}$ as a function of the pulsation phase.}
   {The current findings are a good viaticum to trace the chemical enrichment of the Galactic thin disk by using classical Cepheids and a fundamental stepping stone for further investigations into the more metal-poor regime typical of Magellanic Cepheids.}


   \maketitle
%

\section{Introduction}

The current empirical evidence indicates that the estimates of the Hubble constant based on Cosmic Microwave Background (CMB) and on the standard cosmological model ($\Lambda$ Cold Dark Matter, $\Lambda$CDM) are at odds with the direct estimates based on primary (classical Cepheids) and secondary (type Ia Supernovae) distance indicators \citep{PlanckCollaboration2018,Riessetal2019}. The current tension between early and late Universe measurements is at a 5-$\sigma$ level. It is not clear whether it might be explained by unknown systematics or whether this disagreement is opening the path for new physics beyond the standard cosmological model \citep[see, e.g.,][and references therein]{NiedermannSloth2020}. In this context, it is worth mentioning that the use of the Tip of the Red Giant Branch (TRGB) as primary distance indicators gives values of the Hubble constant that are intermediate between those based on the CMB and those based on the Cepheid distance scale \citep{Freedmanetal2020}.

There are solid reasons to believe that the difference between the TRGB and the Cepheid distance scale is caused by the dependence of the metal content on the zero-point and/or the slope of the adopted diagnostics \citep{Bonoetal2010,Pietrzynskietal2013}. The dependence of optical and near-infrared diagnostics on metallicity is still a controversial issue for both theoretical \citep{Marconietal2005,Bonoetal2010} and empirical \citep{Groenewegen2018,Groenewegen2020,Ripepietal2019,Ripepietal2021,Breuvaletal2020,Breuvaletal2021} approaches. However, the improved accuracy of geometrical distances from Gaia DR3, and the homogeneity of spectroscopic abundances will provide solid constraints on this longstanding problem \citep[][and references therein]{Ripepietal2022}.

\begin{table*}
\caption{Sample of 20 calibrating Cepheids with high-resolution spectra covering either a substantial portion or the entire pulsation cycle.}
\label{table:calib_sample}
\centering
{\small
\begin{tabular}{l c c c r@{ }l c c c c c c}
\noalign{\smallskip}\hline\hline\noalign{\smallskip}
Name & $\alpha_{\rm ICRS}$ & $\delta_{\rm ICRS}$ & Mode &
\multicolumn{2}{c}{${\rm [Fe/H]}_{\rm lit}$ $\pm$ $\sigma$} &
Ref.\tablefootmark{a} &
$N_{\rm F}$ & $N_{\rm H}$ & $N_{\rm U}$ & $N_{\rm S}$ & $N_{\rm tot}$ \\
\noalign{\smallskip}\hline\noalign{\smallskip}
\object{R\,TrA}        & 15:19:45.712 & $-$66:29:45.742 & 0  & $-$0.01 & $\pm$ 0.03 & 2 &   1 &  14 & ... & ... &  15 \\
\object{T\,Vul}        & 20:51:28.238 &   +28:15:01.817 & 0  &    0.07 & $\pm$ 0.10 & 1 & ... & ... & ... &  26 &  26 \\
\object{FF\,Aql}       & 18:58:14.748 &   +17:21:39.296 & 1  &    0.10 & $\pm$ 0.10 & 1 & ... & ... & ... &  27 &  27 \\
\object{S\,Cru}        & 12:54:21.997 & $-$58:25:50.215 & 0  &    0.09 & $\pm$ 0.04 & 2 & ... & ... & ... &  13 &  13 \\
\object{$\delta$\,Cep} & 22:29:10.265 &   +58:24:54.714 & 0  &    0.09 & $\pm$ 0.10 & 1 & ... & ... & ... &  18 &  18 \\
\object{Y\,Sgr}        & 18:21:22.986 & $-$18:51:36.002 & 0  &    0.00 & $\pm$ 0.06 & 2 & ... &  20 & ... &   4 &  24 \\
\object{XX\,Sgr}       & 18:24:44.500 & $-$16:47:49.820 & 0  &    0.06 & $\pm$ 0.04 & 2 &   4 & ... &   5 &   3 &  12 \\
\object{$\eta$\,Aql}   & 19:52:28.368 &   +01:00:20.370 & 0B &    0.24 & $\pm$ 0.09 & 2 & ... & ... & ... &  11 &  11 \\
\object{S\,Sge}        & 19:56:01.261 &   +16:38:05.236 & 0B &    0.14 & $\pm$ 0.10 & 1 & ... & ... & ... &  21 &  21 \\
\object{V500\,Sco}     & 17:48:37.501 & $-$30:28:33.461 & 0B & $-$0.07 & $\pm$ 0.08 & 1 & ... & ... &   4 &   3 &   7 \\
\object{$\beta$\,Dor}  & 05:33:37.512 & $-$62:29:23.323 & 0B & $-$0.03 & $\pm$ 0.05 & 2 &   1 &  46 & ... & ... &  47 \\
\object{$\zeta$\,Gem}  & 07:04:06.530 &   +20:34:13.074 & 0B &    0.16 & $\pm$ 0.05 & 2 & ... &  47 & ... &  81 & 128 \\
\object{VY\,Sgr}       & 18:12:04.566 & $-$20:42:14.465 & 0B &    0.25 & $\pm$ 0.08 & 2 &  29 & ... &   4 & ... &  33 \\
\object{UZ\,Sct}       & 18:31:22.367 & $-$12:55:43.339 & 0B &    0.11 & $\pm$ 0.09 & 2 &  28 & ... &   6 & ... &  34 \\
\object{Y\,Oph}        & 17:52:38.702 & $-$06:08:36.875 & 0  &    0.08 & $\pm$ 0.05 & 2 & ... &   8 & ... & ... &   8 \\
\object{RY\,Sco}       & 17:50:52.345 & $-$33:42:20.411 & 0  &    0.01 & $\pm$ 0.06 & 1 & ... & ... &   5 &   3 &   8 \\ 
\object{RZ\,Vel}       & 08:37:01.303 & $-$44:06:52.844 & 0  &    0.08 & $\pm$ 0.06 & 2 &   1 &  11 & ... & ... &  12 \\
\object{V340\,Ara}     & 16:45:19.112 & $-$51:20:33.394 & 0  &    0.23 & $\pm$ 0.07 & 2 &  26 & ... &   6 & ... &  32 \\
\object{WZ\,Sgr}       & 18:16:59.716 & $-$19:04:32.989 & 0  &    0.28 & $\pm$ 0.08 & 1 &   1 & ... &   5 &   3 &   9 \\
\object{RS\,Pup}       & 08:13:04.216 & $-$34:34:42.692 & 0  &    0.14 & $\pm$ 0.07 & 2 & ... &  15 & ... &   3 &  18 \\
\hline
\end{tabular}
}
\tablefoot{The first four columns give the star name, the right ascension and declination, and the pulsation mode (0: fundamental; 1: first overtone; 0B: fundamental with bump). The Cols.~5 and 6 give the iron abundance from literature and the corresponding references. Columns from 7 to 10 show the number of optical spectra used for each spectrograph: $N_{\rm F}$: FEROS; $N_{\rm H}$: HARPS; $N_{\rm U}$: UVES; $N_{\rm S}$: STELLA. The last column lists the total number of spectra per target.
\tablebib{
\tablefoottext{a}{
1: \citet{Genovalietal2014}; for \object{T\,Vul}, \object{FF\,Aql}, \object{$\delta$\,Cep}, and \object{S\,Sge} we assumed a typical error of 0.1~dex since they were not provided by the original authors;
2: \citet{Proxaufetal2018}, from which the uncertainty adopted is the largest value between $\sigma$ and std.
}}}
\end{table*}

Classical Cepheids are bright ($-2 \leq\ M_V \leq\ -7$~mag) radially pulsating stars with periods ranging from roughly one day to more than one hundred days \citep[see, e.g.,][and references therein]{Skowronetal2019}. Their iron abundances range from solar values in the solar vicinity to $\sim$0.5~dex more metal-rich in the inner Galactic disk and to $-$0.5~dex more metal-poor in the outer disk \citep{Innoetal2019}. Cepheids in the Large Magellanic Cloud (LMC) are on average a factor of two more metal-poor \citep{Romanielloetal2021}, and those in the Small Magellanic Cloud (SMC) a factor of four more metal-poor than solar \citep{Lemasleetal2017}. Cepheids are intermediate-mass stars crossing the instability strip during central helium burning, the so-called blue loops, and their progenitor mass ranges from $\sim$3 to $\sim$10~$M_\odot$ -- the exact limit depends on the chemical composition \citep{Bonoetal2000a,DeSommaetal2021}. Classical Cepheids are solid stellar beacons to trace young stellar populations even in the far side of the Galactic thin disk \citep{Minnitietal2020,Minnitietal2021} and in stellar systems experiencing recent star formation events such as dwarf irregulars and star forming galaxies
\citep{Neeleyetal2021}. Their period-luminosity relations \citep{Breuvaletal2020,Ripepietal2020} provide accurate (at the level of $\sim$1-2\%) measurements of individual distances. Individual distances together with detailed information concerning their elemental abundances (iron peak, $\alpha$, neutron capture), provide the unique opportunity to investigate the radial gradients and, in turn, the recent (t~$\le$~200-300~Myr) chemical enrichment history of the thin disk and nearby stellar systems \citep{Romanielloetal2008,Lemasleetal2013,Genovalietal2015,daSilvaetal2016}.

Furthermore, classical Cepheids are fundamental laboratories for investigating not only the advanced evolutionary phases of intermediate-mass stars \citep{Bonoetal2000b,Neilsonetal2011,PradaMoronietal2012,DeSommaetal2021}, but also for constraining the physical mechanisms driving their variability \citep{Bonoetal1999,Marconietal2013}.

Although classical Cepheids are the cross-road of several recent and long-standing astrophysical problems, we still lack accurate investigations concerning the variation of their physical properties along the pulsation cycle. The current studies are mainly based on both optical and near-infrared (NIR) light curves, but we still lack detailed spectroscopic analysis concerning the variation of effective temperature, surface gravity, and microturbulent velocity \citep[see, e.g.,][and references therein]{Lemasleetal2020,Kovtyukh2007,Proxaufetal2018}.

\begin{figure*}
\centering
\begin{minipage}[t]{0.49\textwidth}
\centering
\resizebox{\hsize}{!}{\includegraphics{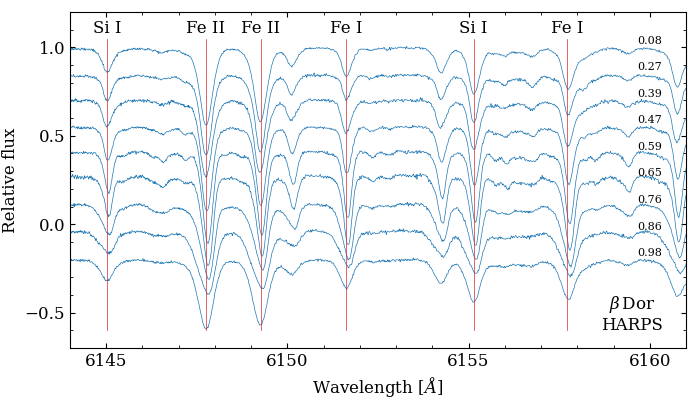}}
\end{minipage}
\begin{minipage}[t]{0.49\textwidth}
\centering
\resizebox{\hsize}{!}{\includegraphics{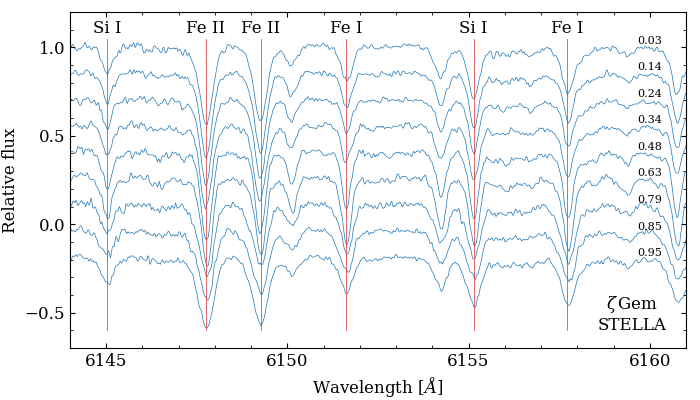}}
\end{minipage}
\caption{Examples of high-resolution spectra for the stars \object{$\beta$\,Dor} and \object{$\zeta$\,Gem} in different pulsation phases (values on the right side of each panel). The vertical red lines indicate the position in the rest-frame wavelength of atomic lines either only used for the abundance determination or also adopted by the LDR method. For the sake of clarity the continuum
of each spectrum was shifted to arbitrary levels.}
\label{spectra}
\end{figure*}

Pioneering investigations based on multiple (from a few to almost two dozen) spectra along the pulsation cycle were provided more than ten years ago by \citet{LuckAndrievsky2004}, \citet{Kovtyukhetal2005}, \citet{Andrievskyetal2005}, and \citet{Lucketal2008}, in which they evaluated the variation of the atmospheric parameters and the abundances in both short- and long-period Cepheids. More recently, we performed a detailed investigation, based on a large dataset of high-resolution spectra, for a sample of 14 calibrating Cepheids \citep{Proxaufetal2018}. In that work, the number of spectra per object ranges from five to more than 120, and the spectra covered a significant portion of the pulsation cycle. In the same investigation we also used a new and more homogeneous list of Line Depth Ratios (LDR) to estimate the effective temperature directly from the spectra, an approach that is independent of the photometric properties. The reader interested in a more detailed discussion is referred to \citet[][and references therein]{Kovtyukh2007}. However, all these investigations were hampered by three limitations:

\begin{itemize}

\item[a)] Line list -- We compiled a list of iron lines using line lists available in the literature. The sources of atomic transition parameters were inhomogeneous, because they did not pass a detailed scrutiny.

\item[b)] Phase coverage -- In spite of the large number of spectra adopted in the spectroscopic analysis, several objects had spectra collected at similar phases and, consequently, they had a modest phase coverage.

\item[c)] Intrinsic variation -- The inhomogeneity of the atomic transition parameters introduces uncertainties in the spectroscopic surface gravity and in the microturbulent velocity. This caused an increase in the standard deviation at fixed pulsation phase, and in turn, some limitation in the analysis of their variation along the pulsation cycle. 

\end{itemize}

To overcome these limitations, the current investigation is based on a new approach including five steps:
$i)$ We updated the atomic transition parameters of our line lists with more recent and accurate laboratory measurements available in the literature;
$ii)$ We collected new high signal-to-noise ratio (S/N) and high-resolution spectra for our sample of calibrating Cepheids, and we also performed a detailed search for similar quality spectra available in online science archives. This translates into an increase in the number of calibrating Cepheids by almost 50\% (from 14 to 20). Moreover, we also added new spectra to Cepheids already included in the former sample. The current sample includes 19 stars pulsating in the fundamental mode and, for the first time used by us as a calibrating Cepheid, one star (\object{FF\,Aql}) pulsating in the first overtone mode, and their metallicities range from $-$0.08 (\object{$\beta$\,Dor}) to 0.19~dex (\object{VY\,Sgr});
$iii)$ We performed a detailed visual check of hundreds of lines along the pulsation cycle to create a new homogeneous line list. Special attention was paid in the identification of lines showing blends (or poor definition of the continuum), or having any correlation with the effective temperature (due to possible NLTE effects) or with the equivalent widths (due to saturated lines), or presenting any systematic over- or under-abundance;
$iv)$ We improved the algorithm that we use for the estimate of the atmospheric parameters, and in turn, for the estimate of the elemental abundances; 
$v)$ In addition to the atmospheric parameters and iron abundances, the current study also includes the abundances of five $\alpha$ elements: Mg, Si, S, Ca, and Ti.

We notice that the star \object{AV\,Sgr}, which was one of the calibrating Cepheids in \citet{Proxaufetal2018}, is not included in the current work. The reason is that, after the revision of the Fe line list, we were not able to derive accurate estimates of the atmospheric parameters for many of the spectra of this variable. The number of spectra with accurate measurements was not enough to provide a good coverage of the pulsation cycle, and therefore, it was removed from the list of calibrating Cepheids. The parameters published in our previous paper can still be used, but keeping in mind that they are based on a different version of the line list.

The structure of the paper is the following: in Sect.~\ref{datasets} we discuss the spectroscopic dataset and provide detailed information concerning the current sample of calibrating Cepheids. In Sect.~\ref{section:data_reduction} we discuss the data reduction process and introduce the atomic line lists used. Sect.~\ref{section:rv_photometry} includes a description of the radial-velocity and photometric data used. In Sect.~\ref{section:atm_params} we explain how we derived the atmospheric parameters and their variation along the pulsation cycle. The iron and $\alpha$-element abundances and their variation along the pulsation cycle are discussed in Sect.~\ref{section:ab_iron_alpha}. A summary of the results and some final remarks concerning this project are given in Sect.~\ref{summary}.

\section{Spectroscopic datasets}
\label{datasets}

The spectroscopic datasets analyzed in the current paper are spectra collected using several different instruments. Three of them are mounted on telescopes of the European Southern Observatory (ESO): the High Accuracy Radial velocity Planet Searcher \citep[HARPS;][]{Mayoretal2003} at the 3.6~m, the Fiber-fed Extended Range Optical Spectrograph \citep[FEROS;][]{Kauferetal1999} at the 2.2~m MPG/ESO, and the Ultraviolet and Visual Echelle Spectrograph \citep[UVES;][]{Dekkeretal2000} at the Very Large Telescope. The other instrument is the STELLA Echelle Spectrograph \citep[SES;][]{Strassmeieretal2004,Strassmeieretal2010} located at the Iz\~ana Observatory on Tenerife in the Canary islands.

The spectral resolution of the quoted spectrographs for the instrument settings used are R$\sim$40\,000 (UVES), R$\sim$115\,000 (HARPS), R$\sim$48\,000 (FEROS), and R$\sim$55\,000 (STELLA). For details on the wavelength ranges of our spectra we refer the reader to Sect.~2 of the papers \citet{Proxaufetal2018} and \citet{Crestanietal2021}.

The entire spectroscopic sample includes reduced spectra downloaded from the ESO and the STELLA archives, for a total of 1383 spectra of 285 stars: 199 HARPS spectra of 9 stars, 419 FEROS spectra of 151 stars, 450 STELLA spectra of 68 stars, and 315 UVES spectra of 123 stars. Their S/N ranges from a few tens to more than 400 (about 70\% of our spectra have S/N of at least 100). From this sample we selected the spectra of 20 targets that cover either a significant part (about half or more) or the entire pulsation cycle. From now on we refer to this sub-sample as our sample of 20 calibrating Cepheids. Details on the number of spectra are given in Table~\ref{table:calib_sample}.

\begin{figure*}
\centering
\begin{minipage}[t]{0.33\textwidth}
\centering
\resizebox{\hsize}{!}{\includegraphics{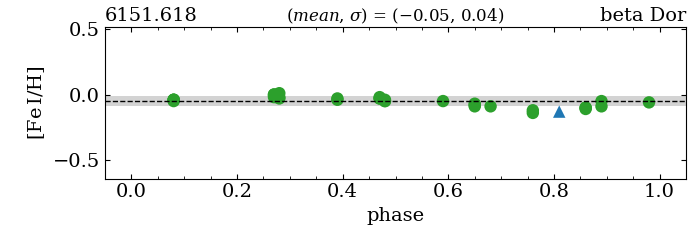}}
\end{minipage}
\begin{minipage}[t]{0.33\textwidth}
\centering
\resizebox{\hsize}{!}{\includegraphics{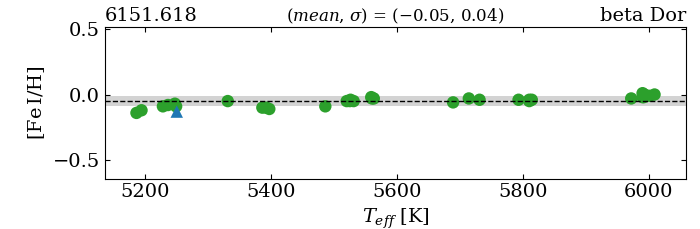}}
\end{minipage}
\begin{minipage}[t]{0.33\textwidth}
\centering
\resizebox{\hsize}{!}{\includegraphics{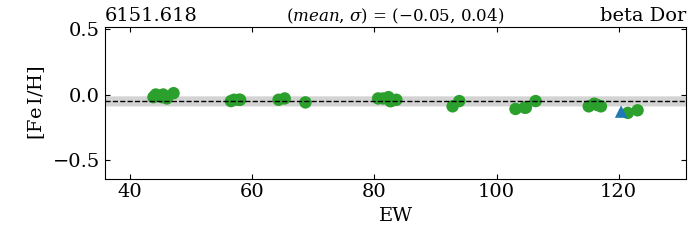}}
\end{minipage} \\
\begin{minipage}[t]{0.33\textwidth}
\centering
\resizebox{\hsize}{!}{\includegraphics{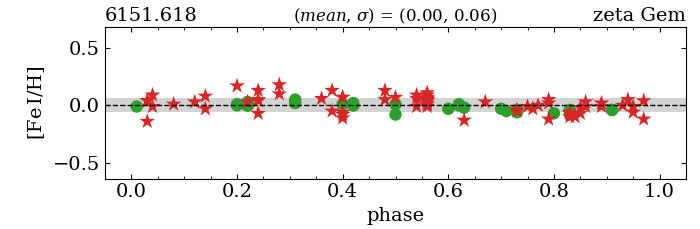}}
\end{minipage}
\begin{minipage}[t]{0.33\textwidth}
\centering
\resizebox{\hsize}{!}{\includegraphics{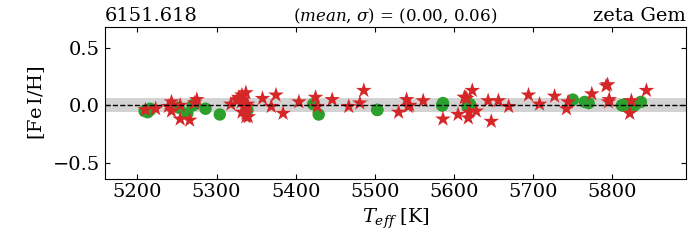}}
\end{minipage}
\begin{minipage}[t]{0.33\textwidth}
\centering
\resizebox{\hsize}{!}{\includegraphics{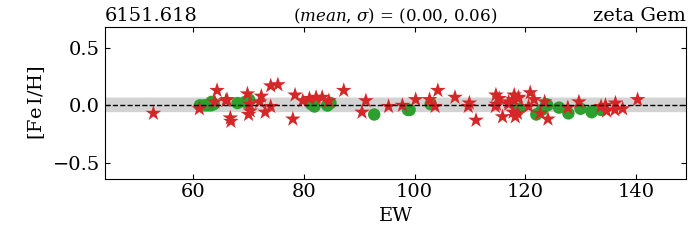}}
\end{minipage}
\caption{Iron abundances as a function of the pulsation phase, effective temperature, and equivalent width. These are examples of good lines that were kept in our initial line list for HARPS (green circles), FEROS (blue triangle), and STELLA (red stars) spectra of \object{$\beta$\,Dor} and \object{$\zeta$\,Gem}. The hatched regions around the dashed lines indicate the 1-$\sigma$ uncertainty around the mean.}
\label{figure:good_lines}
\end{figure*}

\section{Data reduction and analysis}
\label{section:data_reduction}

As mentioned in \citet{Proxaufetal2018}, the spectra from UVES and HARPS (Phase 3) were already pre-reduced (i.e., reduced up to the wavelength calibration step) by their own pipeline. As previously done, the FEROS spectra were reduced using a modified version of the Data Reduction System (DRS) pipeline. A few tens of FEROS spectra collected before 2004, which initially could not be reduced, were later on reduced and delivered by one of the Phase 3 Data Releases (DR3 and DR3.1). However, they are all low-quality spectra that were not included in our analysis. 

After the pre-reduction steps, all the selected spectra were normalized to the continuum using the Image Reduction and Analysis Facility (IRAF\,\footnote{The IRAF package is distributed by the National Optical Astronomy Observatories (NOAO), USA.}) by fitting cubic spline functions to a set of continuum windows. An exception is the case of the STELLA spectra, which were reduced by a dedicated pipeline \citep{Weberetal2012} based on IRAF routines and  that includes several reduction steps up to the continuum normalization.

For more details about the quality of the spectra we refer the reader to Sect.~3 of \citet{Proxaufetal2018}, which also includes a figure (their Fig.~1) showing a few examples of HARPS, FEROS, UVES, and STELLA spectra of different metallicities and S/N. In our Fig.~\ref{spectra}, we plot an example of HARPS spectra for \object{$\beta$\,Dor} and of STELLA spectra for \object{$\zeta$\,Gem} in different pulsation phases, showing how the profile of the absorption lines may change (in depth and in shape) along the pulsation cycle. Studies of line asymmetry changes over the pulsation cycle of Cepheid stars were also reported by \citet{Nardettoetal2006} and \citet{Nardettoetal2008}, in a series of papers on high-resolution spectroscopy, for \ion{Fe}{i} and H$\alpha$ lines in the optical and, more recently, by \citet{Nardettoetal2018} for the \ion{Na}{i} line at 2208.969~nm.

\subsection{Line lists and equivalent widths}
\label{section:linelists_ew}

Similarly to what we did in our previous works, here we use three different line lists:

\begin{enumerate}

\item[a)] one containing atomic lines of N, Si, S, Ca, Ti, V, Cr, Mn, Fe, Co, and Ni, created from a combination of four line lists received from V. Kovtyukh (for a total of 153 atomic lines). These lines are used to derive the effective temperature (\teff) of our sample stars according to the procedure described in Sect.~4.2 of \citet{Proxaufetal2018};

\item[b)] one containing 424 \ion{Fe}{i} and 97 \ion{Fe}{ii} lines, created from a combination of lines from \citet{Genovalietal2013} with lines from the Gaia-ESO Survey \citep[GES,][]{Gilmoreetal2012,Randichetal2013}. It is used as a reference line list in the determination of both the stellar surface gravity (\logg), as described in Sect.~\ref{section:teff_logg_vmic}, and the stellar metallicity, as discussed in Sect.~\ref{section:ab_iron_alpha};

\item[c)] one containing lines of $\alpha$ elements (Mg: 9 lines; Si: 11 lines; S: 1 line; Ca: 36 lines; and Ti: 72 lines), based on line lists from \citet{ForSneden2010}, \citet{Vennetal2012}, \citet{Lemasleetal2013}, and \citet[][and references therein]{McWilliametal2013}.

\end{enumerate}

\begin{table}
\centering
\caption{Excerpt from the list of \ion{Fe}{i} and \ion{Fe}{ii} atomic lines.}
\label{table:linelist_iron}
\begin{tabular}{c c c c c c}
\noalign{\smallskip}\hline\hline\noalign{\smallskip}
$\lambda$ [\AA] &
Species &
$\xi$ [eV] &
$\log{gf}$ &
\parbox[c]{1.1cm}{\centering Quality flag} &
Ref. \\
\noalign{\smallskip}\hline\noalign{\smallskip}
3763.789 & \ion{Fe}{i}  & 0.989 & $-$0.220 & 1   & 1 \\
3787.880 & \ion{Fe}{i}  & 1.010 & $-$0.840 & 1   & 1 \\
...      & ...          & ...   & ...      & ... &   \\
7655.488 & \ion{Fe}{ii} & 3.892 & $-$3.560 & 1   & 2 \\
7711.724 & \ion{Fe}{ii} & 3.903 & $-$2.500 & 1   & 2 \\
\hline
\end{tabular}
\tablefoot{The quality flag column indicates if the line was used (1) or not used (0) in the current study. The complete table is available at the CDS.
\tablebib{
1: \citet{OBrianetal1991};
2: \citet{MelendezBarbuy2009};
3: NIST: \citet{Kramidaetal2020}.
}}
\end{table}

\begin{table}
\centering
\caption{The same as in Table~\ref{table:linelist_iron} but showing an excerpt of atomic lines from the list of $\alpha$ elements.}
\label{table:linelist_alpha}
\begin{tabular}{c c c c c c}
\noalign{\smallskip}\hline\hline\noalign{\smallskip}
$\lambda$ [\AA] &
Species &
$\xi$ [eV] &
$\log{gf}$ &
\parbox[c]{1.1cm}{\centering Quality flag} &
Ref. \\
\noalign{\smallskip}\hline\noalign{\smallskip}
3829.355 & \ion{Mg}{i}  & 2.709 & $-$0.227 & 0   & 3 \\
4571.096 & \ion{Mg}{i}  & 0.000 & $-$5.620 & 1   & 3 \\
...      & ...          & ...   & ...      & ... &   \\
6606.956 & \ion{Ti}{ii} & 2.061 & $-$2.790 & 1   & 3 \\
7214.729 & \ion{Ti}{ii} & 2.590 & $-$1.750 & 1   & 3 \\
\hline
\end{tabular}
\end{table}

The Cepheids variables are bright stars and thus we have at our disposal very high S/N, broad wavelength range spectra. The precision of the atmospheric parameters and of the chemical abundances, however, depends not only on the quality of the spectra, but also on the quality of the adopted line list. With this in mind, a large portion of the present work was dedicated to finding the best atomic transitions for iron and $\alpha$ elements that can be detected in Cepheids. In practice, this means $i)$ collecting the most precise transition parameters, $ii)$ removing absorption lines that are blended with other lines, and $iii)$ eliminating lines that for any reason deviate significantly from the average value for their chemical species, or that display a dependency on any quantity that changes across the pulsation cycle, such as the effective temperature.

In order to address the first point, we departed from a list of hundreds of atomic transitions for iron and $\alpha$ elements commonly employed in the literature. We updated their transition parameters whenever possible with the updated laboratory measurements from \citet{Ruffonietal2014}, \citet{DenHartogetal2014}, and \citet{Belmonteetal2017} for \ion{Fe}{i}, \citet{DenHartogetal2019} for \ion{Fe}{ii}, \citet{Lawleretal2013} for \ion{Ti}{i}, and \citet{Woodetal2013}  for \ion{Ti}{ii}. We also made ample use of the astrophysical, but homogeneous and precise compilation of \citet{MelendezBarbuy2009} for \ion{Fe}{ii}. For the remaining lines, we adopted the transition parameters collected and updated by the National Institute of Standards and Technology (NIST) Atomic Spectra Database \citep{Kramidaetal2020}. If a line is not available in any of these sources, it is eliminated from the preliminary list and not used for the computation of atmospheric parameters nor chemical abundances.

With this preliminary list at hand, we addressed the second point mentioned above by removing any blended lines, using as reference the Solar spectrum table by \citet{Mooreetal1966} and synthetic spectra. Then we measured their equivalent width (EW) using the Automatic Routine for line Equivalent widths in stellar Spectra \citep[ARES,][]{Sousaetal2007,Sousaetal2015}.

\begin{figure}
\centering
\includegraphics[width=8.5cm]{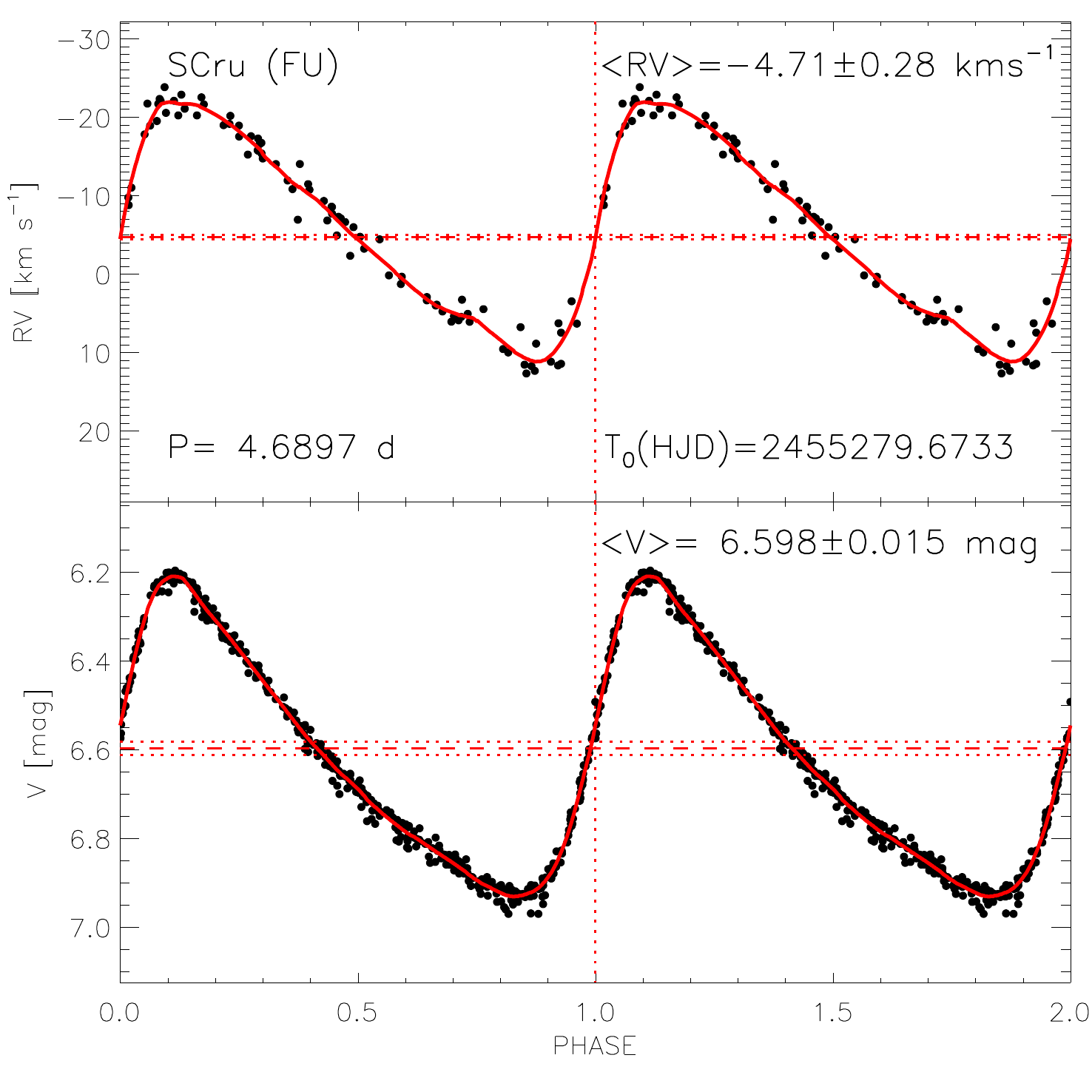}
\caption{Top: RV curve of the fundamental pulsator S Cru. The red solid line displays the PLOESS fit to the RV curve. The horizontal red dashed and dotted lines display the mean RV ($\langle$RV$\rangle$) and its uncertainty. The vertical red dotted line displays the phase where RV equals $\langle$RV$\rangle$ on the rising branch. The name, $\langle$RV$\rangle$, period and reference epoch are labelled. Bottom: same as in the top panel but for the $V$-band light curve, its PLOESS fit and its mean value. We note that, within the uncertainty, the light curve is co-phased with the RV curve.}
\label{figure:tmeanrise}
\end{figure}

Finally, we addressed the third point by removing deviant lines. In a first iteration, we estimated initial values for the atmospheric parameters and the chemical abundances. Lines are considered deviant if they result in abundances diverging by at least 3\,$\sigma$ from the mean for a given element, or if their behavior across the pulsation cycle is irregular. This latter situation is investigated by plotting all line-by-line measurements for each chemical species and individual exposure of our calibrating stars versus phase, effective temperature, and equivalent width, and removing lines that show strong trends in any of those planes. The calibrating stars mostly used are those with the largest number of individual spectra covering the whole pulsation cycle, such as \object{$\beta$\,Dor}, \object{$\zeta$\,Gem}, and \object{FF\,Aql}.

The final clean lists, shown in Tables~\ref{table:linelist_iron} and \ref{table:linelist_alpha} with quality flag 1, were used to compute the final atmospheric parameters and chemical abundances for the whole sample. The remaining lines, shown in the same tables with quality flag 0, were measured only for comparison purposes.

An example of a good Fe line is shown in Fig.~\ref{figure:good_lines} for \object{$\beta$\,Dor} and \object{$\zeta$\,Gem}. Examples and a description of the typical behavior of irregular (less reliable) lines are found in the Appendix~\ref{appendix:bad_lines}, in the plots of Figs.~\ref{figure:bad_lines_blends}-\ref{figure:bad_lines_loggf}. These same irregular lines, however, might provide good results for other stars, for instance, of different spectral types, or having spectra of different quality or different resolution. This is the reason why we preferred to exclude some lines only for this particular analysis, and keep the initial line lists as they are to be used in future studies. It is worth mentioning that we only excluded lines that were clearly having problems in most of the stars in our sample.

\section{Radial velocity and photometric data}
\label{section:rv_photometry}

The radial velocity (RV) measurements for our sample were done using the following methods: $i)$ for HARPS and UVES spectra, we used IRAF packages to cross-correlate the target spectrum with a solar template spectrum \citep[Solar flux atlas from 296 to 1300 nm,][]{Kuruczetal1984} degraded to the UVES resolution;
$ii)$ for FEROS spectra we adopted the RVs derived by ARES, the routine we used for the equivalent width measurements (see Sect.~\ref{section:linelists_ew});
$iii)$ the STELLA radial velocities are written in the header of the FITS files and they are based on cross-correlations with a template spectrum, performed by a dedicated reduction pipeline also using IRAF. In case of the Cepheids, a G-type dwarf template is used.

\begin{figure}
\centering
\resizebox{\hsize}{!}{\includegraphics{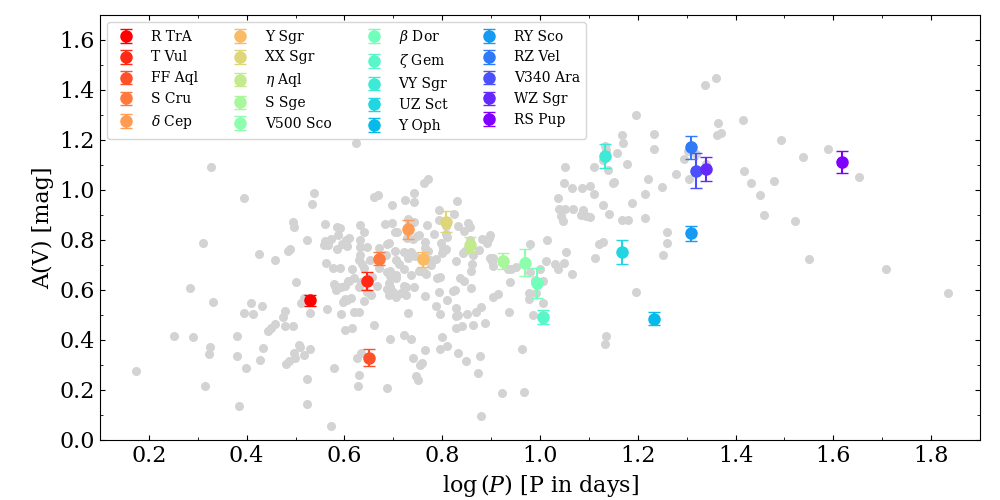}}
\caption{Photometric amplitude as a function of the logarithmic period (Bailey diagram). Our calibrating Cepheids are represented by colored symbols whereas the Cepheids from literature \citep{Lucketal2011,LuckLambert2011,Lemasleetal2013,Yongetal2006} are shown in light gray circles.}
\label{figure:bailey_diagram}
\end{figure}

\begin{table*}
\centering
\caption{Excerpt from the list of atmospheric parameters, Fe abundances, and radial velocities for each spectrum of the 20 calibrating Cepheids.}
\label{table:atm_params_rv_spectra}
{\scriptsize
\begin{tabular}{lcc r@{ }l cc r@{ }l c r@{ }l c r@{ }l}
\noalign{\smallskip}\hline\hline\noalign{\smallskip}
Name & Dataset &
\parbox[c]{0.5cm}{\centering MJD [d]} &
\multicolumn{2}{c}{\parbox[c]{0.8cm}{\centering \teff\ $\pm$ $\sigma$ [K]}} &
\logg\ &
\parbox[c]{0.9cm}{\centering \vmic\ [\kms]} &
\multicolumn{2}{c}{\ion{Fe}{i}  $\pm$ $\sigma$} & $N_{\rm{\ion{Fe}{i}}}$ &
\multicolumn{2}{c}{\ion{Fe}{ii} $\pm$ $\sigma$} & $N_{\rm{\ion{Fe}{ii}}}$ &
\multicolumn{2}{c}{\parbox[c]{0.9cm}{\centering $RV$ $\pm$ $\sigma$ [\kms]}} \\
\noalign{\smallskip}\hline\noalign{\smallskip}
\object{R\,TrA}  & HARPS  & 53150.1353086 & 5918 & $\pm$ 154 & 2.3 & 3.5 & $-$0.02 & $\pm$ 0.09 &  86 & $-$0.03 & $\pm$ 0.03 &   7 &  2.57 & $\pm$ 1.38 \\
\object{R\,TrA}  & HARPS  & 53150.1453265 & 5892 & $\pm$ 128 & 2.3 & 3.6 & $-$0.04 & $\pm$ 0.09 &  81 & $-$0.05 & $\pm$ 0.02 &   8 &  2.40 & $\pm$ 1.37 \\
...              & ...    & ...           &      & ...       & ... & ... &         & ...        & ... &         & ...        & ... &       & ...        \\
\object{RS\,Pup} & STELLA & 57708.2167938 & 5508 & $\pm$  98 & 0.9 & 3.4 &    0.12 & $\pm$ 0.11 & 107 &    0.12 & $\pm$ 0.15 &   7 &  7.40 & $\pm$ 0.20 \\
\object{RS\,Pup} & STELLA & 57713.2628277 & 5298 & $\pm$ 111 & ... & ... &         & ...        & ... &         & ...        & ... & 16.25 & $\pm$ 0.20 \\
\hline
\end{tabular}}
\tablefoot{The first three columns give the target name, spectroscopic dataset, and Modified Julian Date at which the spectrum was collected. Columns~4, 5, and 6 give, respectively, the effective temperature and its standard deviation, surface gravity, and microturbulent velocity. The Cols.~7-8 and 9-10 list both the \ion{Fe}{i} and \ion{Fe}{ii} abundances derived from individual lines together with the standard deviations and the number of lines used (for the \ion{Fe}{ii} abundances, a minimal uncertainty of 0.11~dex was adopt when computing the mean abundances, as described in the text). The last column gives the radial velocities with their uncertainties. The complete table is available at the CDS.}
\end{table*}

\begin{table*}
\caption{Photometric and radial-velocity parameters for our sample of 20 calibrating Cepheids.}
\label{table:physical_params}
\centering
{\small
\begin{tabular}{l r@{ }l r@{ }l r@{ }l r@{ }l c r@{}l c c}
\noalign{\smallskip}\hline\hline\noalign{\smallskip}
Name &
\multicolumn{2}{c}{\parbox[c]{1.0cm}{\centering V [mag]}} &
\multicolumn{2}{c}{\parbox[c]{0.8cm}{\centering $\Delta$V [mag]}} &
\multicolumn{2}{c}{\parbox[c]{1.0cm}{\centering $\gamma$ [km/s]}} &
\multicolumn{2}{c}{\parbox[c]{0.9cm}{\centering $\Delta$RV [km/s]}} &
\parbox[c]{1.0cm}{\centering LC source\tablefootmark{a}} &
\multicolumn{2}{c}{\parbox[c]{0.9cm}{\centering Period [days]}} &
\parbox[c]{1.8cm}{\centering $T_0 - 2\,400\,000$ [days]} &
\parbox[c]{1.7cm}{\centering Ephemerides source\tablefootmark{b}} \\
\noalign{\smallskip}\hline\noalign{\smallskip}
\object{R\,TrA}       &  6.66 & $\pm$ 0.02 & 0.56 & $\pm$ 0.02 & --12.44 & $\pm$ 2.54  & 28.1  & $\pm$ 2.2 & 1 &  3&.3892582  & 44\,423.59377 & 1 \\
\object{T\,Vul}       &  5.75 & $\pm$ 0.03 & 0.64 & $\pm$ 0.04 &    0.85 & $\pm$ 0.11  & 34.5  & $\pm$ 3.5 & 1 &  4&.435462   & 57\,632.18950 & 0 \\
                      &       &            &      &            &         &             &       &           &   &   &          & 50\,735.20310 & 2 \\
\object{FF\,Aql}      &  5.19 & $\pm$ 0.01 & 0.33 & $\pm$ 0.03 & --11.36 & $\pm$ 0.10  & 16.1  & $\pm$ 1.5 & 2 &  4&.470916   & 56\,888.66207 & 0 \\
\object{S\,Cru}       &  6.60 & $\pm$ 0.02 & 0.73 & $\pm$ 0.03 &  --4.70 & $\pm$ 0.28  & 33.8  & $\pm$ 4.3 & 1 &  4&.68973    & 55\,279.67331 & 0 \\
\object{$\delta$\,Cep}&  3.94 & $\pm$ 0.01 & 0.84 & $\pm$ 0.04 & --16.32 & $\pm$ 0.20  & 38.9  & $\pm$ 4.9 & 1 &  5&.3663     & 56\,917.54900 & 0 \\
\object{Y\,Sgr}       &  5.74 & $\pm$ 0.02 & 0.72 & $\pm$ 0.03 &  --2.84 & $\pm$ 3.02  & 34.2  & $\pm$ 3.0 & 1 &  5&.7733866  & 54\,946.78504 & 1 \\
\object{XX\,Sgr}      &  8.86 & $\pm$ 0.02 & 0.87 & $\pm$ 0.04 &   12.30 & $\pm$ 0.46  & 41.6  & $\pm$ 1.2 & 1 &  6&.4243013  & 54\,946.67631 & 1 \\
\object{$\eta$\,Aql}  &  3.89 & $\pm$ 0.02 & 0.78 & $\pm$ 0.03 & --14.47 & $\pm$ 0.63  & 42.1  & $\pm$ 2.2 & 1 &  7&.1768572  & 50\,724.43545 & 1 \\
\object{S\,Sge}       &  5.61 & $\pm$ 0.03 & 0.72 & $\pm$ 0.03 & --16.40 & $\pm$ 0.25  & 39.3  & $\pm$ 4.7 & 1 &  8&.3445707  & 57\,636.28926 & 0 \\
                      &       &            &      &            &         &             &       &           &   &   &          & 48\,746.52000 & 3 \\
                      &       &            &      &            &         &             &       &           &   &  8&.3823514  & 50\,339.17710 & 4 \\
\object{V500\,Sco}    &  8.74 & $\pm$ 0.01 & 0.71 & $\pm$ 0.05 &  --7.61 & $\pm$ 3.37  & 32.2  & $\pm$ 5.4 & 1 &  9&.317      & 44\,789.61621 & 1 \\
\object{$\beta$\,Dor} &  5.25 & $\pm$ 0.01 & 0.63 & $\pm$ 0.06 &    8.17 & $\pm$ 0.11  & 32.9  & $\pm$ 2.3 & 2 &  9&.84308    & 53\,284.79472 & 0 \\
                      &       &            &      &            &         &             &       &           &   &  9&.8527371  & 41\,149.27150 & 4 \\
\object{$\zeta$\,Gem} &  3.89 & $\pm$ 0.03 & 0.49 & $\pm$ 0.03 &    6.78 & $\pm$ 0.42  & 26.5  & $\pm$ 1.0 & 1 & 10&.149816   & 54\,805.78476 & 1 \\
\object{VY\,Sgr}      & 11.46 & $\pm$ 0.02 & 1.14 & $\pm$ 0.05 &   14.23 & $\pm$ 1.58  & 50.4  & $\pm$ 4.4 & 1 & 13&.55845    & 54\,171.32722 & 1 \\
\object{UZ\,Sct}      & 11.25 & $\pm$ 0.03 & 0.75 & $\pm$ 0.05 &   39.52 & $\pm$ 0.18  & 44.7  & $\pm$ 3.5 & 0 & 14&.744162   & 56\,171.62381 & 0 \\
\object{Y\,Oph}       &  6.15 & $\pm$ 0.02 & 0.49 & $\pm$ 0.03 &  --7.56 & $\pm$ 2.01  & 16.5  & $\pm$ 2.0 & 1 & 17&.12633896 & 53\,209.96732 & 1 \\
\object{RY\,Sco}      &  8.02 & $\pm$ 0.02 & 0.83 & $\pm$ 0.03 & --17.73 & $\pm$ 2.02  & 30.1  & $\pm$ 3.0 & 1 & 20&.321538   & 45\,077.36644 & 1 \\
\object{RZ\,Vel}      &  7.09 & $\pm$ 0.04 & 1.17 & $\pm$ 0.05 &   24.11 & $\pm$ 4.52  & 54.6  & $\pm$ 5.4 & 1 & 20&.399727   & 50\,776.13608 & 1 \\
\object{V340\,Ara}    & 10.20 & $\pm$ 0.06 & 1.08 & $\pm$ 0.07 & --76.25 & $\pm$ 4.19  & 47.7  & $\pm$ 9.3 & 1 & 20&.811386   & 56\,138.97597 & 0 \\
\object{WZ\,Sgr}      &  8.03 & $\pm$ 0.03 & 1.08 & $\pm$ 0.05 & --17.78 & $\pm$ 1.17  & 54.2  & $\pm$ 2.9 & 1 & 21&.849708   & 50\,691.28534 & 1 \\
\object{RS\,Pup}      &  6.99 & $\pm$ 0.03 & 1.11 & $\pm$ 0.04 &   25.80 & $\pm$ 1.91  & 48.5  & $\pm$ 2.1 & 1 & 41&.488634   & 53\,092.94483 & 1 \\
\hline
\end{tabular}
}
\tablefoot{From left to right the columns give the star name, the mean magnitude in visual bands, the magnitude amplitude, the mean radial velocity, the RV amplitude, the light curve source, the pulsation period and zero-phase reference epoch of mean magnitude (or mean radial velocity), and the Ephemerides source. For the variables \object{VY\,Sgr}, \object{RZ\,Vel}, and \object{WZ\,Sgr}, our RV measurements and those by \citet[][and references therein]{Groenewegen2008} show a shift in phase of $-$0.05, +0.03, and $-$0.03, respectively. To co-phase the two datasets, the latter sample of radial velocities plotted in Fig.~\ref{figure:mag_rv_phase} were shifted.
\tablebib{
\tablefoottext{a}{
0: ASAS-SN \citep{Shappeeetal2014,Kochaneketal2017}; 
1: \citet[][and references therein]{Groenewegen2008};
2: \citet[][]{Pel1976}.}
\tablefoottext{b}{
0: Period and $T_0$ derived from our own radial velocity curves; 
1: Period and $T_0$ derived from the radial velocities by \citet[][and references therein]{Groenewegen2008};
2: $T_0$ derived from the light curve;
3: $T_0$ derived from the radial velocities by \citet[][and references therein]{Groenewegen2008};
4: Period and $T_0$ derived from the light curve.}
}}
\end{table*}

The uncertainties adopted for these RV estimates depend on the method used: for HARPS, UVES, and STELLA spectra, the uncertainties are estimated by IRAF during the cross-correlation; for FEROS spectra, they are estimated by ARES using spectral windows free of lines. Typical values for our HARPS, UVES, and FEROS spectra range from a few m/s in the best cases to a few \kms, depending on the spectral type. For the STELLA spectra, we adopted an uncertainty of 0.2~\kms. This is a conservative estimate considering that typical formal errors for high S/N spectra of Cepheids stars is about 50~\ms\ and that the RV zero point for this spectrograph has RMS variations of 30 to 150~\ms\ over two years. The RV estimates and the corresponding uncertainties for our sample of calibrating Cepheids are listed in Table~\ref{table:atm_params_rv_spectra}.

In addition to our own RV measurements, we adopted literature RV curves and light curves to provide accurate ephemerides, meaning reference epoch ($T_0$) plus pulsation period ($P$) for these variables. More specifically, we collected RV curves and $V$-band light curves from \citet[][]{Pel1976} and \citet[][and references therein]{Groenewegen2008}. For \object{UZ\,Sct}, we adopted the $V$-band light curve from the All-Sky Automated Survey for Supernovae \citep[ASAS-SN,][]{Shappeeetal2014,Kochaneketal2017} because it is the only well-sampled light curve that can be co-phased with our RV time series.

\begin{table*}
\caption{Mean atmospheric parameters derived for the 20 calibrating Cepheids.}
\label{table:atm_params_stars}
\setlength{\tabcolsep}{3pt}
\centering
{\scriptsize
\begin{tabular}{l r@{ }l r@{ }l r@{ }l c r@{ }l r@{ }l r@{ }l r@{ }l r@{ }l c}
\noalign{\smallskip}\hline\hline\noalign{\smallskip}
Name &
\multicolumn{2}{c}{\parbox[c]{1.0cm}{\centering $\langle$\teff$\rangle$ $\pm$ $\sigma$ [K]}} &
\multicolumn{2}{c}{\parbox[c]{1.2cm}{\centering $\langle\theta\rangle\pm\sigma$}} &
\multicolumn{2}{c}{\parbox[c]{0.9cm}{\centering $\Delta\theta\pm\sigma$}} &
\parbox[c]{1.0cm}{\centering $N_{\rm spec}$ (\teff)} &
\multicolumn{2}{c}{\parbox[c]{1.1cm}{\centering $\langle$\logg$\rangle$ $\pm$ $\sigma$}} &
\multicolumn{2}{c}{\parbox[c]{1.2cm}{\centering $\langle$\vmic$\rangle$ $\pm$ $\sigma$ [\kms]}} &
\multicolumn{2}{c}{\parbox[c]{1.2cm}{\centering [\ion{Fe}{i}/H] $\pm$ $\sigma$}} &
\multicolumn{2}{c}{\parbox[c]{1.3cm}{\centering [\ion{Fe}{ii}/H] $\pm$ $\sigma$}} &
\multicolumn{2}{c}{\parbox[c]{1.6cm}{\centering [Fe/H] $\pm$ $\sigma$ (std)}} &
$N_{\rm spec}$ \\
\noalign{\smallskip}\hline\noalign{\smallskip}
\object{R\,TrA}        & 6035 & $\pm$ 25 &  0.840 & $\pm$ 0.010 & 0.098 & $\pm$ 0.016 &  15 & 2.01 & $\pm$ 0.08 & 3.23 & $\pm$ 0.13 & $-$0.03 & $\pm$ 0.02 & $-$0.03 & $\pm$ 0.03 & $-$0.03 & $\pm$ 0.02 (0.02) &  15 \\
\object{T\,Vul}        & 5934 & $\pm$ 20 &  0.855 & $\pm$ 0.010 & 0.122 & $\pm$ 0.024 &  26 & 1.43 & $\pm$ 0.06 & 3.07 & $\pm$ 0.10 & $-$0.04 & $\pm$ 0.02 & $-$0.03 & $\pm$ 0.02 & $-$0.04 & $\pm$ 0.02 (0.05) &  26 \\
\object{FF\,Aql}       & 6182 & $\pm$ 18 &  0.820 & $\pm$ 0.010 & 0.058 & $\pm$ 0.012 &  27 & 1.35 & $\pm$ 0.06 & 3.03 & $\pm$ 0.10 &    0.05 & $\pm$ 0.02 &    0.05 & $\pm$ 0.03 &    0.05 & $\pm$ 0.02 (0.05) &  27 \\
\object{S\,Cru}        & 6018 & $\pm$ 20 &  0.862 & $\pm$ 0.010 & 0.140 & $\pm$ 0.028 &  13 & 1.57 & $\pm$ 0.08 & 3.01 & $\pm$ 0.14 & $-$0.03 & $\pm$ 0.02 & $-$0.01 & $\pm$ 0.03 & $-$0.02 & $\pm$ 0.02 (0.07) &  13 \\
\object{$\delta$\,Cep} & 5905 & $\pm$ 22 &  0.862 & $\pm$ 0.010 & 0.154 & $\pm$ 0.022 &  18 & 1.42 & $\pm$ 0.07 & 3.08 & $\pm$ 0.12 &    0.06 & $\pm$ 0.02 &    0.05 & $\pm$ 0.03 &    0.05 & $\pm$ 0.02 (0.05) &  18 \\
\object{Y\,Sgr}        & 5914 & $\pm$ 25 &  0.877 & $\pm$ 0.020 & 0.128 & $\pm$ 0.027 &  24 & 1.58 & $\pm$ 0.06 & 3.48 & $\pm$ 0.10 &    0.03 & $\pm$ 0.02 &    0.02 & $\pm$ 0.03 &    0.02 & $\pm$ 0.02 (0.08) &  23 \\
\object{XX\,Sgr}       & 5884 & $\pm$ 29 &  0.855 & $\pm$ 0.010 & 0.151 & $\pm$ 0.031 &  12 & 1.25 & $\pm$ 0.09 & 2.92 & $\pm$ 0.14 & $-$0.01 & $\pm$ 0.03 & $-$0.01 & $\pm$ 0.04 & $-$0.01 & $\pm$ 0.02 (0.05) &  12 \\
\object{$\eta$\,Aql}   & 5480 & $\pm$ 40 &  0.891 & $\pm$ 0.030 & 0.113 & $\pm$ 0.035 &  11 & 1.16 & $\pm$ 0.09 & 3.43 & $\pm$ 0.15 &    0.09 & $\pm$ 0.04 &    0.08 & $\pm$ 0.04 &    0.09 & $\pm$ 0.03 (0.10) &  11 \\
\object{S\,Sge}        & 5777 & $\pm$ 21 &  0.883 & $\pm$ 0.012 & 0.134 & $\pm$ 0.019 &  21 & 1.20 & $\pm$ 0.07 & 3.13 & $\pm$ 0.11 &    0.10 & $\pm$ 0.02 &    0.09 & $\pm$ 0.03 &    0.09 & $\pm$ 0.02 (0.05) &  21 \\
\object{V500\,Sco}     & 5797 & $\pm$ 32 &  0.850 & $\pm$ 0.053 &       &             &   7 & 1.21 & $\pm$ 0.11 & 3.04 & $\pm$ 0.19 & $-$0.03 & $\pm$ 0.04 & $-$0.03 & $\pm$ 0.04 & $-$0.03 & $\pm$ 0.03 (0.02) &   7 \\
\object{$\beta$\,Dor}  & 5552 & $\pm$ 13 &  0.903 & $\pm$ 0.020 & 0.132 & $\pm$ 0.023 &  47 & 1.15 & $\pm$ 0.04 & 3.31 & $\pm$ 0.07 & $-$0.07 & $\pm$ 0.01 & $-$0.08 & $\pm$ 0.02 & $-$0.08 & $\pm$ 0.01 (0.04) &  47 \\
\object{$\zeta$\,Gem}  & 5500 & $\pm$  8 &  0.914 & $\pm$ 0.020 & 0.100 & $\pm$ 0.011 & 128 & 1.14 & $\pm$ 0.03 & 3.32 & $\pm$ 0.04 &    0.02 & $\pm$ 0.01 &    0.01 & $\pm$ 0.01 &    0.01 & $\pm$ 0.01 (0.06) & 128 \\
\object{VY\,Sgr}       & 5343 & $\pm$ 27 &  0.954 & $\pm$ 0.020 & 0.214 & $\pm$ 0.048 &  24 & 1.09 & $\pm$ 0.08 & 3.38 & $\pm$ 0.13 &    0.19 & $\pm$ 0.04 &    0.18 & $\pm$ 0.04 &    0.19 & $\pm$ 0.03 (0.11) &  14 \\
\object{UZ\,Sct}       & 5113 & $\pm$ 25 &  0.979 & $\pm$ 0.045 &       &             &  25 & 0.87 & $\pm$ 0.09 & 3.35 & $\pm$ 0.16 &    0.14 & $\pm$ 0.04 &    0.10 & $\pm$ 0.05 &    0.12 & $\pm$ 0.03 (0.07) &  10 \\
\object{Y\,Oph}        & 5609 & $\pm$ 33 &  0.900 & $\pm$ 0.013 &       &             &   8 & 1.20 & $\pm$ 0.11 & 3.79 & $\pm$ 0.18 & $-$0.05 & $\pm$ 0.04 & $-$0.05 & $\pm$ 0.05 & $-$0.05 & $\pm$ 0.03 (0.02) &   8 \\
\object{RY\,Sco}       & 5743 & $\pm$ 36 &  0.886 & $\pm$ 0.071 &       &             &   8 & 0.99 & $\pm$ 0.11 & 3.69 & $\pm$ 0.18 &    0.05 & $\pm$ 0.04 &    0.04 & $\pm$ 0.04 &    0.04 & $\pm$ 0.03 (0.07) &   8 \\
\object{RZ\,Vel}       & 5483 & $\pm$ 29 &  0.919 & $\pm$ 0.020 & 0.232 & $\pm$ 0.037 &  12 & 1.02 & $\pm$ 0.09 & 4.35 & $\pm$ 0.15 &    0.04 & $\pm$ 0.02 &    0.03 & $\pm$ 0.05 &    0.03 & $\pm$ 0.02 (0.11) &  11 \\
\object{V340\,Ara}     & 5170 & $\pm$ 26 &  0.933 & $\pm$ 0.057 &       &             &  32 & 1.10 & $\pm$ 0.09 & 4.60 & $\pm$ 0.16 &    0.12 & $\pm$ 0.04 &    0.08 & $\pm$ 0.05 &    0.10 & $\pm$ 0.03 (0.09) &  10 \\
\object{WZ\,Sgr}       & 5411 & $\pm$ 33 &  0.924 & $\pm$ 0.067 &       &             &   9 & 0.89 & $\pm$ 0.11 & 3.77 & $\pm$ 0.19 &    0.13 & $\pm$ 0.04 &    0.11 & $\pm$ 0.05 &    0.12 & $\pm$ 0.03 (0.07) &   7 \\
\object{RS\,Pup}       & 5432 & $\pm$ 25 &  0.901 & $\pm$ 0.020 & 0.195 & $\pm$ 0.032 &  18 & 0.81 & $\pm$ 0.07 & 3.74 & $\pm$ 0.12 &    0.07 & $\pm$ 0.02 &    0.09 & $\pm$ 0.03 &    0.08 & $\pm$ 0.02 (0.06) &  16 \\
\hline
\end{tabular}
\tablefoot{Columns from 1 to 5 give, respectively, the star name, the mean effective temperature and mean $\theta$, the $\theta$ amplitude (see Section~\ref{sec:teff_templates}), and the number of spectra used to compute these mean values. Columns from 6 to 9 lists the surface gravity, microturbulent velocity, and iron abundances from neutral and ionized lines. These are the weighted mean values and their standard errors computed from the individual measurements listed in Table~\ref{table:atm_params_rv_spectra}. Column~10 gives the weighted iron abundance and its standard error, calculated from Cols.~8 and 9. The standard deviation calculated using all individual measurements from both \ion{Fe}{i} and \ion{Fe}{ii} lines is also shown within parentheses. The last column gives the number of spectra used to derive \logg, \vmic, and the iron abundances.}}
\end{table*}

We estimated the pulsation periods of our targets by using our own interactive method \citep{Bragaetal2016} based on the Lomb-Scargle periodogram \citep{Scargle1982}. Historically, the most commonly used $T_0$ is the time of maximum light ($T_{\rm max}$), which matches the time of minimum velocity. Starting from \citet{Innoetal2015}, we have been promoting the use of another reference for both Cepheids and RR Lyrae, that is, the epoch of the mean magnitude on the rising branch of the light curve ($T_{\rm mean}^{\rm opt}$) or, equivalently, the epoch of mean velocity on the decreasing branch of the RV curve ($T_{\rm mean}^{\rm RV}$). See Fig.~\ref{figure:tmeanrise} for a visualization of this reference epoch on both the light and RV curves. A detailed procedure to derive $T_{\rm mean}^{\rm opt}$ is described in \citet{Bragaetal2021}. The advantages of the $T_{\rm mean}^{\rm opt}$ reference epoch over $T_{\rm max}$ are well-explained in \citet{Innoetal2015} and quantitatively discussed in  \citet{Bragaetal2021}.

The light and radial-velocity curves for all the sample stars are shown in the Appendix~\ref{appendix:variations_with_phase}, in Fig.~\ref{figure:mag_rv_phase}. The estimated ephemerides of our targets are listed in Table~\ref{table:physical_params} together with the mean magnitudes and magnitude amplitudes in the visual band, and the mean radial velocities and RV amplitudes. We noticed that for the stars \object{VY\,Sgr}, \object{RZ\,Vel}, and \object{WZ\,Sgr}, for which the new period and $T_0$ values are based on the radial velocities derived by \citet[][and references therein]{Groenewegen2008}, the new ephemerides caused a shift in phase between the two datasets shown in Fig.~\ref{figure:mag_rv_phase}. To overcome this problem, we applied a shift of $-$0.05, +0.03, and $-$0.03, respectively, to the literature data plotted in these panels.

Figure~\ref{figure:bailey_diagram} shows the distribution of the current calibrating Cepheids (colored symbols), together with Galactic Cepheids available in the literature (light gray circles), in the so-called Bailey diagram (luminosity amplitude as a function of the logarithmic period). The plotted data display the classical "V" shape in which the local minimum around ten days is mainly caused by the Hertzsprung progression \citep{Bonoetal2000a}. By having a look also at Fig.~\ref{figure:mag_rv_phase}, we see a subgroup of classical Cepheids showing a well defined bump on the light curve. The phase of such bumps moves from the decreasing branch for periods between $\sim$6 and $\sim$9~days, passing across the maximum for periods between $\sim$9 and $\sim$12~days, and along the rising branch for periods between $\sim$12 and $\sim$16~days. At the center of the Hertzsprung progression (P$\sim$10 days) the light curve is flat topped and the luminosity amplitude attains a well defined minimum \citep{Bonoetal2000a,Bonoetal2002}. These variables are called Bump Cepheids for avoiding to be mixed-up with the Beat Cepheids, that is, classical Cepheids pulsating simultaneously in two or more modes. The calibrating Cepheids cover both the short- and the long-period range, and in particular, the current sample includes seven Cepheids across the Hertzsprung progression.  

\section{Atmospheric parameters and effective temperature curve templates}
\label{section:atm_params}

\subsection{Effective temperature, surface gravity, and microturbulent velocity}
\label{section:teff_logg_vmic}

The approach we adopted to estimate the atmospheric parameters (i.e., the effective temperature, the surface gravity, and the microturbulent velocity) was already discussed in detail by \citet{Proxaufetal2018}. Here we only recap the key points. The \teff\ along the pulsation cycle was estimated by using the LDR method, which relies on the correlation between the line depth ratios of pairs of absorption lines in the spectra of different stars and the effective temperature of the same stars. The surface gravity was derived through the ionization equilibrium of \ion{Fe}{i} and \ion{Fe}{ii} lines, and the microturbulent velocity was obtained by minimizing the dependence of the abundances provided by single \ion{Fe}{i} lines on their EWs. During this procedure, the effective temperature was kept fixed, whereas \logg\ and $\xi$ were iteratively changed until convergence. The metallicity used as input by our algorithm is updated in each step, and the adopted value is [\ion{Fe}{i}/H], which is the mean iron abundances provided by individual \ion{Fe}{i} lines.

The estimates of \teff, \logg, and $\xi$ for the individual spectrum of our sample are listed in Table~\ref{table:atm_params_rv_spectra}. The uncertainties on \teff\ are the standard deviation calculated using the LDR method, that is, from the effective temperatures provided by the individual pairs of lines. The uncertainties on the individual estimates of \logg\ and $\xi$  are assumed to be of the order of 0.3~dex and 0.5~\kms, respectively \citep[see][for a detailed discussion]{Genovalietal2014}. The weighted mean values of the atmospheric parameters are listed in Table~\ref{table:atm_params_stars} together with the corresponding standard errors calculated for each calibrating Cepheid.

\begin{figure}
\centering
\begin{minipage}[t]{0.49\textwidth}
\centering
\resizebox{\hsize}{!}{\includegraphics{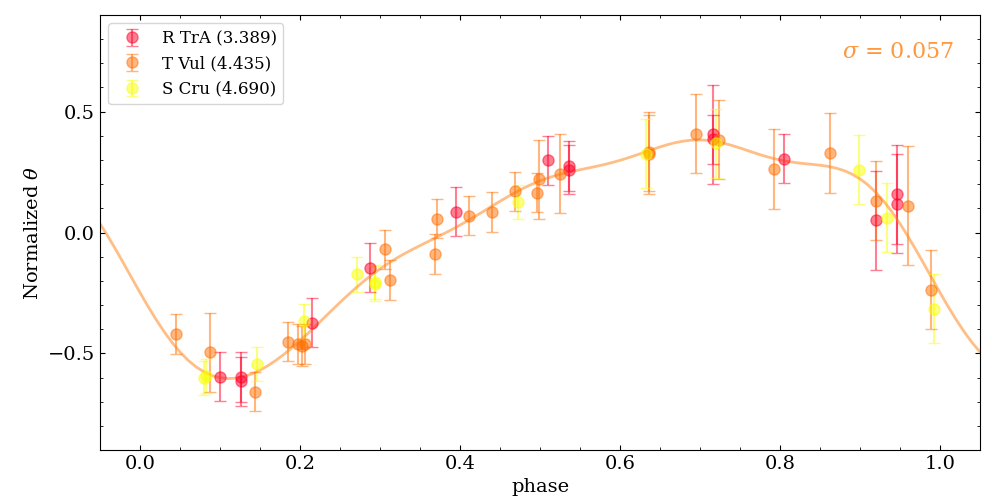}}
\end{minipage} \\
\begin{minipage}[t]{0.49\textwidth}
\centering
\resizebox{\hsize}{!}{\includegraphics{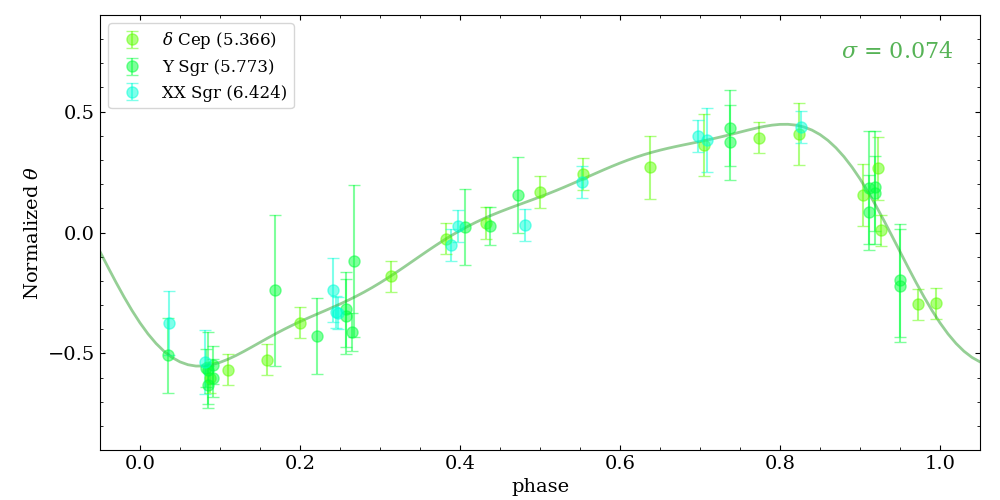}}
\end{minipage} \\
\begin{minipage}[t]{0.49\textwidth}
\centering
\resizebox{\hsize}{!}{\includegraphics{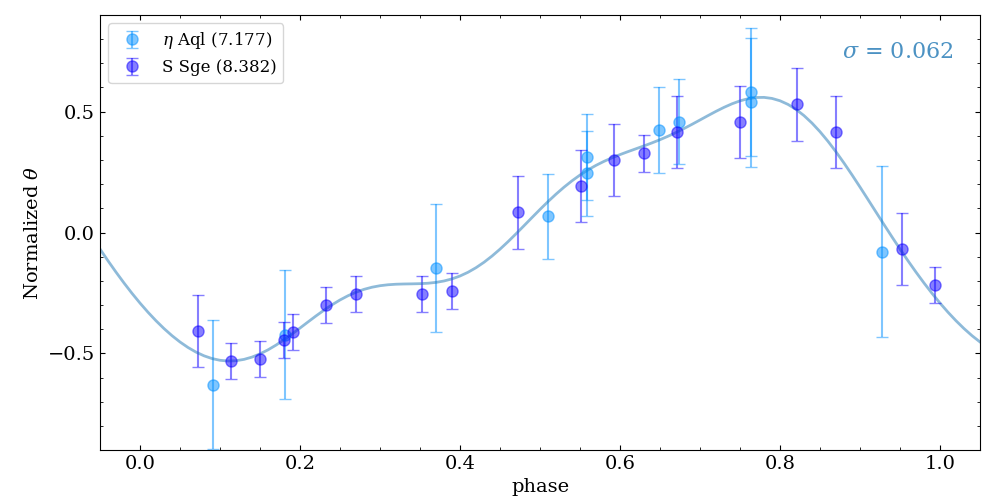}}
\end{minipage} \\
\begin{minipage}[t]{0.49\textwidth}
\centering
\resizebox{\hsize}{!}{\includegraphics{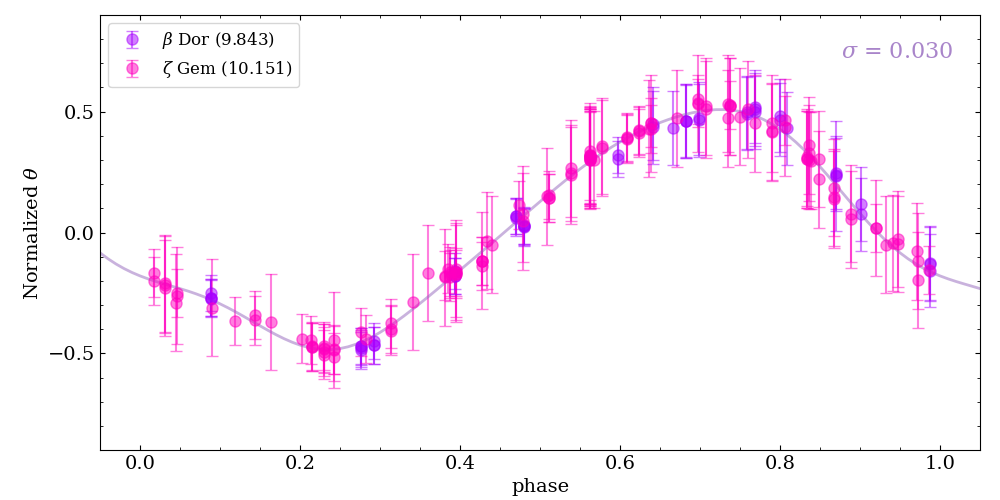}}
\end{minipage}
\caption{Normalized $\theta$ (5040/\teff) as a function of the pulsation phase. The panels show the phase-folded curves separating the Cepheids in different bins of pulsation period (values within parentheses): from 3 to 5~days, from 5 to 7~days, from 7 to 9.5~days, and from 9.5 to 10.5~days. For the variable \object{S\,Sge}, the two outliers shown in Fig.~\ref{figure:teff_phase} were not used in the construction of the \teff\ curve templates. The curves display the fit of the Fourier series to the data (see Table~\ref{table:templates}). The standard deviations of the fit are also shown.}
\label{figure:theta_phase}
\end{figure}

\begin{figure}
\centering
\resizebox{\hsize}{!}{\includegraphics{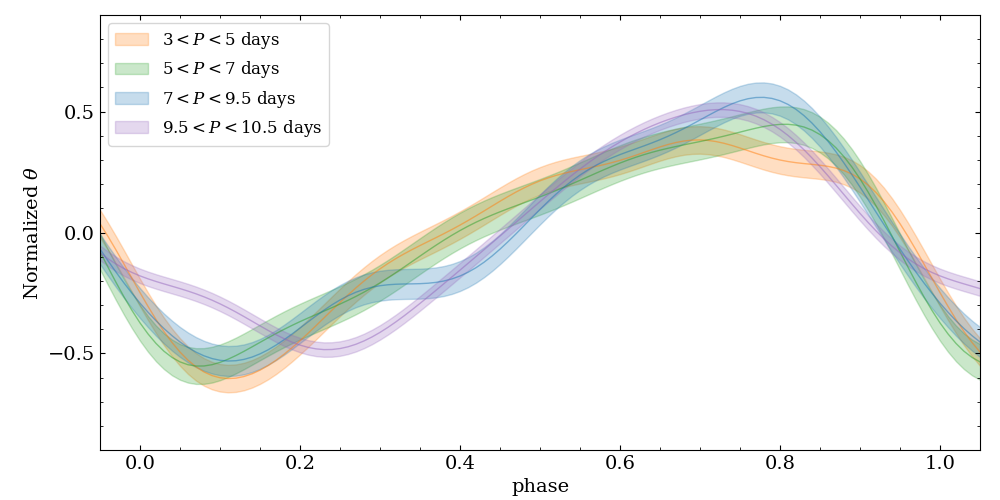}}
\caption{Normalized $\theta$ (5040/\teff) as a function of the pulsation phase showing the fits of the Fourier series to the data of Fig.~\ref{figure:theta_phase}. The shaded areas display the 1-$\sigma$ standard deviation.}
\label{figure:theta_phase_func}
\end{figure}

In the Appendix~\ref{appendix:variations_with_phase}, Figs.~\ref{figure:teff_phase}, \ref{figure:logg_phase}, and \ref{figure:vmic_phase} show the variation of the atmospheric parameters along the pulsation cycle. To provide a quantitative estimate of the differences between the current approach and our previous measurements, we fitted Fourier series to the effective temperature, surface gravity, and microturbulent velocity curves in order to evaluate their amplitudes, similarly to what we did in \citet[][their Figs.~3 and 4]{Proxaufetal2018}. We found that, on average, the errors on the derived amplitudes for the effective temperature curves are a factor 2 smaller than our previous estimates, whereas for the surface gravity the differences are even more significant, with typical errors almost 3 times smaller. For the microturbulent velocity variation, the errors on the current amplitude estimates are more than twice smaller (the dispersions are $\sim$50~K for \teff, $\sim$0.2~dex for \logg, and $\sim$0.2~\kms\ for $\xi$). This means that the variation of the atmospheric parameters along the pulsation cycle is now much better constrained, thanks not only to the larger number of spectra but also to the better determination of such parameters.

The improvement in the estimate of \logg\ and $\xi$ is mainly due to the improvement in the line list, which reduced the number of spurious abundance values provided by unreliable iron lines. As described in the Appendix~\ref{appendix:bad_lines}, we removed lines that are blended, lines that display a well defined trend with effective temperature or with equivalent width, among others. The estimate of \teff\ is based on the same line list that we previously used, therefore the improvement is mainly due to the larger number of spectra analyzed in the current work. All in all, these changes increased the accuracy in the atmospheric parameters. The reader interested in detailed discussion concerning the estimate of the intrinsic parameters of classical Cepheids is referred to \citet{Vasilyevetal2018} and \citet{Lemasleetal2020}.

We note in passing that the improvement in the accuracy and in the sampling of the pulsation cycle allow us to state that the minimum in the effective temperature is often approached in the same phases in which the minimum in surface gravity is also approached. Moreover, and even more importantly, these minimum phases anticipate the increase in microturbulent velocity. This parameter, as expected, attains it maximum just before the maximum in effective temperature, which can be clearly observed by comparing several panels of Figs.~\ref{figure:teff_phase} and \ref{figure:vmic_phase}.

\begin{table}
\centering
\caption{Coefficients of the Fourier series fitted to the data shown in Fig.~\ref{figure:theta_phase} for different bins of pulsation period.}
\label{table:templates}
{\small
\begin{tabular}{l r@{ }l r@{ }l r@{ }l r@{ }l}
\noalign{\smallskip}\hline\hline\noalign{\smallskip}
 &
\multicolumn{2}{c}{\parbox[c]{1.3cm}{\centering $3 < P < 5$ [days]}} &
\multicolumn{2}{c}{\parbox[c]{1.3cm}{\centering $5 < P < 7$ [days]}} &
\multicolumn{2}{c}{\parbox[c]{1.6cm}{\centering $7 < P < 9.5$ [days]}} &
\multicolumn{2}{c}{\parbox[c]{2.0cm}{\centering $9.5 < P < 10.5$ [days]}} \\
\noalign{\smallskip}\hline\noalign{\smallskip}
$c_0$ &    0.44 & $\pm$  0.21 & 0.44 & $\pm$ 0.23 &    0.48 & $\pm$ 0.26 &    0.47 & $\pm$ 0.12 \\
$c_1$ &    2.1  & $\pm$  0.5  & 2.1  & $\pm$ 0.5  & $-$4.4  & $\pm$ 0.6  & $-$4.4  & $\pm$ 0.2  \\
$c_2$ &    0.13 & $\pm$  0.22 & 0.15 & $\pm$ 0.24 &    0.14 & $\pm$ 0.28 &    0.04 & $\pm$ 0.12 \\
$c_3$ & $-$4.5  & $\pm$  1.4  & 2.2  & $\pm$ 1.4  &    2.6  & $\pm$ 2.0  &    4.2  & $\pm$ 2.9  \\
$c_4$ &    0.06 & $\pm$  0.21 & 0.06 & $\pm$ 0.23 &    0.05 & $\pm$	0.27 &    0.04 & $\pm$ 0.11 \\
$c_5$ &    1.4  & $\pm$  3.6  & 2.6  & $\pm$ 3.7  &    2.2  & $\pm$	5.6  &    4.4  & $\pm$ 2.7  \\
$c_6$ &    0.03 & $\pm$  0.21 & 0.03 & $\pm$ 0.22 &    0.03 & $\pm$	0.28 &    0.02 & $\pm$ 0.11 \\
$c_7$ &    1.3  & $\pm$  8.3  & 2.7  & $\pm$ 7.6  & $-$0.93 & $\pm$	9.56 &    4.2  & $\pm$ 6.9  \\
$c_8$ &    0.01 & $\pm$  0.19 &      & ... 		  &		    & ...		 &         & ...		\\
$c_9$ &    2.8  & $\pm$ 16.7  & 	 & ... 		  &		    & ...		 &         & ...		\\
\hline
\end{tabular}
}
\tablefoot{The Fourier series are of the form $\sum c_{2i} \cos{\left[2(i+1)\pi x + c_{2i+1}\right]}$ for $i = 0, 1, \dots, N-1$, where $N$ is the number of terms.}
\end{table}

\subsection{Effective temperature curve templates}
\label{sec:teff_templates}

By taking advantage of the substantial phase coverage of our sample of calibrating Cepheids, we used the effective temperature measurements to compute new templates for different bins of pulsation periods. However, instead of using the \teff\ curves directly, we preferred to provide templates for the theta parameter --- defined as $\theta = 5040/{T_{\rm eff}}$ --- given its linear dependency on the Cousins R-I color index \citep{Taylor1994}.

The approach is similar to the NIR light-curve templates provided by \citet{Innoetal2015}: first, we adopted the ephemerides in Table~\ref{table:physical_params} to fold the $\theta$ curves ($\theta$Cs). Subsequently, we normalized the folded $\theta$Cs by subtracting their average and dividing by their amplitudes. Our $\theta$Cs are well sampled, meaning that we have enough phase points to separate the Cepheids into different period bins and to provide analytical relations for the $\theta$ curve templates in each bin. We adopted the same period thresholds introduced by \citet{Innoetal2015}. Therefore, based on their Table~1, we generated four cumulative and normalized theta curves for the bins 2 (3-5~days, three Cepheids), 3 (5-7~days, three Cepheids), 4 (7-9.5~days, two Cepheids), and 5 (9.5-10.5~days, two Cepheids). By adopting the same period bins we are able to provide $\theta$C templates that are homogeneous with those from the NIR light curves. The reader interested in a more detailed and quantitative discussion concerning the use of cumulative and normalized curves to derive the analytical fits, together with the adopted thresholds for the different period bins, is referred to \citet{Innoetal2015}.

\begin{figure}
\centering
\resizebox{\hsize}{!}{\includegraphics{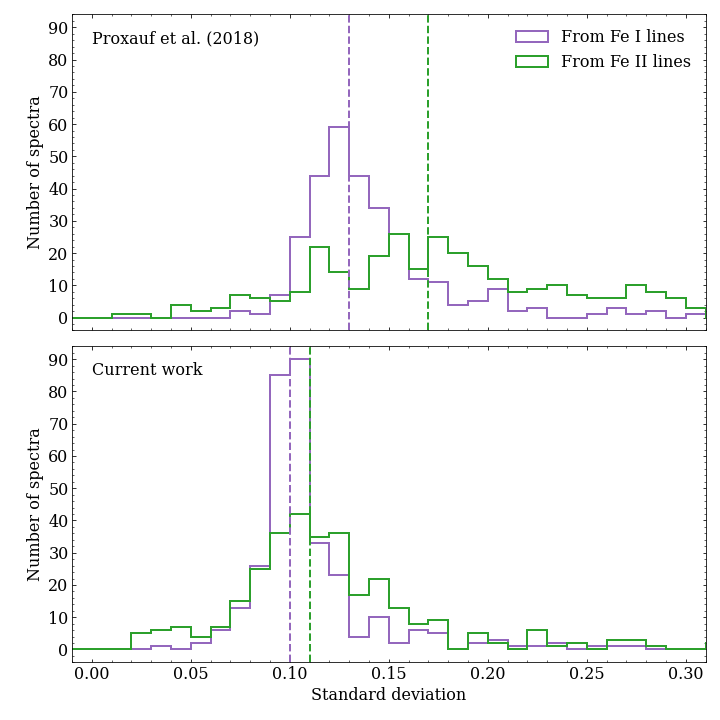}}
\caption{Distribution of standard deviations computed from individual abundances of \ion{Fe}{i} and \ion{Fe}{ii} lines. The panels show the histogram for spectra of stars in common with \citet{Proxaufetal2018}.}
\label{figure:std_feh_vs_proxauf}
\end{figure}

\begin{figure}
\centering
\resizebox{\hsize}{!}{\includegraphics{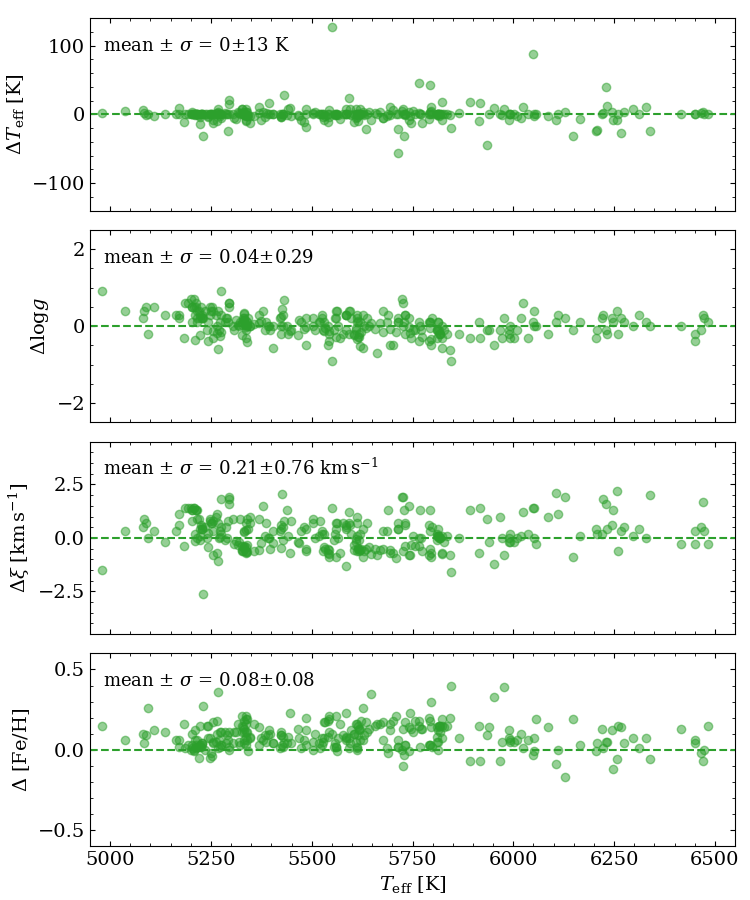}}
\caption{Atmospheric parameters differences for stars in common with \citet{Proxaufetal2018}. Each panel shows our previous determinations subtracted by the current values.}
\label{figure:atm_param_vs_proxauf}
\end{figure}

The four normalized $\theta$ curves are separately shown in Fig.~\ref{figure:theta_phase}, and in Fig.~\ref{figure:theta_phase_func} we compare the Fourier series fitted to the data of each bin. The corresponding coefficients of the fitted functions are listed in Table~\ref{table:templates}. The separation of the Cepheids into different period bins is a mandatory step because not only the amplitude of the theta curves (listed in Table~\ref{table:atm_params_stars}), but also their shape changes with period (as shown in Figs.~\ref{figure:theta_phase} and \ref{figure:theta_phase_func}). Although we do have data for Cepheids with periods outside of the selected period bins, they are either not as well sampled as the data for the other Cepheids, or the theta curves do not overlap well because the number of measurements is too limited. More accurate and homogeneous data are required to overcome this limitation.

In order to allow the reader to use these curve templates to derive the mean effective temperature by using the ephemerides, the luminosity amplitude ($\Delta V$), and a single spectroscopic measurement of the \teff, we also provide a linear relation between $\Delta V$ and $\Delta\theta$:
\begin{equation}
\Delta\theta = (0.184 \pm 0.014) \Delta V + (0.000 \pm 0.011); \, (\sigma = 0.013)
\end{equation}
The estimate of the $\Delta\theta$ amplitude has to be applied to the normalized template as a multiplicative factor. Only after this rescaling operation the template can be anchored to the empirical data and used to derive the mean \teff.

\begin{table*}
\centering
\caption{Excerpt from the list of abundances of $\alpha$ elements for each spectrum of the 20 calibrating Cepheids.}
\label{table:alpha_spectra}
\begin{tabular}{lcc r@{ }l cc r@{ }l c r@{ }l c r@{ }l}
\noalign{\smallskip}\hline\hline\noalign{\smallskip}
Name & Dataset &
\parbox[c]{0.7cm}{\centering MJD [d]} &
\multicolumn{2}{c}{[Mg/H] $\pm$ $\sigma$} &
$N_{\rm{Mg}}$ & ... &
\multicolumn{2}{c}{\ion{Ti}{i}  $\pm$ $\sigma$} & $N_{\rm{\ion{Ti}{i}}}$ &
\multicolumn{2}{c}{\ion{Ti}{ii} $\pm$ $\sigma$} & $N_{\rm{\ion{Ti}{ii}}}$ &
\multicolumn{2}{c}{[Ti/H] $\pm$ $\sigma$} \\
\noalign{\smallskip}\hline\noalign{\smallskip}
\object{R\,TrA}  & HARPS  & 53150.1453265 &    0.20 & $\pm$ 0.11 &   1 & ... & $-$0.21 & $\pm$ 0.15 &   8 & 0.05 & $\pm$ 0.10 &   2 & $-$0.12 & $\pm$ 0.19 \\
\object{R\,TrA}  & HARPS  & 53150.1353086 &    0.17 & $\pm$ 0.11 &   1 & ... & $-$0.17 & $\pm$ 0.27 &  12 & 0.01 & $\pm$ 0.10 &   3 & $-$0.14 & $\pm$ 0.25 \\
...              & ...    & ...           &         & ...        & ... & ... &         & ...        & ... &      & ...        & ... &         & ...        \\
\object{RS\,Pup} & STELLA & 57708.2167938 &    0.24 & $\pm$ 0.11 &   2 & ... & $-$0.06 & $\pm$ 0.28 &  15 & 0.22 & $\pm$ 0.10 &   4 &    0.09 & $\pm$ 0.19 \\
\object{RS\,Pup} & STELLA & 57713.2628277 &         & ...        & ... & ... &         & ...        & ... &      & ...        & ... &         & ...        \\
\hline
\end{tabular}
\tablefoot{The first three columns give the target name, spectroscopic dataset, and Modified Julian Date at which the spectrum was collected. The other columns give the abundances from both neutral and ionized lines, their standard deviations, and the number of lines used. For each element X, the weighted mean of \ion{X}{i} and \ion{X}{ii} abundances (weighted by 1/$\sigma^2$) and its intrinsic error is also shown. Magnesium and Sulfur abundances are from neutral lines only. The complete table is available at the CDS.}
\end{table*}

\begin{table*}
\caption{Excerpt from the list of mean abundances of $\alpha$ elements derived for the 20 calibrating Cepheids.}
\label{table:alpha_stars}
\centering
\begin{tabular}{l r@{ }l c c r@{ }l r@{ }l r@{ }l c}
\noalign{\smallskip}\hline\hline\noalign{\smallskip}
Name &
\multicolumn{2}{c}{\parbox[c]{1.7cm}{\centering [Mg/H] $\pm$ $\sigma$} (std)} &
$N_{\rm spec}$ & ... &
\multicolumn{2}{c}{\parbox[c]{1.7cm}{\centering [\ion{Ti}{i}/H] $\pm$ $\sigma$}} &
\multicolumn{2}{c}{\parbox[c]{1.8cm}{\centering [\ion{Ti}{ii}/H] $\pm$ $\sigma$}} &
\multicolumn{2}{c}{\parbox[c]{1.6cm}{\centering [Ti/H] $\pm$ $\sigma$} (std)} &
$N_{\rm spec}$ \\
\noalign{\smallskip}\hline\noalign{\smallskip}
\object{R\,TrA}  & 0.10 & $\pm$ 0.03 (0.08) &  15 & ... & $-$0.19 & $\pm$ 0.04 & $-$0.09 & $\pm$ 0.03 & $-$0.18 & $\pm$ 0.04 (0.10) &  15 \\
\object{T\,Vul}  & 0.07 & $\pm$ 0.03 (0.15) &  26 & ... & $-$0.11 & $\pm$ 0.03 & $-$0.07 & $\pm$ 0.02 & $-$0.10 & $\pm$ 0.03 (0.10) &  26 \\
...              &      & ...               & ... & ... &         & ...        &         & ...        &         & ...               & ... \\
\object{WZ\,Sgr} & 0.12 & $\pm$ 0.06 (0.03) &   4 & ... &    0.08 & $\pm$ 0.07 &    0.01 & $\pm$ 0.05 &    0.06 & $\pm$ 0.07 (0.12) &   7 \\
\object{RS\,Pup} & 0.09 & $\pm$ 0.03 (0.14) &  10 & ... &    0.02 & $\pm$ 0.04 &    0.16 & $\pm$ 0.03 &    0.04 & $\pm$ 0.04 (0.14) &  16 \\
\hline
\end{tabular}
\tablefoot{From left to right the columns give the star name, the abundances from both neutral and ionized lines, their uncertainties, and the number of spectra used. These are the weighted mean values and their standard errors computed from the individual measurements listed in Table~\ref{table:alpha_spectra}. The standard deviation calculated using all individual abundances from both neutral and ionized lines is also shown. The complete table is available at the CDS.}
\end{table*}

Optical and NIR light curves of the calibrating Cepheids, together with effective temperature and radial velocity curves, will be adopted by our group to perform a detailed comparison with nonlinear, convective hydrodynamical models of classical Cepheids. Dating back to \citet{Nataleetal2008} it has been found that the simultaneous fit of both luminosity and radial-velocity variations provides solid constraints on the physical assumptions adopted to build pulsation models \citep{Marconietal2013}. Moreover, the use of the effective temperature curves (shapes and amplitudes), covering a broad range of pulsation periods, brings forward the opportunity to constrain, on a quantitative basis, the efficiency of the convective transport over the entire pulsation cycle.

\begin{figure}
\centering
\resizebox{\hsize}{!}{\includegraphics{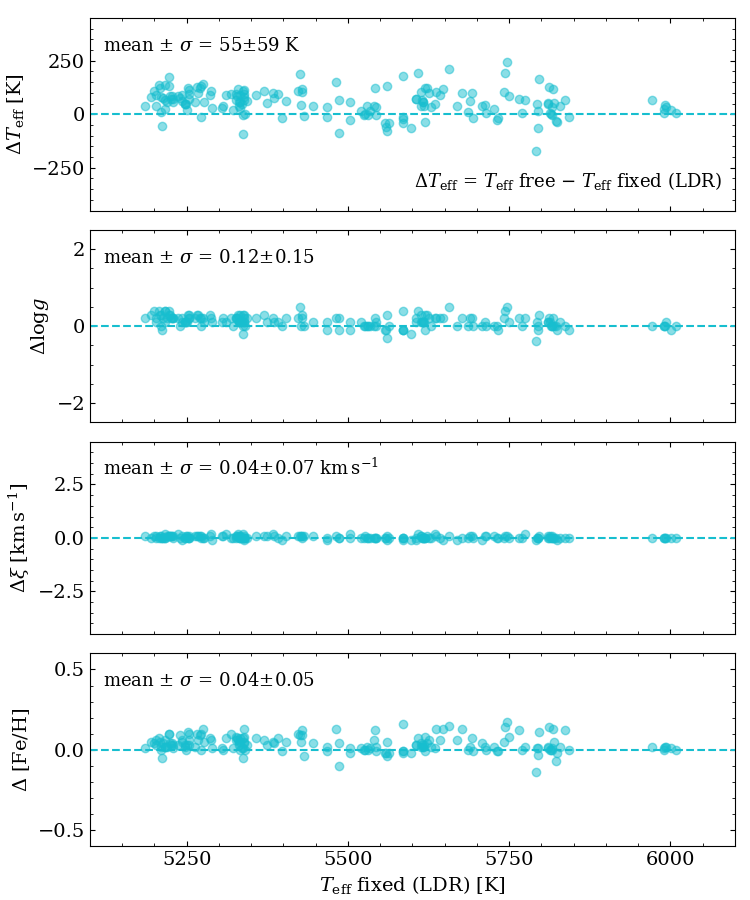}}
\caption{Atmospheric parameters derived by keeping fixed the effective temperature (previously obtained from the LDR method) in comparison with the same parameters derived with \teff\ as a free parameter. The panels show the comparison for the \object{$\beta$\,Dor} and \object{$\zeta$\,Gem} stars.}
\label{figure:teff_free_fixed}
\end{figure}

\section{Iron and $\alpha$-element abundances}
\label{section:ab_iron_alpha}

\subsection{Iron abundance}
\label{section:ab_iron}

The determination of the atmospheric parameters was done with the Python wrapper\,\footnote{pyMOOGi: https://github.com/madamow/pymoogi} of the MOOG code \citep{Sneden2002}. This is a LTE radiative code that we applied to model atmospheres interpolated on the grids of \citet{CastelliKurucz2004}. Once a convergence is achieved in the determination of \teff, \logg, $\xi$, and [Fe/H], as described in Sect.~\ref{section:teff_logg_vmic}, MOOG also provides the final iron abundance derived from individual \ion{Fe}{i} and \ion{Fe}{ii} lines. Such lines are those included in the input line list and that passed the selection criteria that we applied to choose the best atomic transitions.

Table~\ref{table:atm_params_rv_spectra} lists, for each spectrum in our sample, the mean \ion{Fe}{i} and \ion{Fe}{ii} abundances derived from individual lines together with the standard deviations and the number of lines used. The mean \ion{Fe}{i} and \ion{Fe}{ii} abundances derived for each star from the individual spectra, weighted by the standard deviations aforementioned, and the corresponding standard errors are shown in Table~\ref{table:atm_params_stars}. Column~8 gives our final determination for the stellar metallicity, which is the weighted mean calculated from both the [\ion{Fe}{i}/H] and the [\ion{Fe}{ii}/H] abundances. The adopted uncertainties are the largest values between the standard error computed from the weighted mean ($\sigma$) and the standard deviation (std).

\begin{figure*}
\centering
\begin{minipage}[t]{\textwidth}
\centering
\resizebox{\hsize}{!}{\includegraphics{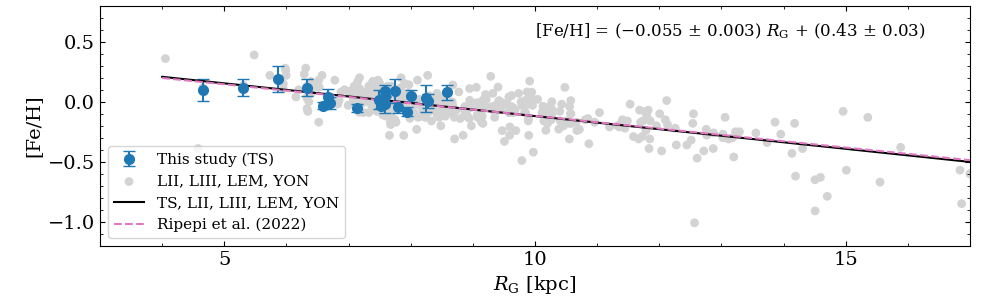}}
\end{minipage} \\
\begin{minipage}[t]{\textwidth}
\centering
\resizebox{\hsize}{!}{\includegraphics{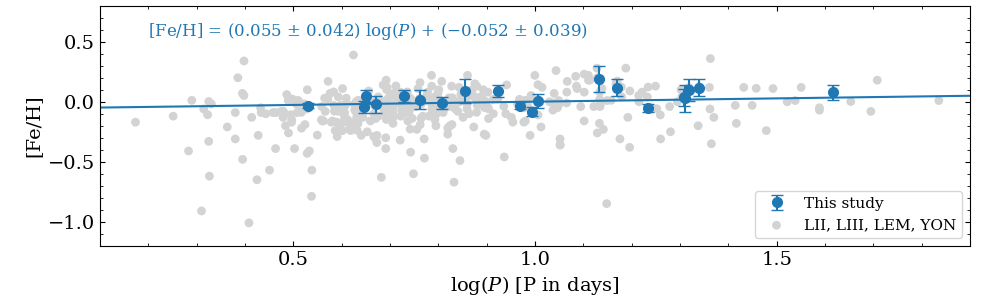}}
\end{minipage}
\caption{Stellar metallicity measured for our sample of 20 calibrating Cepheids compared with results from literature. Top panel: metallicity as a function of the Galactocentric distance from the current work (blue circles) and from literature (gray circles): \citet[][LII]{Lucketal2011}; \citet[][LIII]{LuckLambert2011}; \citet[][LEM]{Lemasleetal2013}; \citet[][YON]{Yongetal2006}. A linear regression (solid black line plus equation) fitted to the entire sample is compared with the radial gradient provided by \citet{Ripepietal2022} (dashed magenta line). The latter was artificially shifted to coincide with the current radial gradient at ${\rm R}_G$ = 10~kpc. The ${\rm R}_G$ values are from \citet{Genovalietal2014}. Bottom panel: same as the top, but as a function of the logarithmic pulsation period. A linear regression (solid blue line plus equation) fitted to the current sample is also show.}
\label{figure:feh_period}
\end{figure*}

Figure~\ref{figure:feh_phase} shows the derived metallicities as a function of the pulsation phase. Data plotted in this figure show that the iron abundances are quite stable along the pulsation cycle. This outcome applies to short-/long-period Cepheids and to Bump Cepheids. We note that the current approach improves the results obtained by \citet{Proxaufetal2018}, since in their iron abundances was still present a mild variation along the rising branch of large amplitude Cepheids (in our current results the dispersions are in most cases smaller than 0.05~dex). The improvement is mainly due to the very careful selection of the lines adopted to estimate the iron abundance. The difference is soundly supported by the distribution of the standard deviations displayed in Fig.~\ref{figure:std_feh_vs_proxauf} for the Cepheids in common. The current standard deviations for both [\ion{Fe}{i}/H] (violet) and [\ion{Fe}{ii}/H] (green) lines (bottom panel) are smaller by 30\%-40\% when compared with our previous investigation. Moreover and even more importantly, they also attain similar values.

To further constrain the difference between the current and our previous investigation, Fig.~\ref{figure:atm_param_vs_proxauf} shows the comparison for both the atmospheric parameters and the iron abundance. The agreement is remarkable over the entire temperature range. The mean and the standard deviations for the atmospheric parameters are well within the current uncertainties. The new iron abundances are, on average, $\sim$0.1 dex more metal-poor. Owing to the similarity in the atmospheric parameters, this difference seems an obvious consequence of the new line list.  

Several different approaches have been suggested in the literature to estimate the atmospheric parameters of variable stars \citep{Fukueetal2015,Jianetal2020,Lemasleetal2020,Matsunagaetal2021,Taniguchietal2021,Romanielloetal2021}. In the current investigation, the effective temperature is firstly derived using the LDR method, then the other atmospheric parameters are estimated at fixed effective temperature. The advantages of this approach have already been discussed in several papers. To provide a more quantitative estimate of the differences compared to another method normally used, in which the minimization process also includes \teff\ as a free parameter, we performed a new and independent determination of the three atmospheric parameters plus abundances simultaneously for two variables in our sample (\object{$\beta$\,Dor} and \object{$\zeta$\,Gem}). The differences between our canonical approach and the literature approach are shown in Fig.~\ref{figure:teff_free_fixed}. The overall agreement is quite good over the entire temperature range. Indeed, the mean and the standard deviations attain tiny values. The mean difference in effective temperature is 55~K, and for the surface gravity it is 0.12~dex only, while the microturbulent velocities attain almost identical values. The quoted differences cause a small increase of 0.04~dex in the iron abundance when \teff\ is a free parameter. We note that we adopted the same line list in the tests performed either with fixed or with free \teff.

\begin{figure*}
\centering
\begin{minipage}[t]{0.49\textwidth}
\centering
\resizebox{\hsize}{!}{\includegraphics{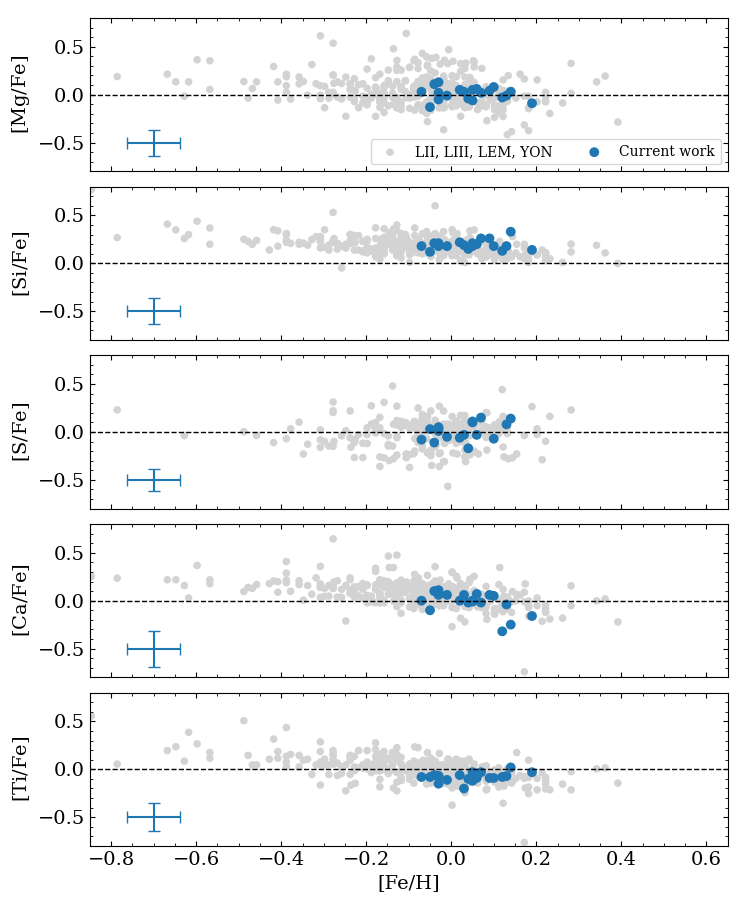}}
\end{minipage}
\begin{minipage}[t]{0.49\textwidth}
\centering
\resizebox{\hsize}{!}{\includegraphics{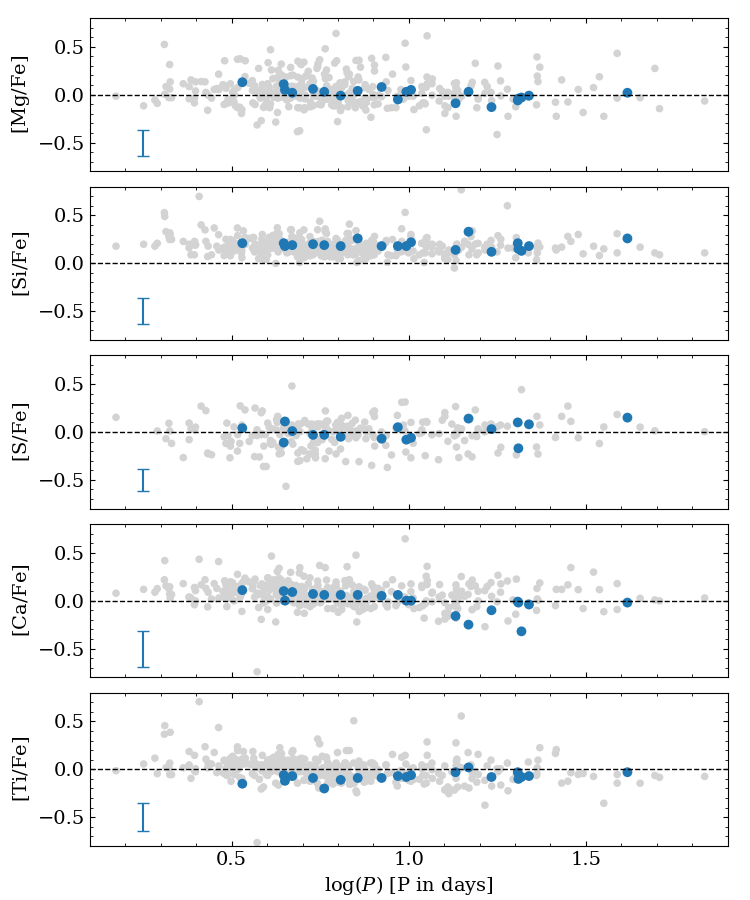}}
\end{minipage}
\caption{[X/Fe] abundances as a function of metallicity (left panels) and of the logarithmic pulsation period (right panels) comparing our 20 calibrating Cepheids with stars from literature: LII: \citet{Lucketal2011}; LIII: \citet{LuckLambert2011}; LEM: \citet{Lemasleetal2013}; YON: \citet{Yongetal2006}. The error bars indicate our typical errors.}
\label{figure:xfe_feh_period}
\end{figure*}

In closing this subsection, we also investigate whether the current iron abundances display any clear trend with the Galactocentric distance (${\rm R}_G$). The top panel of Fig.~\ref{figure:feh_period} shows the iron abundance as a function of ${\rm R}_G$ for the calibrating Cepheids (blue circles) and for Cepheids available in the literature (gray circles). We performed a linear fit over the entire sample and we found the following relation:
\begin{equation}
{\rm [Fe/H]} = (-0.055 \pm 0.003) {\rm R}_G + (0.43 \pm 0.03)
\end{equation}
The current slope agrees quite well with similar estimates available in the literature and, in particular, with the recent estimate of the iron radial gradient (dashed line) provided by \citet{Ripepietal2022}. We note that, to overcome variations in the zero-point mainly introduced by innermost and outermost Cepheids, the gradient from \citet{Ripepietal2022} was artificially shifted to coincide with the current radial gradient at ${\rm R}_G$ = 10~kpc.

The bottom panel of Fig.~\ref{figure:feh_period} shows the iron abundance as a function of the logarithmic pulsation period. Data plotted in this panel agree quite well with similar estimates available in the literature. Moreover, they do not show any significant variation when moving from young (long-period) to less young (short-period) classical Cepheids.

\subsection{$\alpha$-element abundances}
\label{section:ab_alpha}

\begin{figure*}
\centering
\begin{minipage}[t]{0.49\textwidth}
\centering
\resizebox{\hsize}{!}{\includegraphics{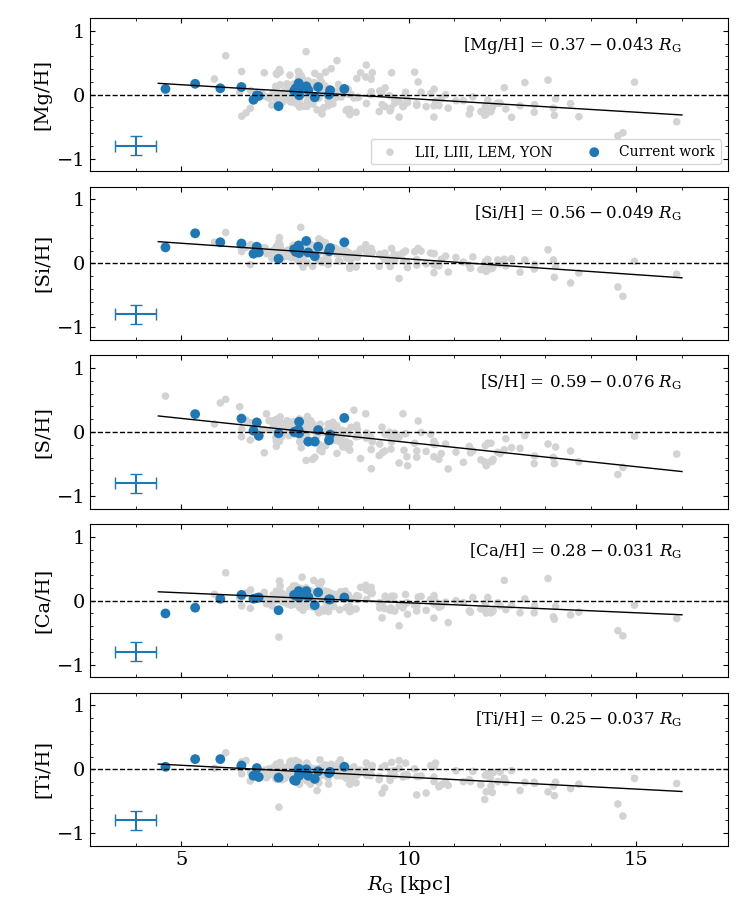}}
\end{minipage}
\begin{minipage}[t]{0.49\textwidth}
\centering
\resizebox{\hsize}{!}{\includegraphics{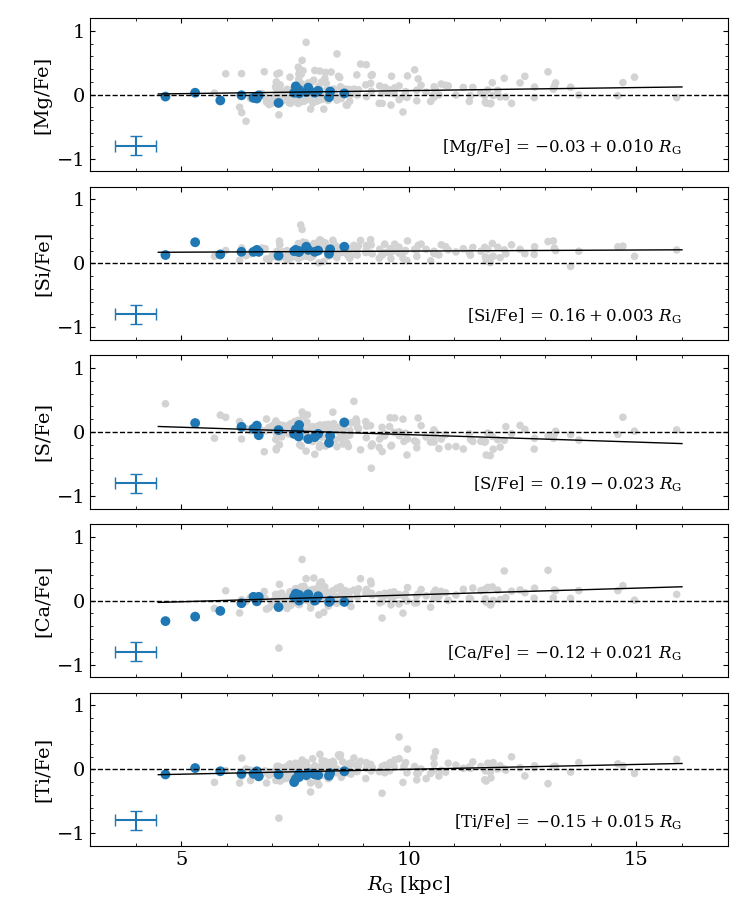}}
\end{minipage}
\caption{[X/H] and [X/Fe] abundance ratios as a function of the Galactocentric distance ($R_{\rm G}$) for our 20 calibrating Cepheids compared with stars from literature: LII: \citet{Lucketal2011}; LIII: \citet{LuckLambert2011}; LEM: \citet{Lemasleetal2013}; YON: \citet{Yongetal2006}. The error bars indicate our typical errors. Linear regressions fitted to the entire sample are also shown (solid lines)}
\label{figure:xh_xfe_RGal}
\end{figure*}

The abundances of the $\alpha$ elements were derived using \mbox{pyMOOGi} as well (except sulfur, for which the abundances were estimated using the spectral synthesis method - see Sect.~\ref{section:ab_sulfur}). We used the $abfind$ driver, which requires as input $i)$ the model atmospheres, calculated from the atmospheric parameters obtained for each star, and $ii)$ a list of lines containing the wavelength, the atomic number, the lower-level excitation potential, the \loggf, and the measured equivalent widths. Table~\ref{table:alpha_spectra} lists the individual $\alpha$-element abundances obtained for each spectrum together with the standard deviations and the number of lines used for each element. The mean abundances, weighted by the standard deviations, and the corresponding standard errors are shown in Table~\ref{table:alpha_stars}. The same mean data are plotted in Figs.~\ref{figure:xfe_feh_period}, \ref{figure:xh_xfe_RGal}, and \ref{figure:xh_phase} for our 20 calibrating Cepheids (blue circles) and for similar data available in the literature (gray circles).

For cases in which the abundance of a given element at a given phase is based on only one or two lines, or if the estimated error is lower than a typical value for each element/species, the uncertainty is set to that typical value, namely: $\sigma$([Mg/H]) = 0.11, $\sigma$([\ion{Si}{i}/H]) = 0.10, $\sigma$([\ion{Ca}{i}/H]) = 0.18, $\sigma$([\ion{Ca}{ii}/H]) = 0.18, $\sigma$([\ion{Ti}{i}/H]) = 0.14, and $\sigma$([\ion{Ti}{ii}/H]) = 0.10~dex (for the sulfur abundances, the uncertainties come from the spectral synthesis). The estimate of these typical errors was done by calculating the median of the standard deviations listed in Table~\ref{table:alpha_spectra} for spectra having at least 2 lines of the same element/species.

The adopted standard solar abundances for both iron and $\alpha$ elements are from \citet{Asplundetal2009}, namely: $A({\rm Fe})_\odot$ = 7.50, $A({\rm Mg})_\odot$ = 7.60, $A({\rm Si})_\odot$ = 7.51, $A({\rm S})_\odot$ = 7.12, $A({\rm Ca})_\odot$ = 6.34, and $A({\rm Ti})_\odot$ = 4.95, with $A$ representing the typical logarithmic notation where H is defined to be $A({\rm H})_\odot$ = 12.00. These values are in good agreement with the very recent determinations done by \citet{Asplundetal2021}, which were obtained using the most up-to-date atomic and molecular data.

For both iron and $\alpha$ abundances, we evaluated the systematic differences for stars in common between the current investigation and measurements from literature: \citet{Lucketal2011}, \citet{LuckLambert2011}, \citet{Lemasleetal2013}, and \citet{Yongetal2006} for iron and $\alpha$ elements, and \citet{Proxaufetal2018} for iron only. The differences for all the investigated elements are listed in Table~\ref{table:zero_points}. In order to perform a direct comparison, we applied these zero-point differences to the literature datasets by putting the iron and the $\alpha$-element abundances in our metallicity scale. The spectroscopic samples for which there are no variables in common between our investigation and those available in the literature, the homogenization of the abundances was performed in two steps. To be more specific, we have no variables in common with \citet{Yongetal2006}, therefore, we first scaled their abundances to match those from \citet{LuckLambert2011}, and then we scaled them to match our metallicity scale.

Data plotted in Fig.~\ref{figure:xh_phase} display, once again, the remarkable agreement of the $\alpha$-element abundances over the entire pulsation cycle (the dispersions in most cases are smaller than 0.10~dex). This outcome applies to all the investigated $\alpha$ elements (except sulfur, for which we have abundances at a single phase only), and to both neutral and ionized species. We note in passing that this finding seems more relevant when compared with the variation of the iron abundances along the pulsation cycle. Indeed, the $\alpha$-element abundances are only based on a modest number or, in some cases, just a few lines. Therefore, they are more prone to possible systematics.

To further investigate the internal consistency of the current $\alpha$-element abundances with similar estimates available in the literature, we checked the behavior of the [X/Fe] abundance ratios as a function of the stellar metallicity. Data plotted in the left panels of Fig.~\ref{figure:xfe_feh_period} show that the current results agree, within the errors, quite well with similar determinations done in previous investigations \citep{LuckLambert2011,Lucketal2011,Lemasleetal2013,Yongetal2006}. The key difference seems to be the smaller dispersion at fixed iron abundance, thus suggesting that a fraction of the spread currently observed might still be caused by intrinsic errors. This finding is further supported by the evidence that the current estimates do not show any clear trend when plotted against the pulsation period (see the right panels of Fig.~\ref{figure:xfe_feh_period}). We note that classical Cepheids, when moving from short to long periods, become systematically 
cooler and more expanded (lower surface gravity).        

\subsubsection{Sulfur abundances}
\label{section:ab_sulfur}

To determine the sulfur abundances, we carried out spectral synthesis calculations by exploiting the multiplet at 6757~\AA, which is proven to be reliable and not affected by NLTE effects \citep[see, e.g.,][and references therein]{Takedaetal2016,Duffauetal2017}. Synthetic spectra were computed using the $synth$ driver of MOOG (2019 version), the same grid of model atmospheres by \citet{CastelliKurucz2004} used for the other elements, and atomic parameters from \citet{Caffauetal2007}. We synthesized a 30~\AA\ wide region around the \ion{S}{i} multiplet in order to determine the broadening of the spectral lines. Then we degraded the synthetic spectra to match the resolution of the observational datasets. The best fit was found with $\chi$-square minimization methods, as routinely done in the literature. Finally, we estimated the internal errors, with a mean value of $\sigma$([\ion{S}{i}/H]) = 0.10~dex. These errors are due to continuum placement (related to the S/N, quality of the spectra, best fitting procedure) and due to errors on the atmospheric parameters (this was done in the standard way, which is by varying one parameter at a time and inspecting the corresponding variations). We refer the reader to \citet{DOrazietal2017,DOrazietal2020} for details on error computation.

The reasons why we focused our attention on \ion{S}{i} abundances for our sample of calibrating Cepheids are manifold: 

{\em i)}-- Sulfur is a volatile element that is not locked in dust grains, which means that no correction is required to compare the current sulfur abundances with predictions based on chemical evolution models \citep{Chiappinietal2001,Cescuttietal2007}. Therefore, sulfur is a more solid tracer of $\alpha$-element abundances than Si and Ca. These three $\alpha$ elements are produced during the oxygen burning and their abundances should show similar chemical enrichment histories \citep{LimongiChieffi2003}.

{\em ii)}-- Sulfur brings forward several key advantages when compared with other $\alpha$ elements, since it shows several lines at both optical and NIR wavelengths. The most popular sulfur lines are the multiplets at 6757~\AA\ and at 8694~\AA, but they are weak and they are typically measured on stars more metal-rich than [Fe/H] $\sim$ $-$1.5~dex. More recently, the multiplet at 9237~\AA\ has also been used because it is very strong and it can be measured even in very metal-poor stars \citep{Nissenetal2004,Caffauetal2005,Caffauetal2007}. However, these lines are also affected by NLTE effects, making them even stronger, and the sulfur abundances need to be properly treated. Even more recently, high-resolution NIR spectrographs also provided the opportunity to use the \ion{S}{i} multiplet at 1045~nm for both dwarf and giant stars \citep{Caffauetal2016}.

{\em iii)}-- Possibility of comparison with the S abundances in external galaxies, such as Blue Compact Galaxies, Damped Lyman Alpha systems \citep[][]{Dessaugesetal2007}, and Active Galactic Nuclei \citep[][Mizumoto et al., submitted to ApJ]{Liuetal2015}. Indeed, thanks to strong emission lines, the S abundances can be measured even at large redshifts.

{\em iv)}-- In our previous investigation focused on  $\alpha$-element abundances in classical Cepheids \citep{Genovalietal2015} we neglected sulfur. In the current study, we perform accurate measurements of the multiplet at 6757~\AA. The multiplet at 9237~\AA\ is only available in a small subsets of spectra (UVES), but the lines are too strong (EW $\ge$ 300~m\AA) and it was neglected. Data plotted in Fig.~\ref{figure:xh_xfe_RGal} show the [X/H] (left panels) and the [X/Fe] (right panels) $\alpha$-element abundances as a function of the Galactocentric distances, estimated by \citet{Genovalietal2014} using NIR mean magnitudes and predicted Period-Luminosity relations \citep{Innoetal2013}. The radial gradients plotted in the left panels indicate that sulfur is the $\alpha$ element with the steeper slope (see labeled values). Indeed, its slope is at least a factor of two steeper than for Ca (0.076 vs. 0.031~${\rm dex\,kpc}^{-1}$) and $\sim$50\% steeper than those for Mg and Si. In passing, we also note that the radial gradient of the abundance ratios plotted in the right panels of the same figure display, as expected \citep[see][]{Genovalietal2015}, either a flat trend (Si) or a marginal positive slope (Mg, Ti) across the thin disk. There are two exceptions. Ca is steadily becoming overabundant (positive slope) when moving from the innermost (more metal-rich) to the outermost (more metal-poor) disk regions. This is a consequence of the fact that the slope of its radial gradient is shallower than the iron gradient (0.31 vs. 0.55~${\rm dex\,kpc}^{-1}$). S shows a negative slope, because it is the only $\alpha$-element with a slope in the radial gradient steeper than the iron gradient (0.76 vs. 0.55~${\rm dex\,kpc}^{-1}$).

\section{Summary and final remarks}
\label{summary}

We performed a new spectroscopic analysis of almost two dozen of northern and southern calibrating Cepheids. As a whole, we collected more than 500 high-resolution (R$\sim$40\,000--115\,000), high S/N ($\sim$50--370) optical spectra with four different spectrographs (STELLA, FEROS, HARPS, UVES) covering either a substantial portion or the entire pulsation cycle. The main aim of this investigation is twofold: 
$i)$ provide new and accurate constraints on possible systematics affecting the estimate of atmospheric parameters and elemental abundances along the pulsation cycle; 
$ii)$ develop new approaches and/or diagnostics to provide accurate estimates of both atmospheric parameters and chemical abundances. 

\begin{table}
\centering
\caption{Mean abundance differences for stars in common between the current sample of calibrating Cepheids and other datasets.}
\label{table:zero_points}
\begin{tabular}{c c r@{ }l c}
\noalign{\smallskip}\hline\hline\noalign{\smallskip}
\parbox[c]{1.6cm}{\centering Abundance ratio} &
Datasets$^1$ &
\multicolumn{2}{c}{\parbox[c]{1.6cm}{\centering Zero-point difference}} &
$N_{\rm Common}$ \\
\noalign{\smallskip}\hline\noalign{\smallskip}
 {[Fe/H]} & P18$-$TS & 0.07 & $\pm$ 0.06 & 13 \\
 {[Fe/H]} & LII$-$TS & 0.05 & $\pm$ 0.07 & 20 \\
 {[Fe/H]} & LIII$-$TS & 0.11 & $\pm$ 0.08 & 6 \\
 {[Fe/H]} & LII$-$LEM & 0.07 & $\pm$ 0.12 & 55 \\
 {[Fe/H]} & YON$-$LIII & $-$0.33 & $\pm$ 0.15 & 21 \\[0.1cm]
 {[Mg/H]} & LII$-$TS & $-$0.14 & $\pm$ 0.11 & 16 \\
 {[Mg/H]} & LIII$-$TS & 0.11 & $\pm$ 0.16 & 5 \\
 {[Mg/H]} & LII$-$LEM & $-$0.26 & $\pm$ 0.24 & 35 \\
 {[Mg/H]} & YON$-$LIII & $-$0.09 & $\pm$ 0.14 & 16 \\[0.1cm]
 {[Si/H]} & LII$-$TS & $-$0.13 & $\pm$ 0.07 & 20 \\
 {[Si/H]} & LIII$-$TS & $-$0.01 & $\pm$ 0.09 & 6 \\
 {[Si/H]} & LII$-$LEM & $-$0.06 & $\pm$ 0.11 & 55 \\
 {[Si/H]} & YON$-$LIII & $-$0.12 & $\pm$ 0.07 & 20 \\[0.1cm]
 {[S/H]} & LII$-$TS & 0.11 & $\pm$ 0.13 & 17 \\
 {[S/H]} & LII$-$LEM & $-$0.13 & $\pm$ 0.19 & 45 \\[0.1cm]
 {[Ca/H]} & LII$-$TS & $-$0.07 & $\pm$ 0.10 & 20 \\
 {[Ca/H]} & LIII$-$TS & $-$0.01 & $\pm$ 0.10 & 6 \\
 {[Ca/H]} & LII$-$LEM & $-$0.06 & $\pm$ 0.17 & 55 \\
 {[Ca/H]} & YON$-$LIII & $-$0.16 & $\pm$ 0.11 & 21 \\[0.1cm]
 {[Ti/H]} & LII$-$TS & 0.12 & $\pm$ 0.09 & 20 \\
 {[Ti/H]} & LIII$-$TS & 0.17 & $\pm$ 0.12 & 6 \\
 {[Ti/H]} & YON$-$LIII & 0.33 & $\pm$ 0.20 & 20 \\
\hline
\end{tabular}
\tablefoot{$^{(1)}$ TS: this study; P18: \citet{Proxaufetal2018}; LII: \citet{Lucketal2011}; LIII: \citet{LuckLambert2011}; LEM: \citet{Lemasleetal2013}; YON: \citet{Yongetal2006}.}
\end{table}

To accomplish the former goal, we assembled three different line lists. By using different line lists available in the literature, we compiled a very detailed list of both \ion{Fe}{i} and \ion{Fe}{ii} lines. Moreover, we also collected a sizable sample of lines for $\alpha$-elements: Mg, Si, S, Ca, and Ti. Special care was also paid to the line list of both iron-peak and $\alpha$-elements used for the estimate of the effective temperature using the LDR method.

In order to provide homogeneous and accurate estimates of both atmospheric parameters and elemental abundances, we undertook a lengthy and detailed analysis of the quoted line lists. Our approach moved along three different paths: $i)$ we collected the three most precise atomic transition parameters available in the literature; $ii)$ we removed absorption lines that are blended with other lines; $iii)$ we removed lines that either show relevant abundance variations from the average value of their chemical species or that display a steady variation along the pulsation cycle with atmospheric parameters. Concerning the transition parameters, we gave the priority to recent laboratory measurements, then to homogeneous astrophysical estimates and, for the remaining lines, to the transition parameters available on the NIST Atomic Spectra Database. To identify blended lines, we adopted both the solar spectrum by \citet{Mooreetal1966} and a grid of synthetic spectra. Finally, we applied an iterative approach to remove lines showing significant variations. This latter step was done based on a preliminary estimate of the atmospheric parameters and the abundances. We were then able to identify lines that show either a variation in abundance at the 3-$\sigma$ level from the mean value of the specific element and/or a steady variation along the pulsation cycle. In more detail, to properly identify robust lines, we plotted all the lines available in our initial line lists, for each chemical species, individual exposure, and calibrating Cepheid, as a function of the pulsation phase, eﬀective temperature, and equivalent width. Lastly, we performed a new estimate of the atmospheric parameters and the individual elemental abundances, using an improved version of our algorithm and using only the line lists based on optimal lines.

The comparison between the current atmospheric parameters and similar estimated provided by our group, using a similar approach but old line lists, show smoother variation along the pulsation cycle and, in particular, smaller standard deviations ($\sim$50~K for the effective temperature, $\sim$0.2~dex for the surface gravity, and $\sim$0.2~\kms\ for the microturbulent velocity). The estimate of both iron and $\alpha$-element abundances is also significantly improved, since they display an almost constant value as a function of the pulsation phase. The dispersions are $\lesssim$ 0.05~dex for iron and $\lesssim$ 0.10~dex for the $\alpha$ elements. We note that this is the first time that $\alpha$-element abundances are critically investigated for possible variations along the pulsation cycle.

This finding is further supported by the comparison with similar abundance estimates available in the literature. We found that the current calibrating Cepheids display, at fixed either iron abundance or pulsation period, smaller standard deviations. This result suggests that a good fraction of the spread normally observed in elemental abundances is mainly caused by the adopted line list and by the algorithm used to estimate the atmospheric parameters.  

Moreover and even more importantly, we also derived for the first time new and accurate effective temperature templates. Following \citet{Innoetal2015}, the calibrating Cepheids were split into four different period bins, ranging from 3 to 10.5~days in period. For each period bin, we performed an analytical fit with Fourier series providing $\theta = 5040~/{T_{\rm eff}}$ as a function of the pulsation phase. A similar approach was already applied to fundamental RR Lyrae by \citet{ForSneden2010} and by \citet{Magurnoetal2018}, but the variations along the pulsation cycle are, at fixed pulsation phase, larger due to the larger luminosity amplitudes and probably to the possible occurrence of nonlinear phenomena (shocks). The current findings appear quite promising, because the new templates can provide accurate estimates of the effective temperature once the pulsation period and the phase (and in turn the reference epoch) of a Cepheid are known. We note that this information is crucial in the abundance analysis of NIR spectra, because in the metal-poor regimes only a modest number of ionized and neutral iron lines are present.

The current investigation was focused on Galactic Cepheids with iron abundances close to the solar value. In a forthcoming paper we plan to apply the same approach to more metal-poor Magellanic Cloud Cepheids. 

\begin{acknowledgements}
It is a real pleasure to thank the anonymous referee for her/his positive appreciation of the current investigation and for her/his pertinent suggestions that improved the content and the readability of the paper. It is also a pleasure to thank F. Matteucci and E. Spitoni for several detailed discussions concerning the role of $\alpha$ elements in constraining chemical evolution models. R.P.K. gratefully acknowledges support by the Munich Excellence Cluster Origins Funded by the Deutsche Forschungsgemeinschaft (DFG, German Research Foundation) under Germany's Excellence Strategy EXC-2094 39078331. B.L. acknowledges the Deutsche Forschungsgemeinschaft (DFG, German Research Foundation) -- Project-ID 138713538 -- SFB 881 ("The Milky Way System", sub-project A05).
\end{acknowledgements}

\bibliographystyle{aa}
\bibliography{daSilvaetal2021.bib}


\begin{appendix}

\section{Cleaning from irregular atomic lines}
\label{appendix:bad_lines}

\begin{figure*}
\centering
\begin{minipage}[t]{0.33\textwidth}
\centering
\resizebox{\hsize}{!}{\includegraphics{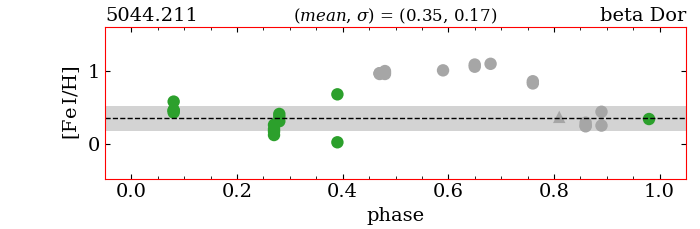}}
\end{minipage}
\begin{minipage}[t]{0.33\textwidth}
\centering
\resizebox{\hsize}{!}{\includegraphics{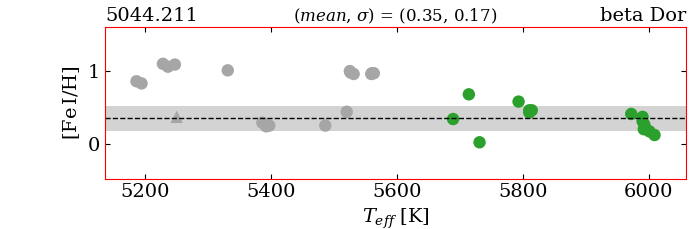}}
\end{minipage}
\begin{minipage}[t]{0.33\textwidth}
\centering
\resizebox{\hsize}{!}{\includegraphics{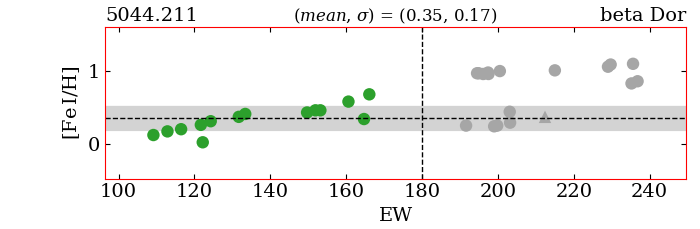}}
\end{minipage} \\
\begin{minipage}[t]{0.33\textwidth}
\centering
\resizebox{\hsize}{!}{\includegraphics{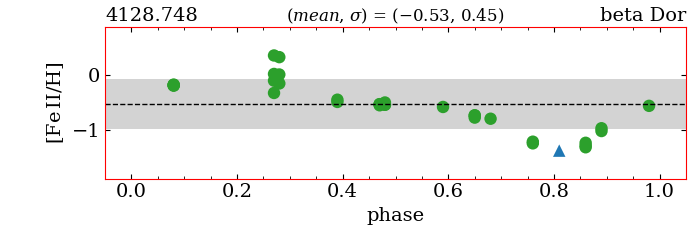}}
\end{minipage}
\begin{minipage}[t]{0.33\textwidth}
\centering
\resizebox{\hsize}{!}{\includegraphics{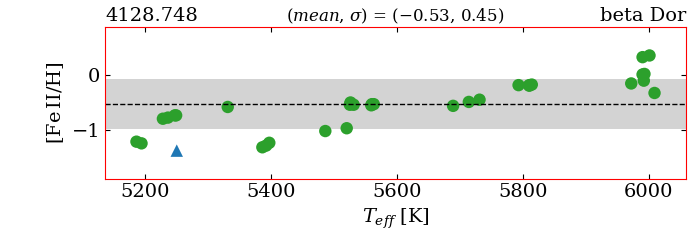}}
\end{minipage}
\begin{minipage}[t]{0.33\textwidth}
\centering
\resizebox{\hsize}{!}{\includegraphics{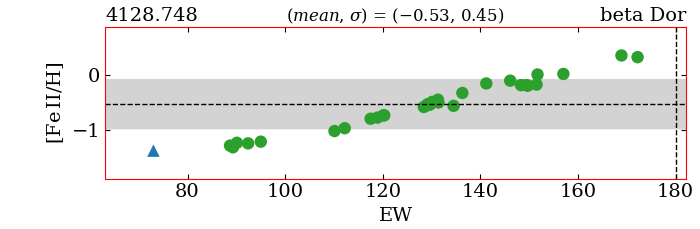}}
\end{minipage} \\
\begin{minipage}[t]{0.33\textwidth}
\centering
\resizebox{\hsize}{!}{\includegraphics{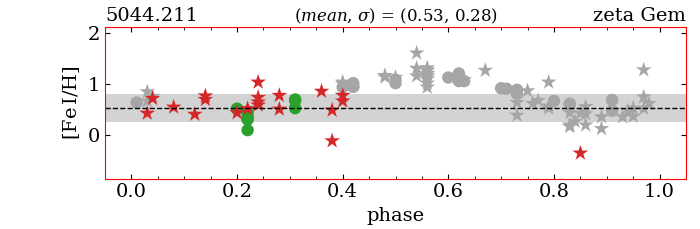}}
\end{minipage}
\begin{minipage}[t]{0.33\textwidth}
\centering
\resizebox{\hsize}{!}{\includegraphics{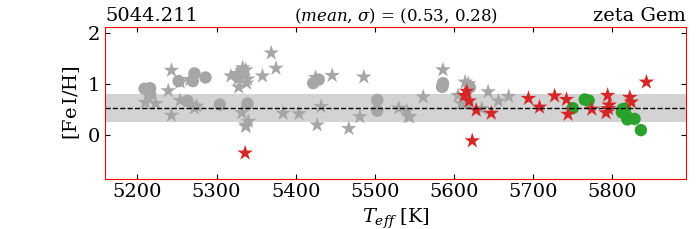}}
\end{minipage}
\begin{minipage}[t]{0.33\textwidth}
\centering
\resizebox{\hsize}{!}{\includegraphics{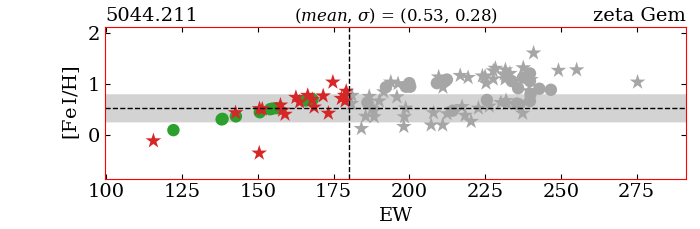}}
\end{minipage} \\
\begin{minipage}[t]{0.33\textwidth}
\centering
\resizebox{\hsize}{!}{\includegraphics{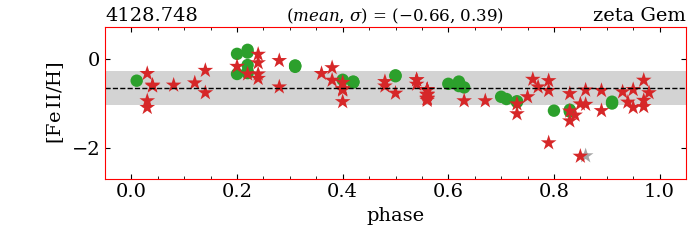}}
\end{minipage}
\begin{minipage}[t]{0.33\textwidth}
\centering
\resizebox{\hsize}{!}{\includegraphics{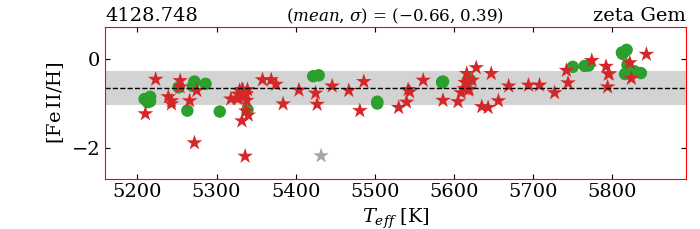}}
\end{minipage}
\begin{minipage}[t]{0.33\textwidth}
\centering
\resizebox{\hsize}{!}{\includegraphics{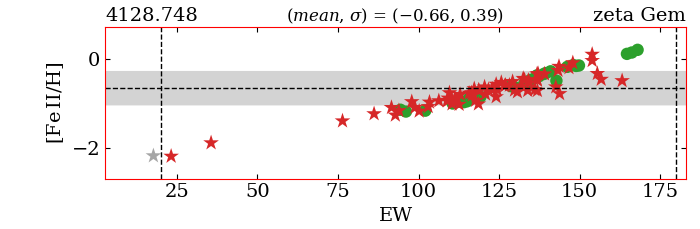}}
\end{minipage}
\caption{Iron abundances as a function of the pulsation phase, effective temperature, and equivalent width for lines excluded from the initial line list. These examples, for HARPS (green circles), FEROS (blue triangle), and STELLA (red stars) spectra of \object{$\beta$\,Dor} and \object{$\zeta$\,Gem}, show a dependence of the abundances on the equivalent widths possibly caused by blends. Lines too weak or too strong (probably saturated) are represented by light gray symbols. The hatched regions around the dashed lines indicate the 1-$\sigma$ uncertainty around the mean.}
\label{figure:bad_lines_blends}
\end{figure*}

\begin{figure*}
\centering
\begin{minipage}[t]{0.33\textwidth}
\centering
\resizebox{\hsize}{!}{\includegraphics{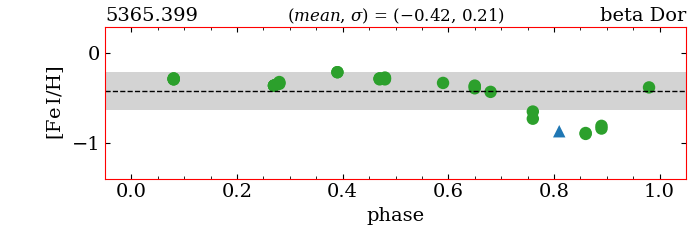}}
\end{minipage}
\begin{minipage}[t]{0.33\textwidth}
\centering
\resizebox{\hsize}{!}{\includegraphics{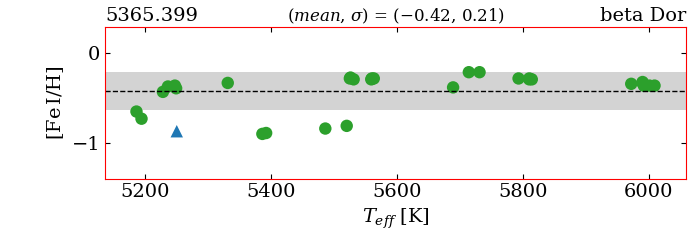}}
\end{minipage}
\begin{minipage}[t]{0.33\textwidth}
\centering
\resizebox{\hsize}{!}{\includegraphics{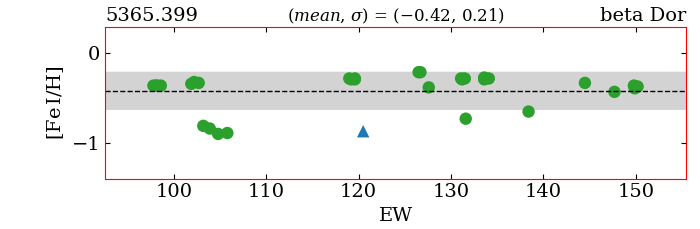}}
\end{minipage} \\
\begin{minipage}[t]{0.33\textwidth}
\centering
\resizebox{\hsize}{!}{\includegraphics{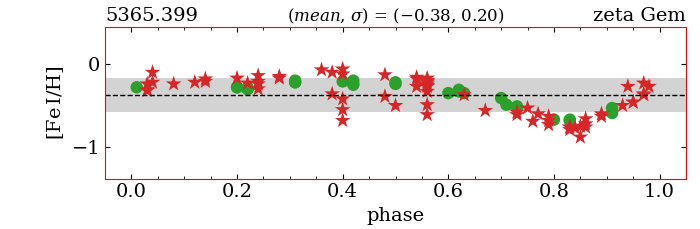}}
\end{minipage}
\begin{minipage}[t]{0.33\textwidth}
\centering
\resizebox{\hsize}{!}{\includegraphics{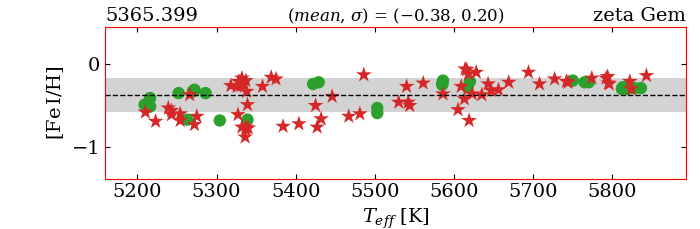}}
\end{minipage}
\begin{minipage}[t]{0.33\textwidth}
\centering
\resizebox{\hsize}{!}{\includegraphics{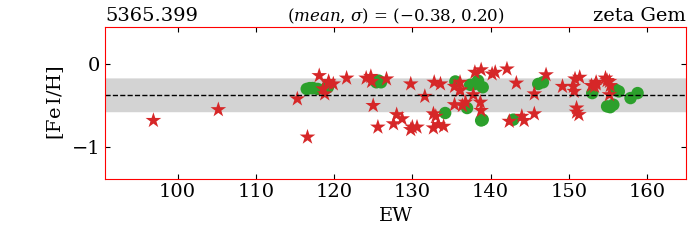}}
\end{minipage}
\caption{Same as in Fig.~\ref{figure:bad_lines_blends} but showing an example of a line excluded because of having a bump around pulsation phases 0.8-0.9, not caused by saturation or by any correlation with the equivalent widths.}
\label{figure:bad_lines_bump}
\end{figure*}

\begin{figure*}
\centering
\begin{minipage}[t]{0.33\textwidth}
\centering
\resizebox{\hsize}{!}{\includegraphics{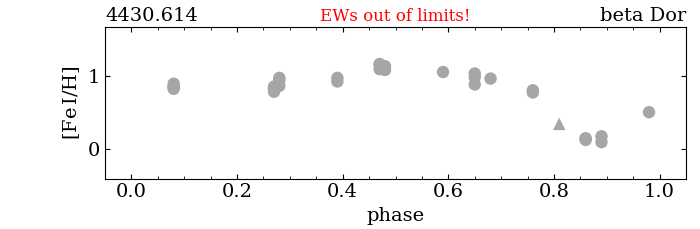}}
\end{minipage}
\begin{minipage}[t]{0.33\textwidth}
\centering
\resizebox{\hsize}{!}{\includegraphics{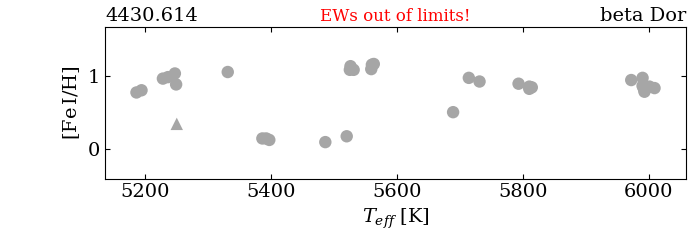}}
\end{minipage}
\begin{minipage}[t]{0.33\textwidth}
\centering
\resizebox{\hsize}{!}{\includegraphics{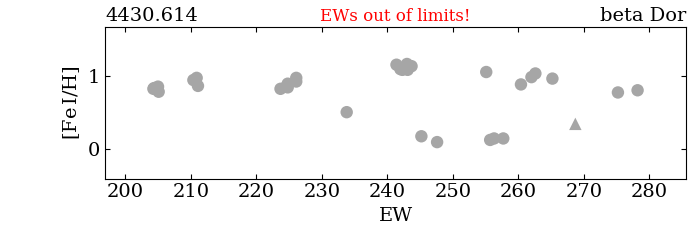}}
\end{minipage} \\
\begin{minipage}[t]{0.33\textwidth}
\centering
\resizebox{\hsize}{!}{\includegraphics{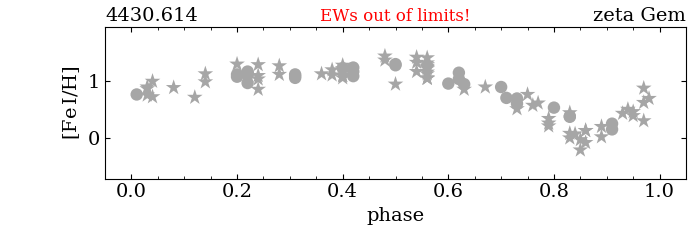}}
\end{minipage}
\begin{minipage}[t]{0.33\textwidth}
\centering
\resizebox{\hsize}{!}{\includegraphics{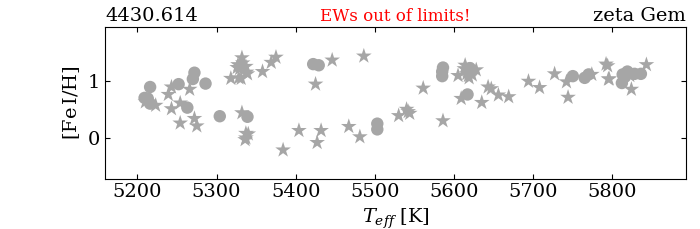}}
\end{minipage}
\begin{minipage}[t]{0.33\textwidth}
\centering
\resizebox{\hsize}{!}{\includegraphics{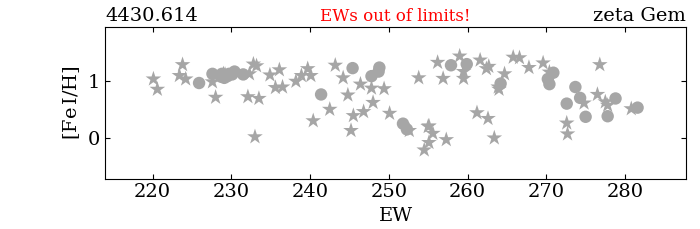}}
\end{minipage} \\
\begin{minipage}[t]{0.33\textwidth}
\centering
\resizebox{\hsize}{!}{\includegraphics{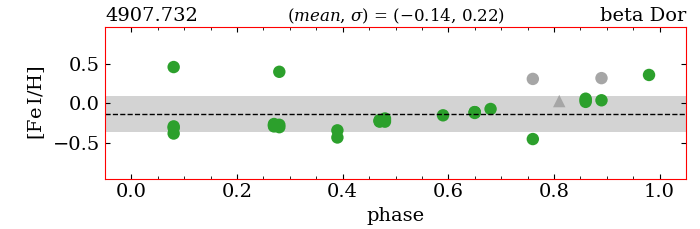}}
\end{minipage}
\begin{minipage}[t]{0.33\textwidth}
\centering
\resizebox{\hsize}{!}{\includegraphics{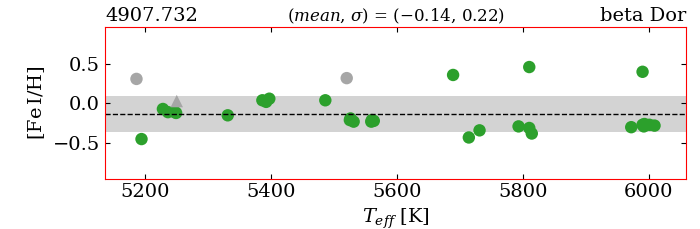}}
\end{minipage}
\begin{minipage}[t]{0.33\textwidth}
\centering
\resizebox{\hsize}{!}{\includegraphics{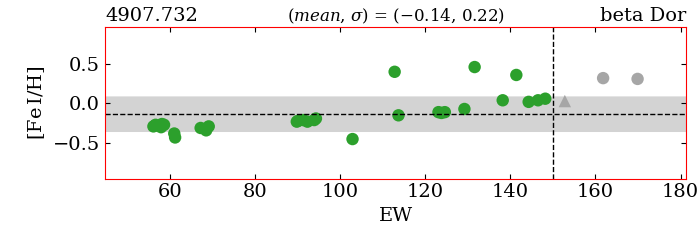}}
\end{minipage} \\
\begin{minipage}[t]{0.33\textwidth}
\centering
\resizebox{\hsize}{!}{\includegraphics{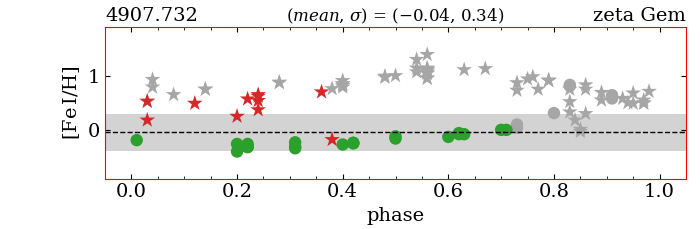}}
\end{minipage}
\begin{minipage}[t]{0.33\textwidth}
\centering
\resizebox{\hsize}{!}{\includegraphics{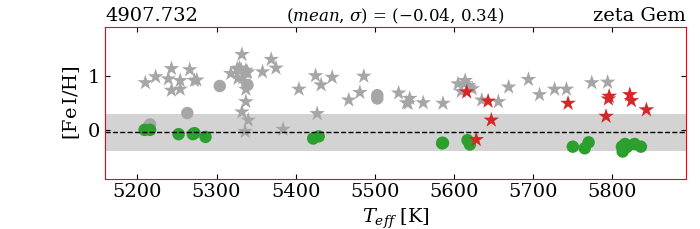}}
\end{minipage}
\begin{minipage}[t]{0.33\textwidth}
\centering
\resizebox{\hsize}{!}{\includegraphics{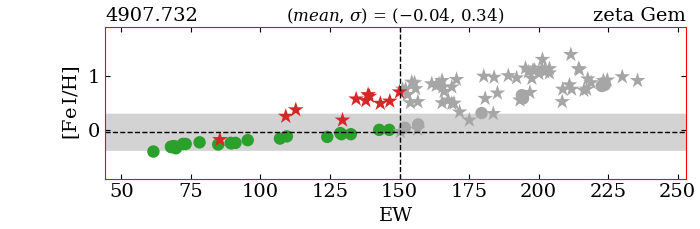}}
\end{minipage}
\caption{Same as in Fig.~\ref{figure:bad_lines_blends} but showing examples of lines excluded either because of being completely saturated or because of having a chaotic distribution, probably caused by blends.}
\label{figure:bad_lines_chaos}
\end{figure*}

\begin{figure*}
\centering
\begin{minipage}[t]{0.33\textwidth}
\centering
\resizebox{\hsize}{!}{\includegraphics{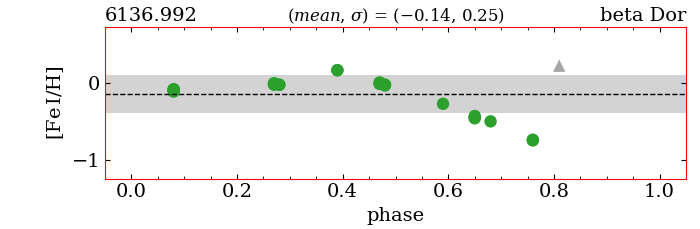}}
\end{minipage}
\begin{minipage}[t]{0.33\textwidth}
\centering
\resizebox{\hsize}{!}{\includegraphics{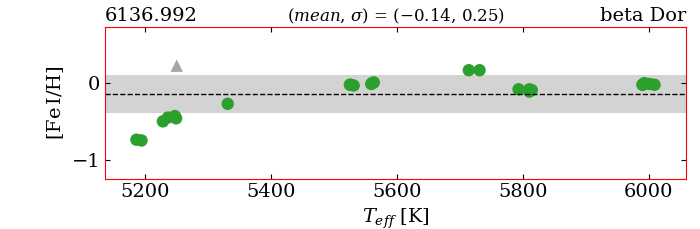}}
\end{minipage}
\begin{minipage}[t]{0.33\textwidth}
\centering
\resizebox{\hsize}{!}{\includegraphics{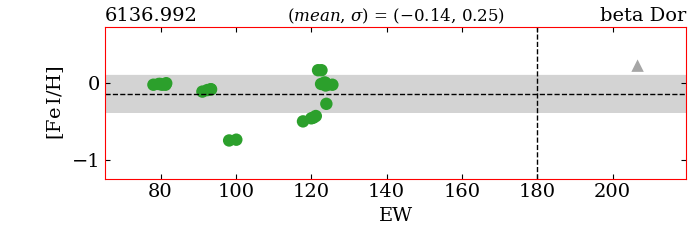}}
\end{minipage} \\
\begin{minipage}[t]{0.33\textwidth}
\centering
\resizebox{\hsize}{!}{\includegraphics{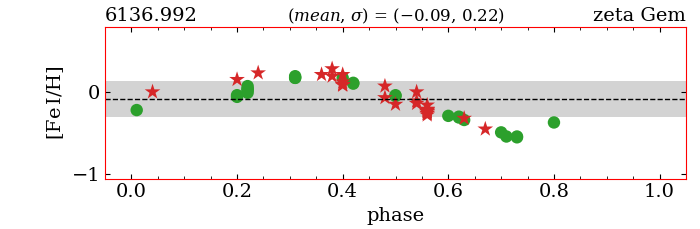}}
\end{minipage}
\begin{minipage}[t]{0.33\textwidth}
\centering
\resizebox{\hsize}{!}{\includegraphics{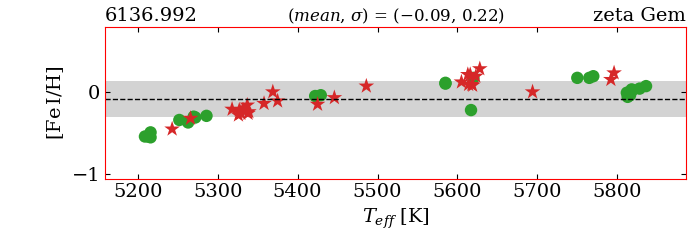}}
\end{minipage}
\begin{minipage}[t]{0.33\textwidth}
\centering
\resizebox{\hsize}{!}{\includegraphics{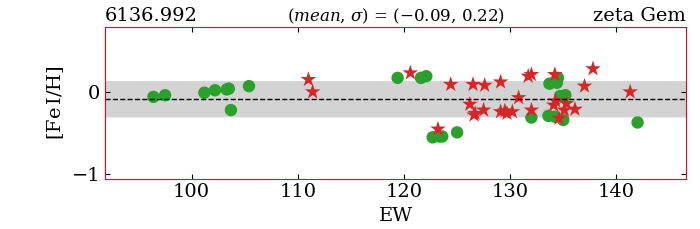}}
\end{minipage}
\caption{Same as in Fig.~\ref{figure:bad_lines_blends} but showing an example of a line excluded because of having some correlation with the effective temperature, possibly caused by NLTE effects.}
\label{figure:bad_lines_nlte}
\end{figure*}

\begin{figure*}
\centering
\begin{minipage}[t]{0.33\textwidth}
\centering
\resizebox{\hsize}{!}{\includegraphics{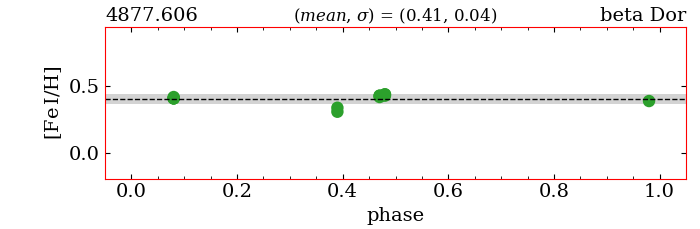}}
\end{minipage}
\begin{minipage}[t]{0.33\textwidth}
\centering
\resizebox{\hsize}{!}{\includegraphics{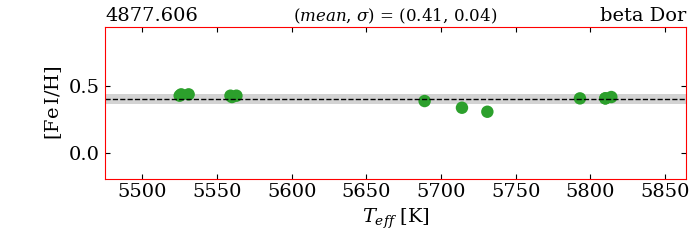}}
\end{minipage}
\begin{minipage}[t]{0.33\textwidth}
\centering
\resizebox{\hsize}{!}{\includegraphics{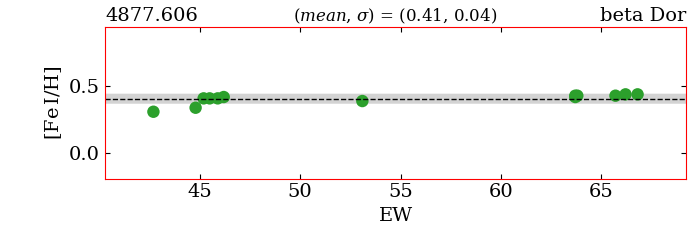}}
\end{minipage} \\
\begin{minipage}[t]{0.33\textwidth}
\centering
\resizebox{\hsize}{!}{\includegraphics{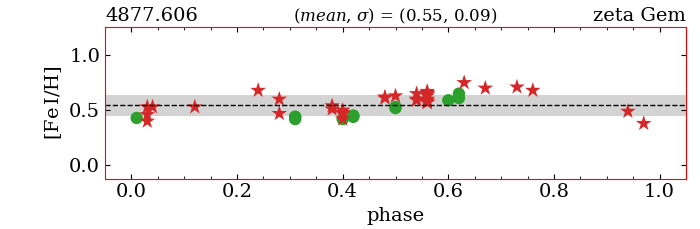}}
\end{minipage}
\begin{minipage}[t]{0.33\textwidth}
\centering
\resizebox{\hsize}{!}{\includegraphics{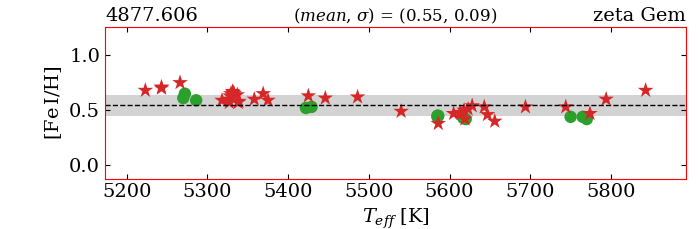}}
\end{minipage}
\begin{minipage}[t]{0.33\textwidth}
\centering
\resizebox{\hsize}{!}{\includegraphics{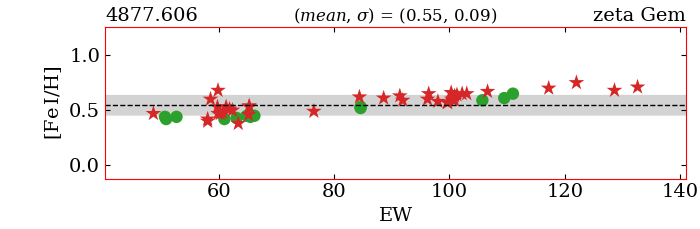}}
\end{minipage}
\caption{Same as in Fig.~\ref{figure:bad_lines_blends} but showing an example of a line that, even if the dispersion in abundance is very small, was excluded from the initial line list because of having an unexpected over-abundance in metallicity, possibly caused by a wrong \loggf\ value.}
\label{figure:bad_lines_loggf}
\end{figure*}

Many of the atomic lines present in our initial line lists of iron and $\alpha$ elements turned out to be not reliable enough. The abundances provided for our spectra by such irregular lines exhibit at least one of the following problems:

\begin{enumerate}

\item {\bf Dependence on EW:} a clear trend, possibly caused by blended lines, is seen in the [X/H] abundance ratios as a function of the equivalent widths. In these cases, we have either eliminated the line or defined a new and acceptable EW range. A few examples for \ion{Fe}{i} and \ion{Fe}{ii}, for the stars \object{$\beta$\,Dor} and \object{$\zeta$\,Gem}, are shown in Fig.~\ref{figure:bad_lines_blends}.

\item {\bf Abundance bump at some phases:} there are cases in which the abundance derived for a given element changes along the pulsation cycle, being under- or over-estimated around phases between 0.8 and 0.9. These bumps are not caused by saturation, so there is no clear correlation with EW. In such situations, we only kept the lines where the bump was not very large in order to avoid a wrong determination of the mean abundance. Bump examples are shown in Fig.~\ref{figure:bad_lines_bump}.

\item {\bf Chaotic behavior:} the [X/H] abundance ratios have a very large dispersion and no clear trend with any other parameter. This might be caused by very close blends where the deblending with two or multiple Gaussian functions not always worked. See examples in Fig.~\ref{figure:bad_lines_chaos}. The same figure also shows examples of lines completely saturated.

\item {\bf Dependence on \teff:} Fig.~\ref{figure:bad_lines_nlte} shows two examples for which the abundances derived from \ion{Fe}{i} lines have a clear trend with the effective temperature, possibly caused by NLTE effects. An option to avoid this problem could be to define, at least for these irregular lines, a range of \teff\ within which the abundances they provide can be accepted. For this work, however, we preferred not to use such lines at all.

\item {\bf Under- or over-abundances:} Some of the atomic lines in our lists, though exhibiting a coherent behavior in phase, \teff, and EW, provide systematic under- or over-abundances. See examples in Fig.~\ref{figure:bad_lines_loggf}. These problems are probably not caused by saturation but by a wrong value of \loggf, or by a wrong continuum identification, which is not always an easy task in regions with too many lines. For several cases we were able to adjust the \loggf\ values using a high-resolution spectrum of Arcturus as reference. For other unsolved cases the lines were simply not used in our analysis.

\end{enumerate}

\FloatBarrier

\section{Variations along the pulsation cycle}
\label{appendix:variations_with_phase}

\subsection{Light and radial-velocity curves}

The panels of Fig.~\ref{figure:mag_rv_phase} show the phase-folded light and radial-velocity curves for our sample of 20 calibrating Cepheids. The stellar magnitude in visual bands were taken from literature, as explained in Sect.~\ref{section:rv_photometry}. The radial velocities were either measured from our spectra or taken from literature. Only \object{FF\,Aql} and \object{$\beta$\,Dor} have RV values derived from our spectra only.

Our RV measurements agree quite well with the literature values. Significant differences are seen only for two stars, \object{T\,Vul} and \object{S\,Sge}, which might be related to instrumental drifts caused by their long baseline observations (for both stars, the two datasets, ours and from literature, are separated by almost 17 years) or, most likely, to the binarity of these objects \citep[][]{EvansNancy1992,Evansetal1993}.

As detailed in Table~\ref{table:physical_params}, for \object{T\,Vul} we derived two different values of reference epoch ($T_0$), one from our own radial velocities and another one from the light curve. For \object{S\,Sge} we have independent measurements of $T_0$ and pulsation period derived from both the RV datasets and from the light curve. Another particular case is \object{$\beta$\,Dor}, one of our bump Cepheids (see Table~\ref{table:calib_sample}), for which we have independent values of $T_0$ and period derived from our RV curve and from the literature light curve.

The main criteria we adopted for selecting the light curves available in the literature were: $i)$ well-sampled photometric data; $ii)$ minimization of the difference in epoch between photometric and spectroscopic data. It goes without saying that well-sampled light curves from Gaia DR3 will allow us to provide accurate and homogeneous ephemerides for the majority of classical Cepheids.

\begin{figure*}
\centering
\begin{minipage}[t]{0.49\textwidth}
\centering
\resizebox{\hsize}{!}{\includegraphics{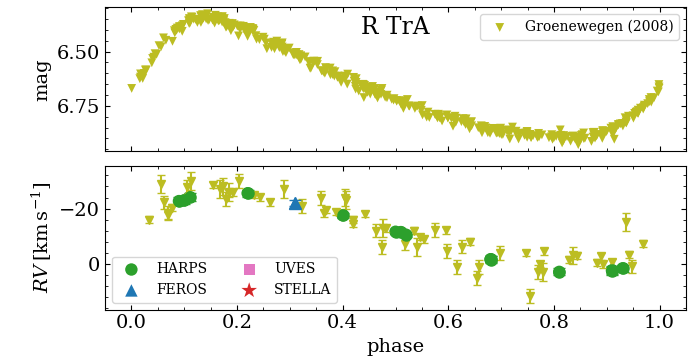}}
\end{minipage}
\begin{minipage}[t]{0.49\textwidth}
\centering
\resizebox{\hsize}{!}{\includegraphics{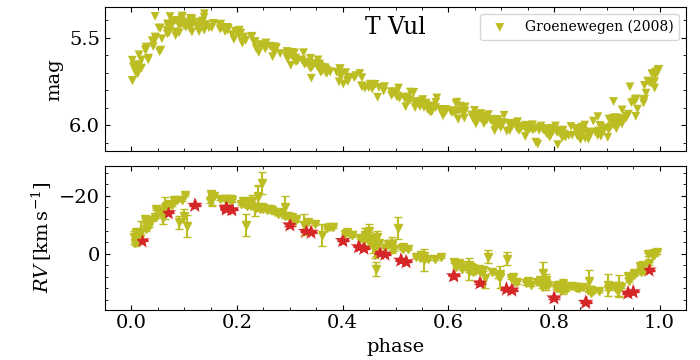}}
\end{minipage} \\
\begin{minipage}[t]{0.49\textwidth}
\centering
\resizebox{\hsize}{!}{\includegraphics{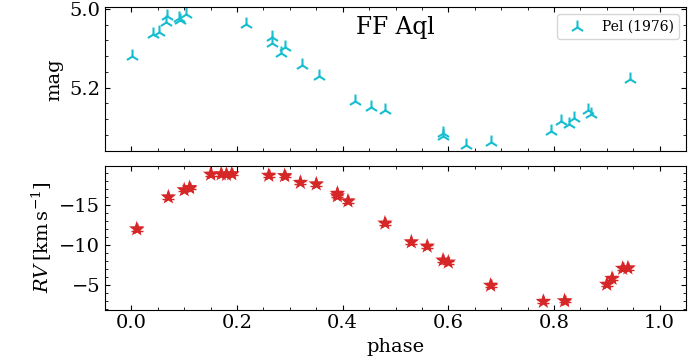}}
\end{minipage}
\begin{minipage}[t]{0.49\textwidth}
\centering
\resizebox{\hsize}{!}{\includegraphics{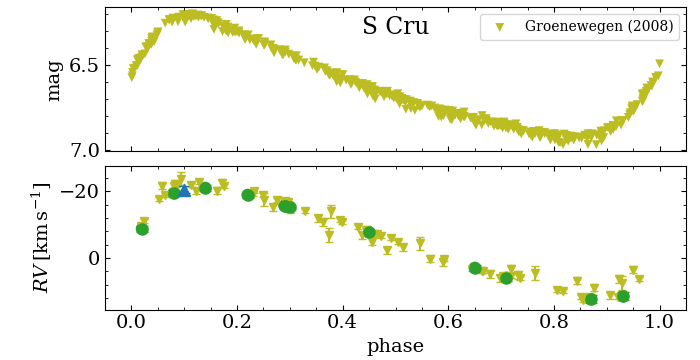}}
\end{minipage} \\
\begin{minipage}[t]{0.49\textwidth}
\centering
\resizebox{\hsize}{!}{\includegraphics{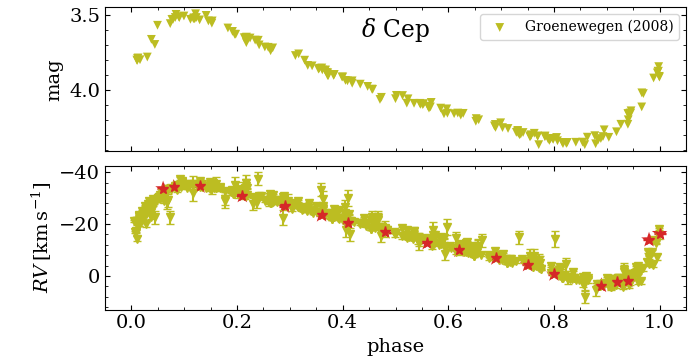}}
\end{minipage}
\begin{minipage}[t]{0.49\textwidth}
\centering
\resizebox{\hsize}{!}{\includegraphics{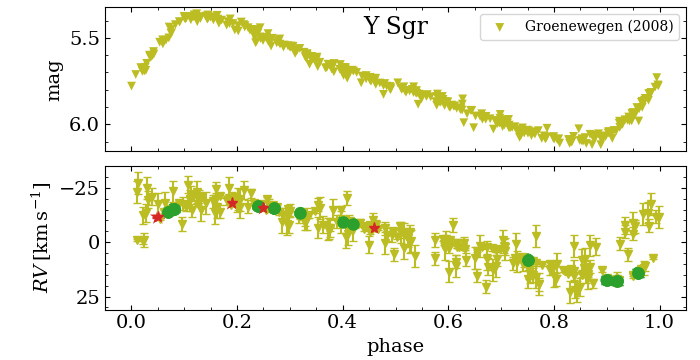}}
\end{minipage} \\
\begin{minipage}[t]{0.49\textwidth}
\centering
\resizebox{\hsize}{!}{\includegraphics{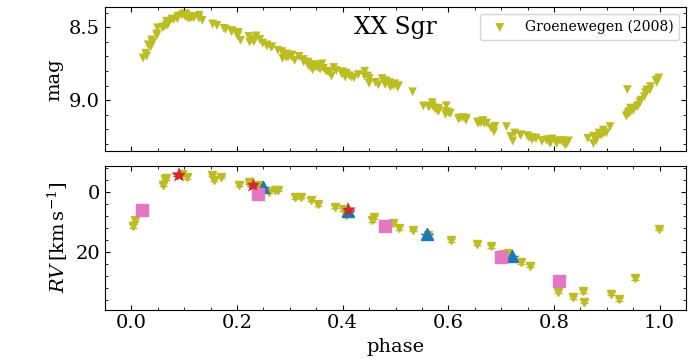}}
\end{minipage}
\begin{minipage}[t]{0.49\textwidth}
\centering
\resizebox{\hsize}{!}{\includegraphics{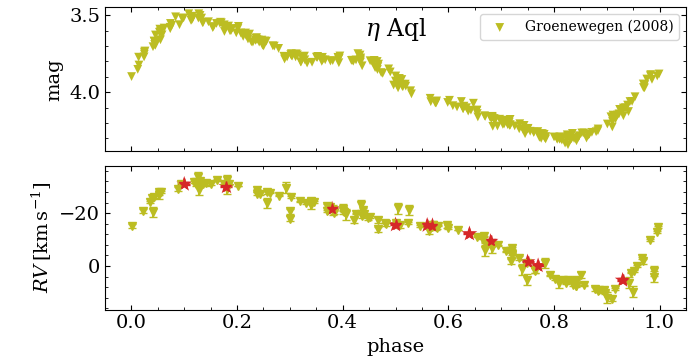}}
\end{minipage} \\
\begin{minipage}[t]{0.49\textwidth}
\centering
\resizebox{\hsize}{!}{\includegraphics{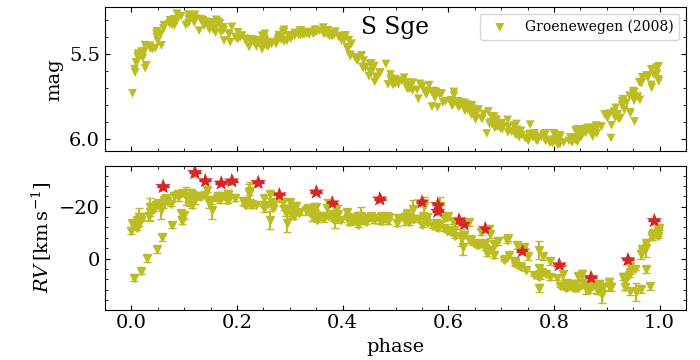}}
\end{minipage}
\begin{minipage}[t]{0.49\textwidth}
\centering
\resizebox{\hsize}{!}{\includegraphics{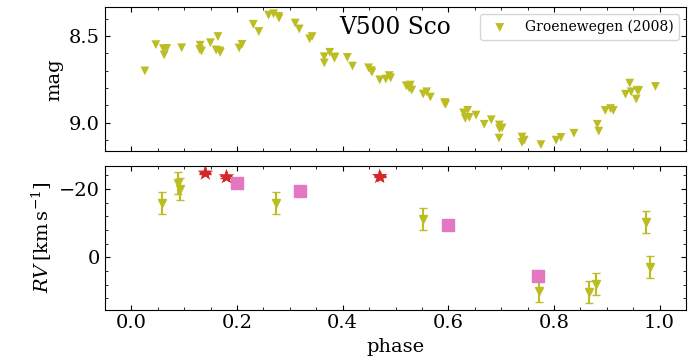}}
\end{minipage}
\caption{Light curves and radial velocities as a function of the pulsation phase. Measurements from different spectrographs are indicated with different colors and symbols. The error bars in some cases are smaller than the symbol size.}
\label{figure:mag_rv_phase}
\end{figure*}
\addtocounter{figure}{-1}
\begin{figure*}
\centering
\begin{minipage}[t]{0.49\textwidth}
\centering
\resizebox{\hsize}{!}{\includegraphics{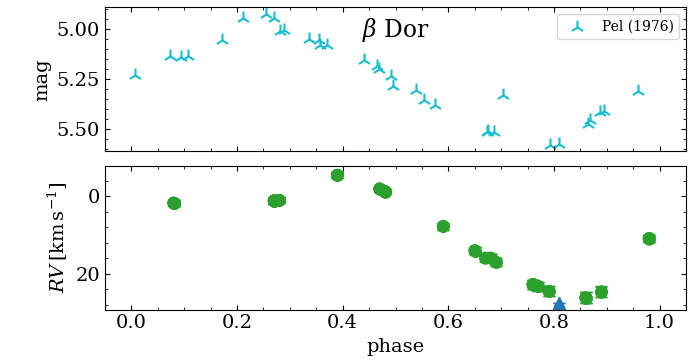}}
\end{minipage}
\begin{minipage}[t]{0.49\textwidth}
\centering
\resizebox{\hsize}{!}{\includegraphics{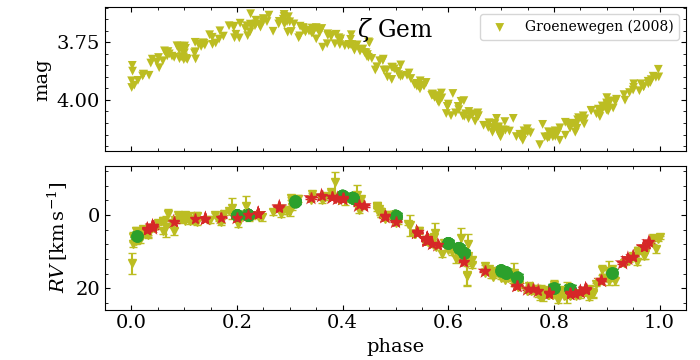}}
\end{minipage} \\
\begin{minipage}[t]{0.49\textwidth}
\centering
\resizebox{\hsize}{!}{\includegraphics{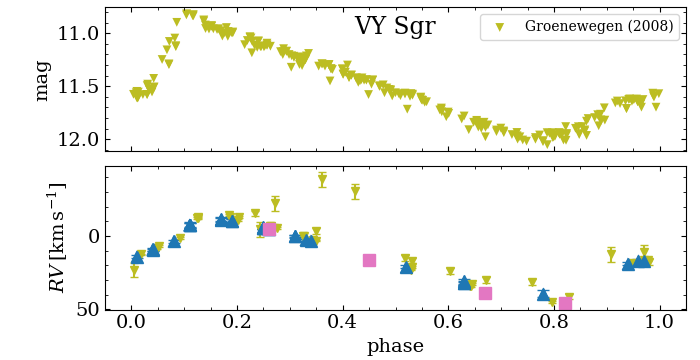}}
\end{minipage}
\begin{minipage}[t]{0.49\textwidth}
\centering
\resizebox{\hsize}{!}{\includegraphics{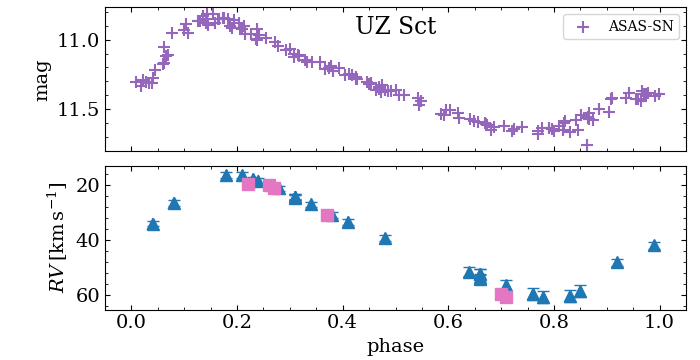}}
\end{minipage} \\
\begin{minipage}[t]{0.49\textwidth}
\centering
\resizebox{\hsize}{!}{\includegraphics{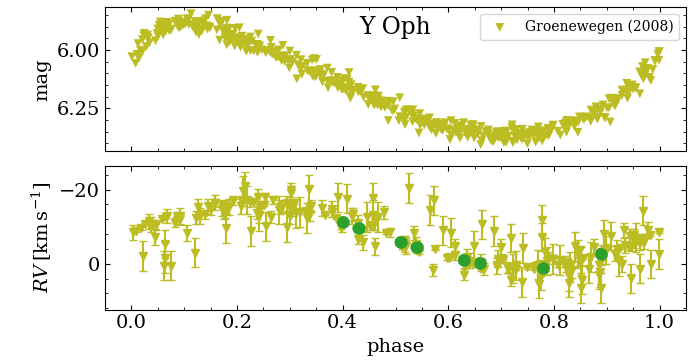}}
\end{minipage}
\begin{minipage}[t]{0.49\textwidth}
\centering
\resizebox{\hsize}{!}{\includegraphics{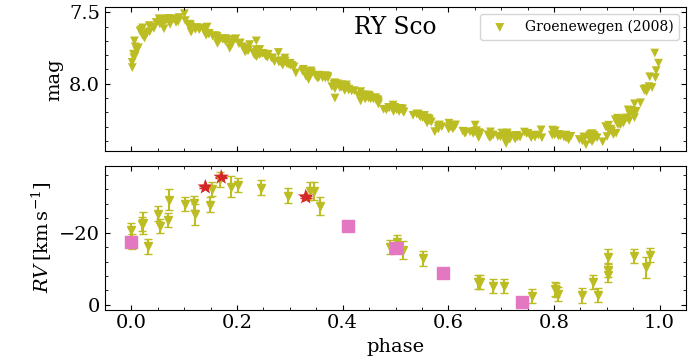}}
\end{minipage} \\
\begin{minipage}[t]{0.49\textwidth}
\centering
\resizebox{\hsize}{!}{\includegraphics{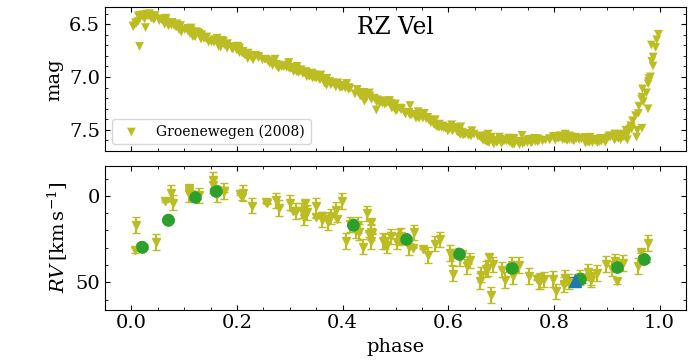}}
\end{minipage}
\begin{minipage}[t]{0.49\textwidth}
\centering
\resizebox{\hsize}{!}{\includegraphics{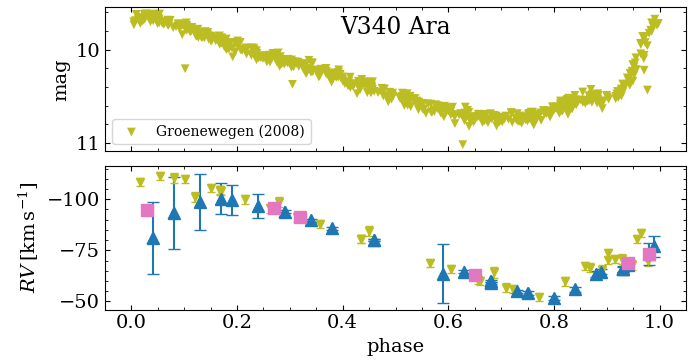}}
\end{minipage} \\
\begin{minipage}[t]{0.49\textwidth}
\centering
\resizebox{\hsize}{!}{\includegraphics{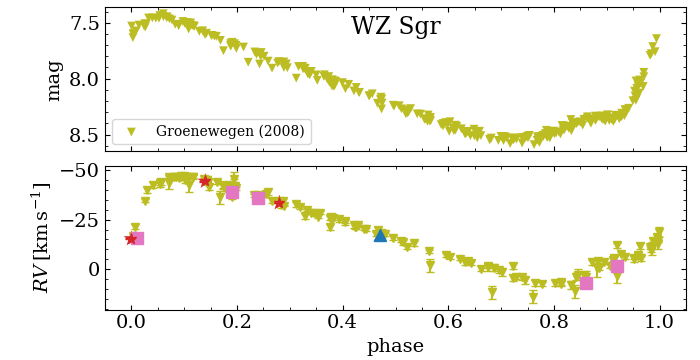}}
\end{minipage}
\begin{minipage}[t]{0.49\textwidth}
\centering
\resizebox{\hsize}{!}{\includegraphics{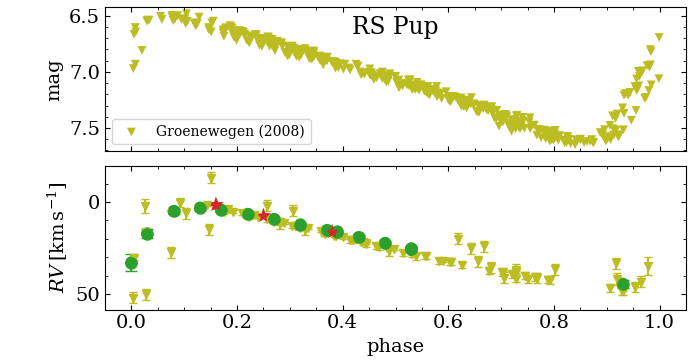}}
\end{minipage}
\caption[]{continued.}
\end{figure*}

\subsection{Atmospheric parameters and metallicity}

\begin{figure*}
\centering
\resizebox{\hsize}{!}{\includegraphics{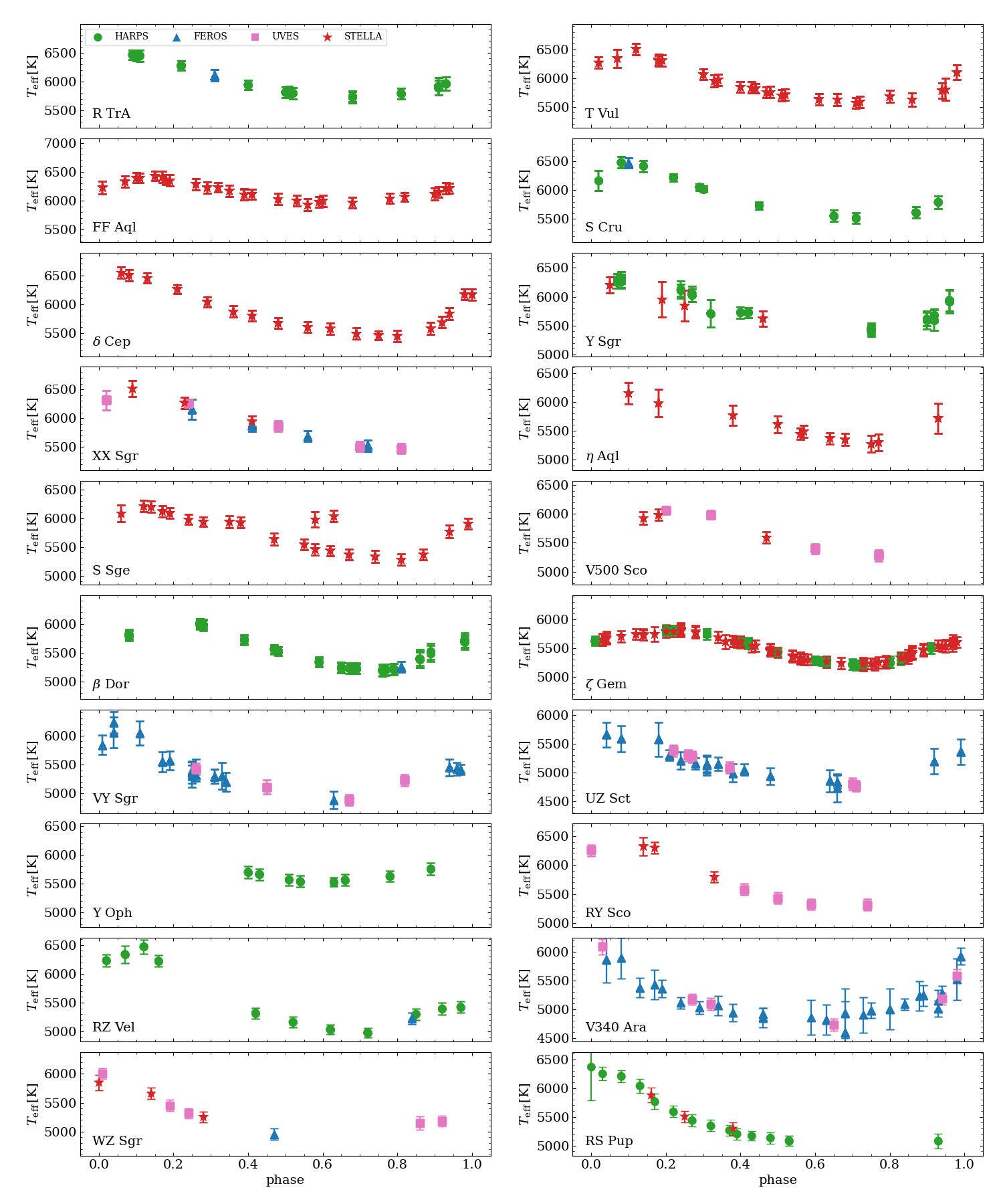}}
\caption{Effective temperature as a function of the pulsation phase. Measurements from different spectrographs are indicated with different colors and symbols. To help with the comparison, the panels are plotted with the same y-axis range.}
\label{figure:teff_phase}
\end{figure*}

\begin{figure*}
\centering
\resizebox{\hsize}{!}{\includegraphics{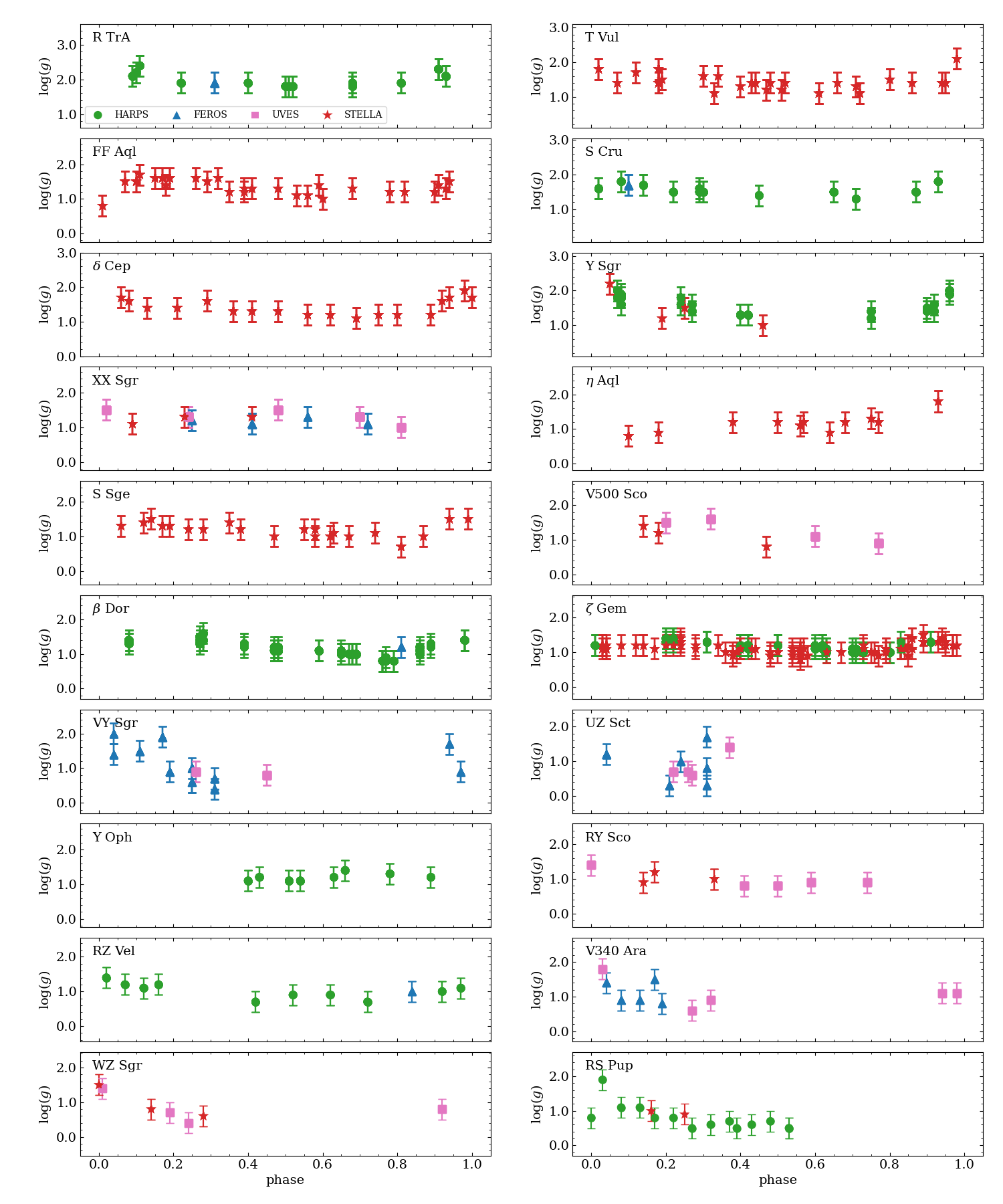}}
\caption{Same as in Fig.~\ref{figure:teff_phase}, but showing the surface gravity as a function of the pulsation phase.}
\label{figure:logg_phase}
\end{figure*}

\begin{figure*}
\centering
\resizebox{\hsize}{!}{\includegraphics{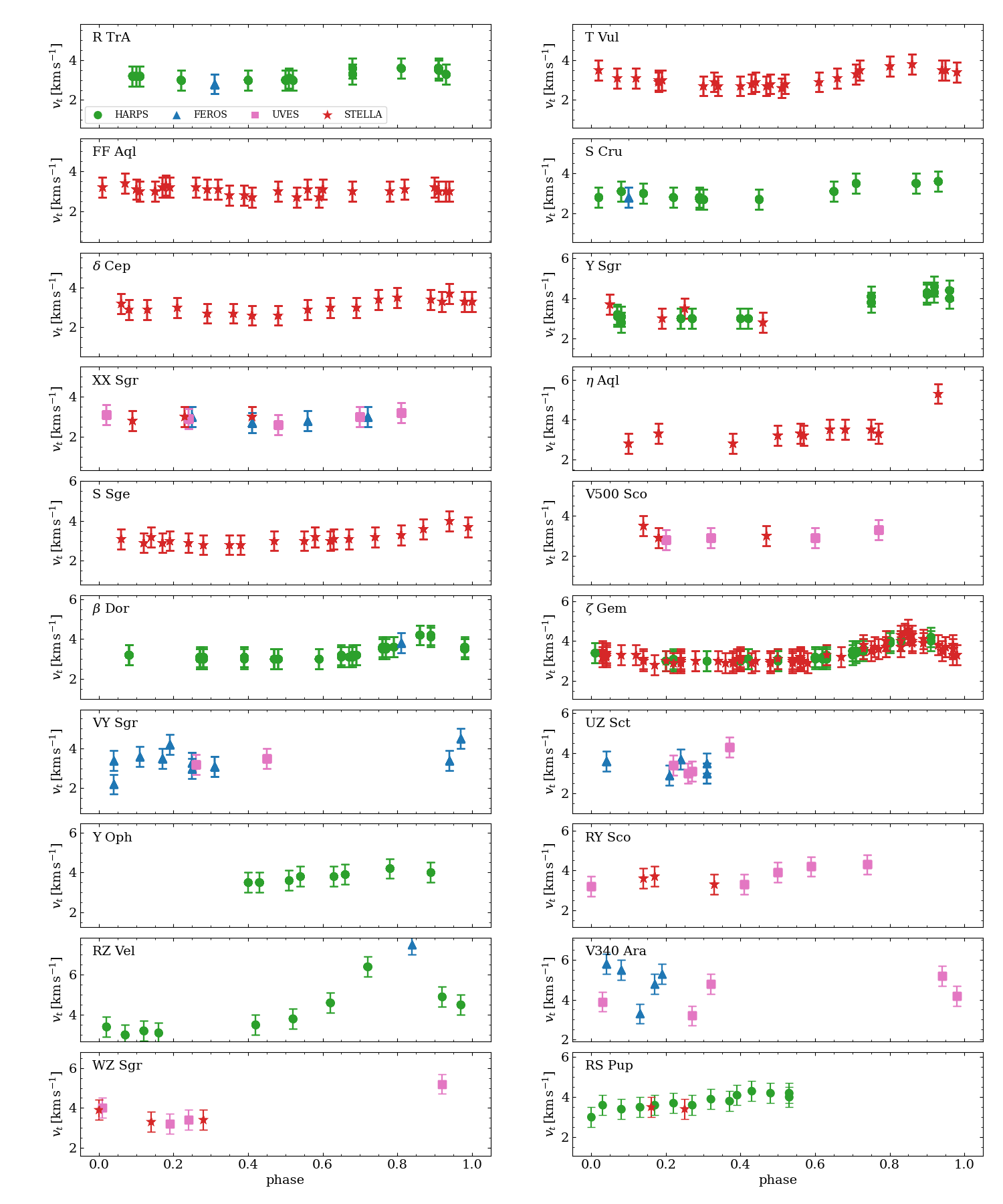}}
\caption{Same as in Fig.~\ref{figure:teff_phase}, but showing the microturbulent velocity as a function of the pulsation phase.}
\label{figure:vmic_phase}
\end{figure*}

The variation of the atmospheric parameters (\teff, \logg, and $\xi$) and the iron abundances along the pulsation cycle are shown in the panels of figures from \ref{figure:teff_phase} to \ref{figure:feh_phase}. It is worth mentioning that the number of points in the \teff\ panels are larger than in the other panels. This is because the determination of the effective temperature and of the other parameters are based on different and independent methods. For many spectra we were only able to derive the effective temperature (using the LDR method), whereas the derivation of the other parameters did not converge when we tried to force the ionization equilibrium of \ion{Fe}{i} and \ion{Fe}{ii} lines and the independence of the abundances on the equivalent widths (very often due to a limited number of good lines).

The variable \object{S\,Sge} shows two outliers in the \teff\ curve plotted in the panel of Fig.~\ref{figure:teff_phase}. However, these values might be correct because their shift with respect to the other spectra could be due to a limited accuracy in their phase determination, caused either by uncertainties in the period estimate or by any variation in period.


\begin{figure*}
\centering
\begin{minipage}[t]{0.49\textwidth}
\centering
\resizebox{\hsize}{!}{\includegraphics{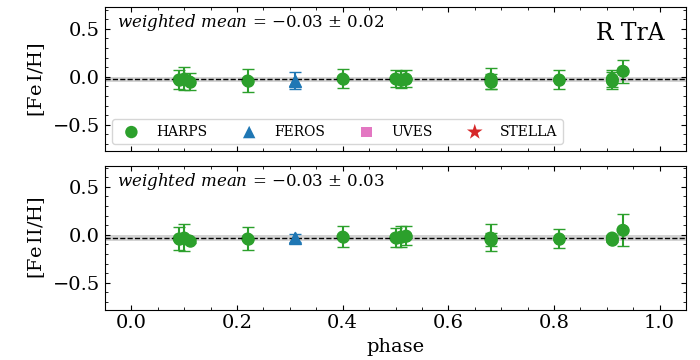}}
\end{minipage}
\begin{minipage}[t]{0.49\textwidth}
\centering
\resizebox{\hsize}{!}{\includegraphics{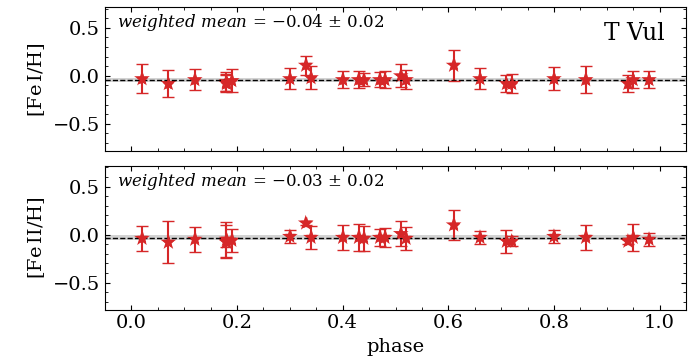}}
\end{minipage} \\
\begin{minipage}[t]{0.49\textwidth}
\centering
\resizebox{\hsize}{!}{\includegraphics{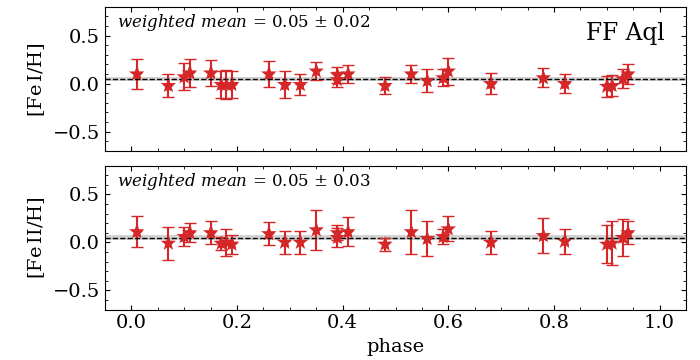}}
\end{minipage}
\begin{minipage}[t]{0.49\textwidth}
\centering
\resizebox{\hsize}{!}{\includegraphics{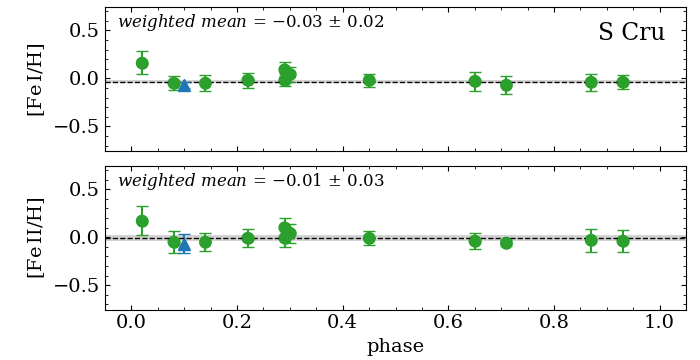}}
\end{minipage} \\
\begin{minipage}[t]{0.49\textwidth}
\centering
\resizebox{\hsize}{!}{\includegraphics{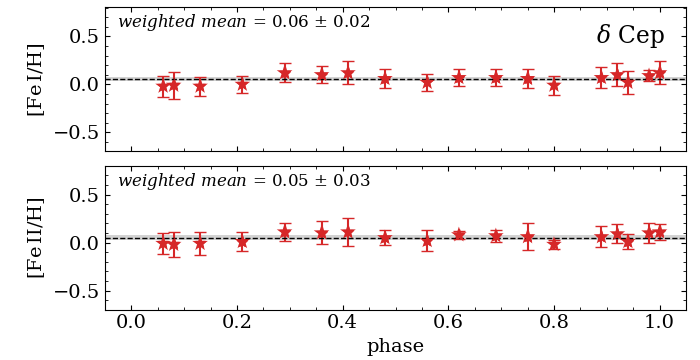}}
\end{minipage}
\begin{minipage}[t]{0.49\textwidth}
\centering
\resizebox{\hsize}{!}{\includegraphics{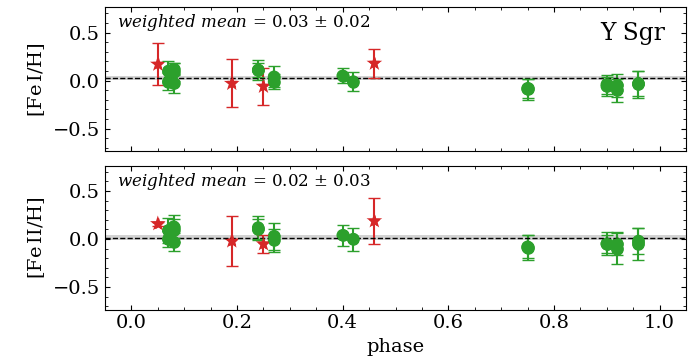}}
\end{minipage} \\
\begin{minipage}[t]{0.49\textwidth}
\centering
\resizebox{\hsize}{!}{\includegraphics{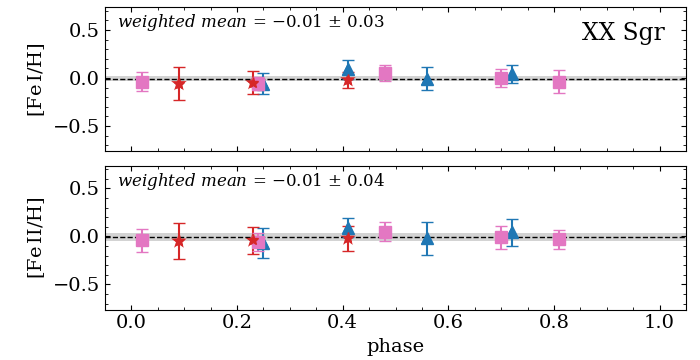}}
\end{minipage}
\begin{minipage}[t]{0.49\textwidth}
\centering
\resizebox{\hsize}{!}{\includegraphics{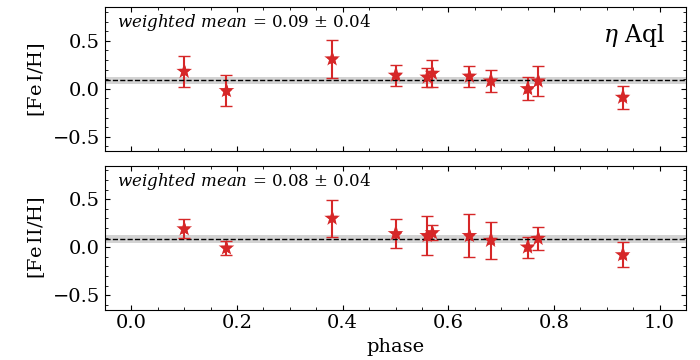}}
\end{minipage} \\
\begin{minipage}[t]{0.49\textwidth}
\centering
\resizebox{\hsize}{!}{\includegraphics{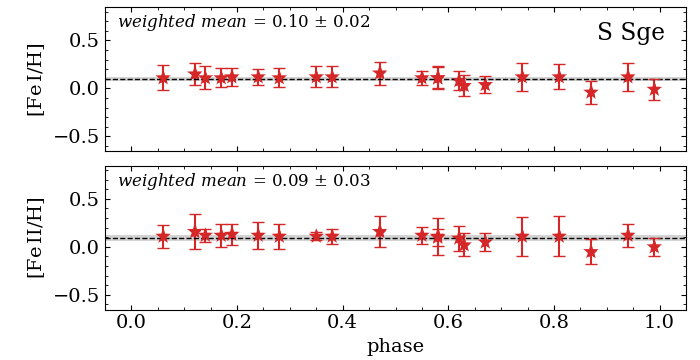}}
\end{minipage}
\begin{minipage}[t]{0.49\textwidth}
\centering
\resizebox{\hsize}{!}{\includegraphics{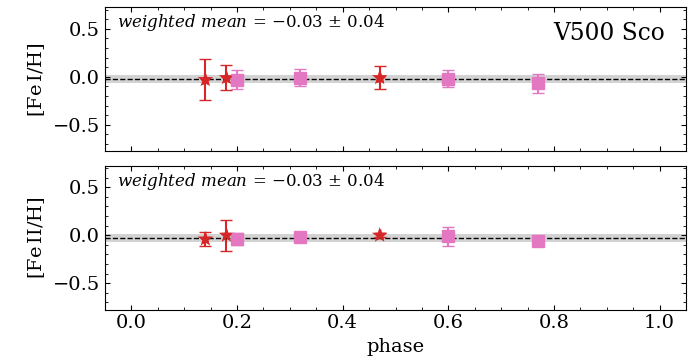}}
\end{minipage}
\caption{Abundances from \ion{Fe}{i} and \ion{Fe}{ii} lines as a function of the pulsation phase. The color coding of the various points is the same as in Fig.~\ref{figure:teff_phase}. To help with the comparison, the panels are plotted with same y-axis range: 1.5~dex for both \ion{Fe}{i} and \ion{Fe}{ii} panels. The light gray shaded regions indicate the $\pm$1-$\sigma$ uncertainty around the weighted mean (from Cols.~5 and 6 of Table~\ref{table:atm_params_stars}).}
\label{figure:feh_phase}
\end{figure*}
\addtocounter{figure}{-1}
\begin{figure*}
\centering
\begin{minipage}[t]{0.49\textwidth}
\centering
\resizebox{\hsize}{!}{\includegraphics{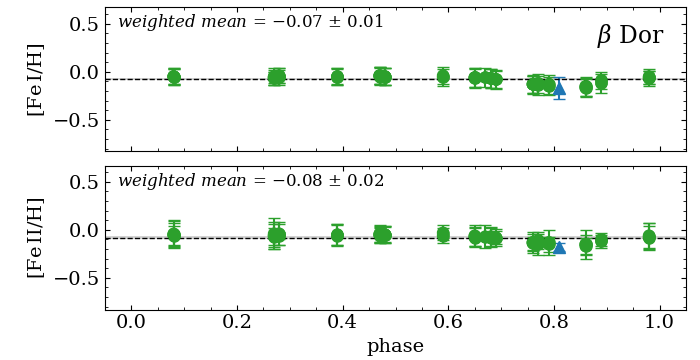}}
\end{minipage}
\begin{minipage}[t]{0.49\textwidth}
\centering
\resizebox{\hsize}{!}{\includegraphics{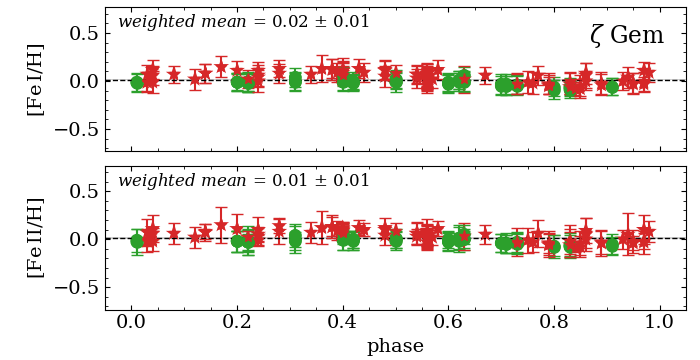}}
\end{minipage} \\
\begin{minipage}[t]{0.49\textwidth}
\centering
\resizebox{\hsize}{!}{\includegraphics{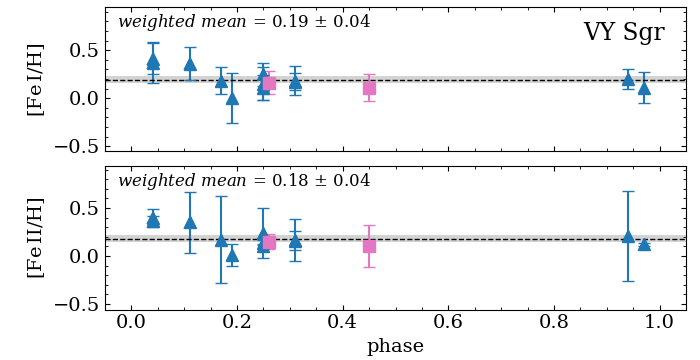}}
\end{minipage}
\begin{minipage}[t]{0.49\textwidth}
\centering
\resizebox{\hsize}{!}{\includegraphics{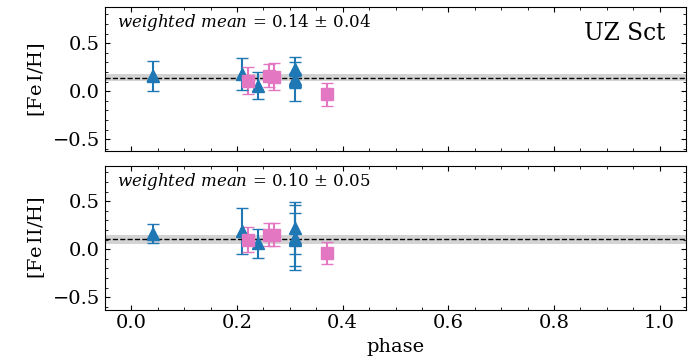}}
\end{minipage} \\
\begin{minipage}[t]{0.49\textwidth}
\centering
\resizebox{\hsize}{!}{\includegraphics{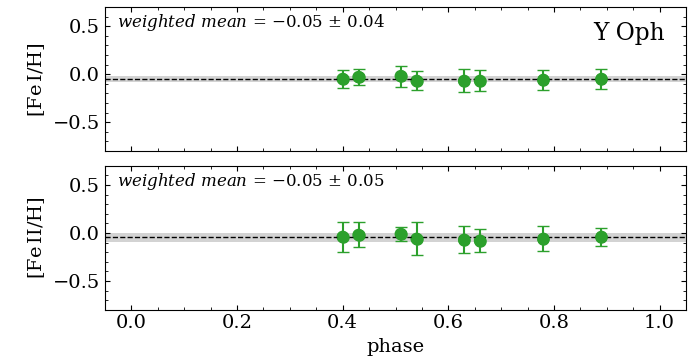}}
\end{minipage}
\begin{minipage}[t]{0.49\textwidth}
\centering
\resizebox{\hsize}{!}{\includegraphics{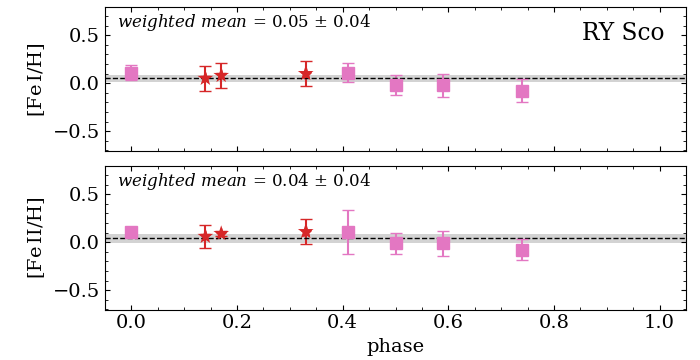}}
\end{minipage} \\
\begin{minipage}[t]{0.49\textwidth}
\centering
\resizebox{\hsize}{!}{\includegraphics{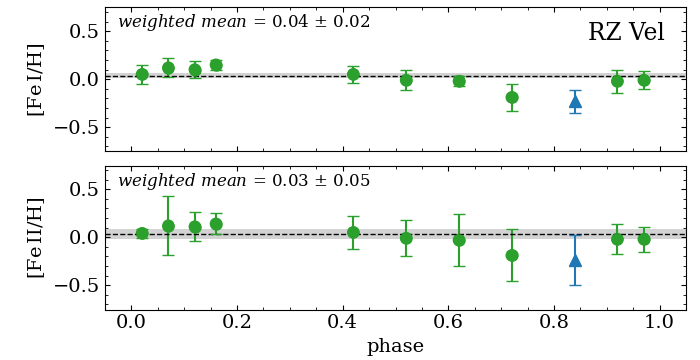}}
\end{minipage}
\begin{minipage}[t]{0.49\textwidth}
\centering
\resizebox{\hsize}{!}{\includegraphics{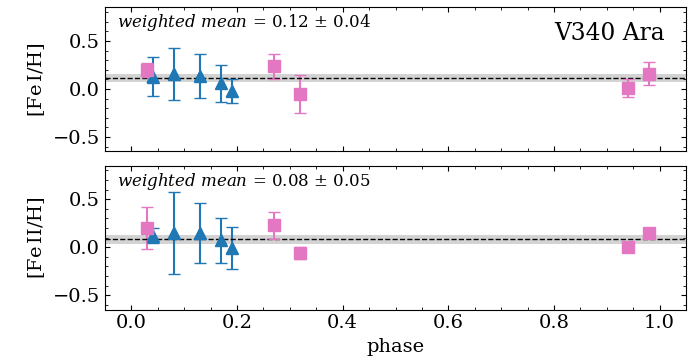}}
\end{minipage} \\
\begin{minipage}[t]{0.49\textwidth}
\centering
\resizebox{\hsize}{!}{\includegraphics{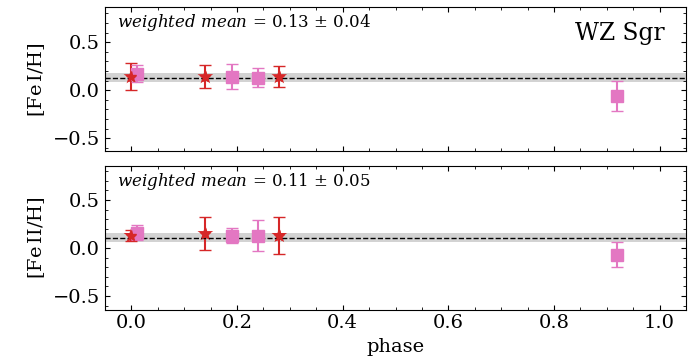}}
\end{minipage}
\begin{minipage}[t]{0.49\textwidth}
\centering
\resizebox{\hsize}{!}{\includegraphics{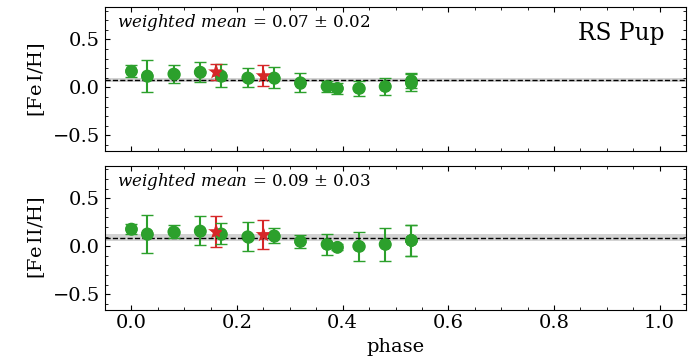}}
\end{minipage}
\caption[]{continued.}
\end{figure*}

\subsection{Abundances of $\alpha$ elements}

Figures~\ref{figure:xh_phase} shows the [$\alpha$/H] abundance ratios for our sample as a function of the pulsation phase, in which we also compare the abundances from neutral (colored symbols) and ionized (light gray symbols) elements, when available. The agreement between the two species is normally within 2~sigma, with only a few exceptions for the Ca abundances (the \ion{Ca}{ii} abundances are very often based on a few weak lines only).

\begin{figure*}
\centering
\begin{minipage}[t]{0.49\textwidth}
\centering
\resizebox{\hsize}{!}{\includegraphics{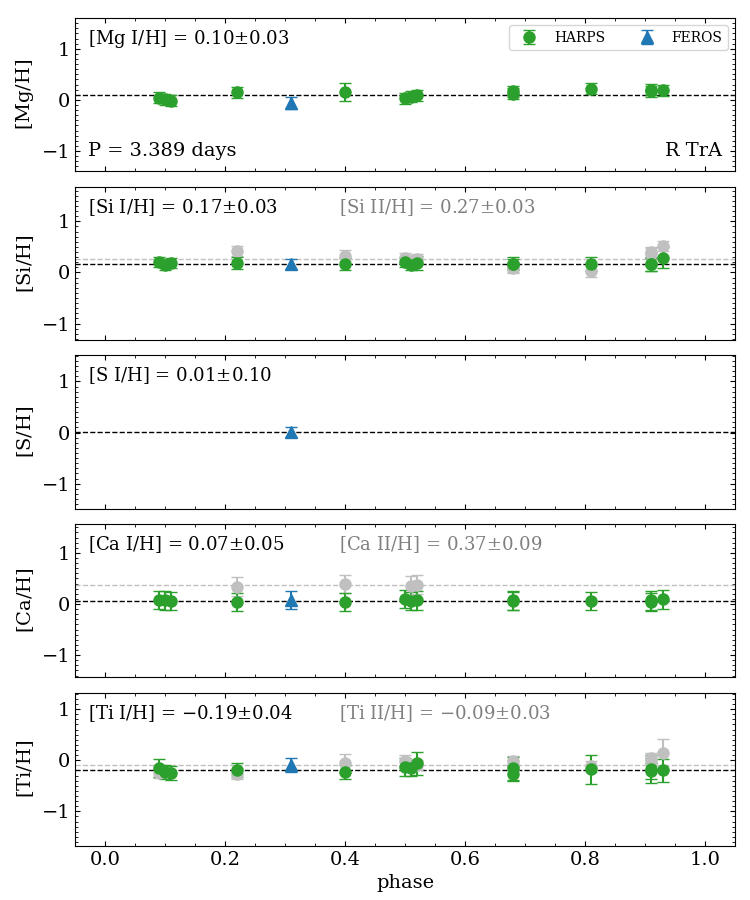}}
\end{minipage}
\begin{minipage}[t]{0.49\textwidth}
\centering
\resizebox{\hsize}{!}{\includegraphics{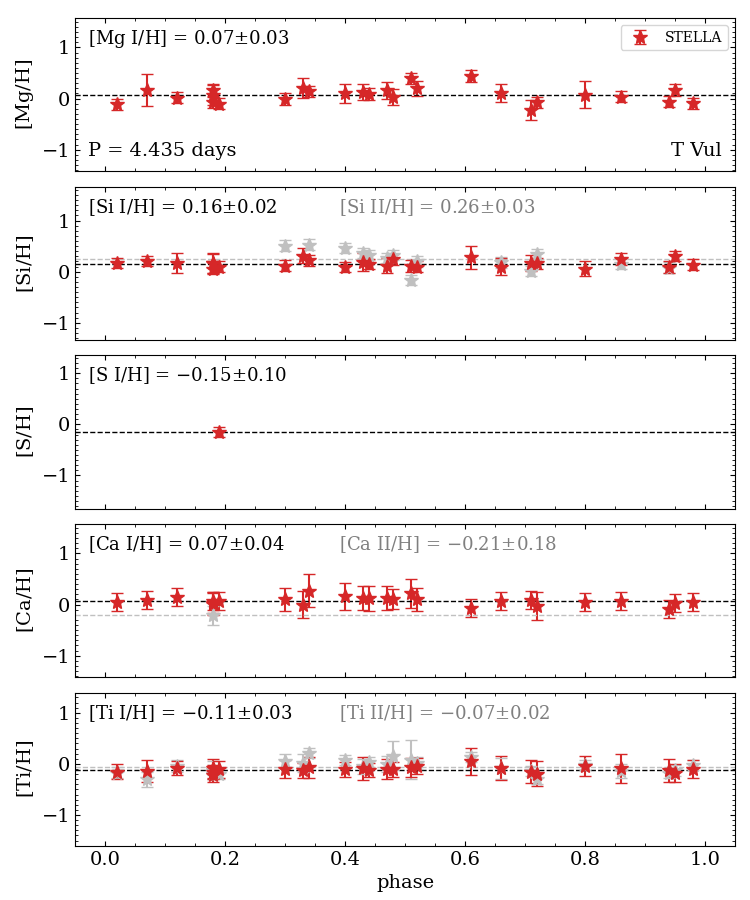}}
\end{minipage} \\
\begin{minipage}[t]{0.49\textwidth}
\centering
\resizebox{\hsize}{!}{\includegraphics{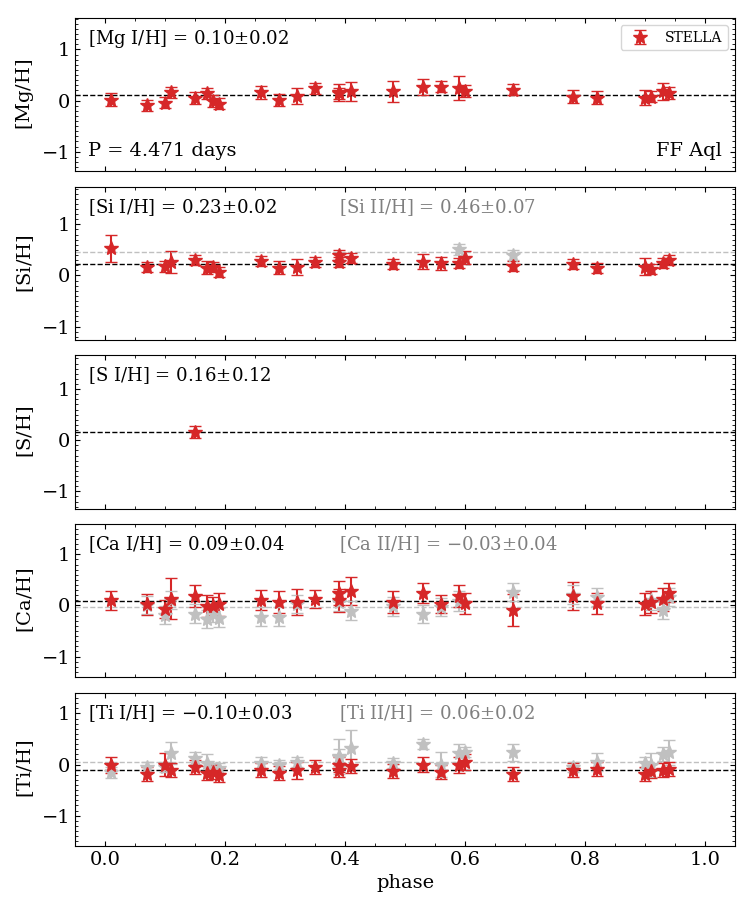}}
\end{minipage}
\begin{minipage}[t]{0.49\textwidth}
\centering
\resizebox{\hsize}{!}{\includegraphics{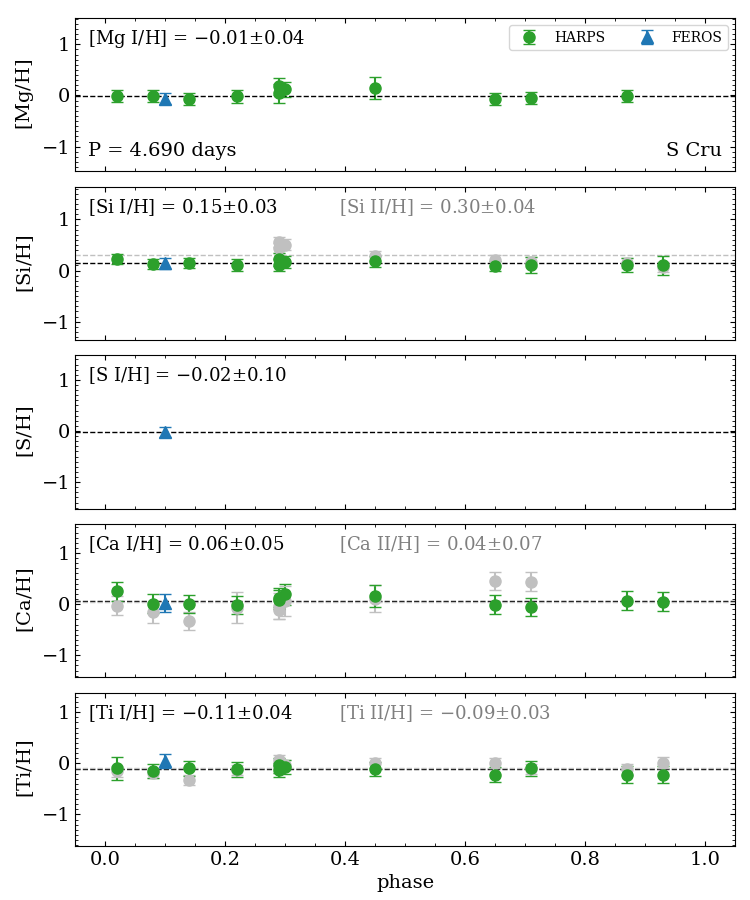}}
\end{minipage}
\caption{[X/H] abundances as a function of the pulsation phase. Each panel shows the abundances of $\alpha$ elements from neutral and ionized lines as listed in Tables~\ref{table:alpha_spectra} and \ref{table:alpha_stars}. As explained in the text, the sulfur abundances were derived using the spectral synthesis method applied to one spectrum of each star (an exception are the stars \object{$\eta$Aql}, \object{VY\,Sgr}, and \object{V340\,Ara}, for which the [S/H] abundances could not be derived due to the noisy profile of the sulfur line used).}
\label{figure:xh_phase}
\end{figure*}
\addtocounter{figure}{-1}
\begin{figure*}
\centering
\begin{minipage}[t]{0.49\textwidth}
\centering
\resizebox{\hsize}{!}{\includegraphics{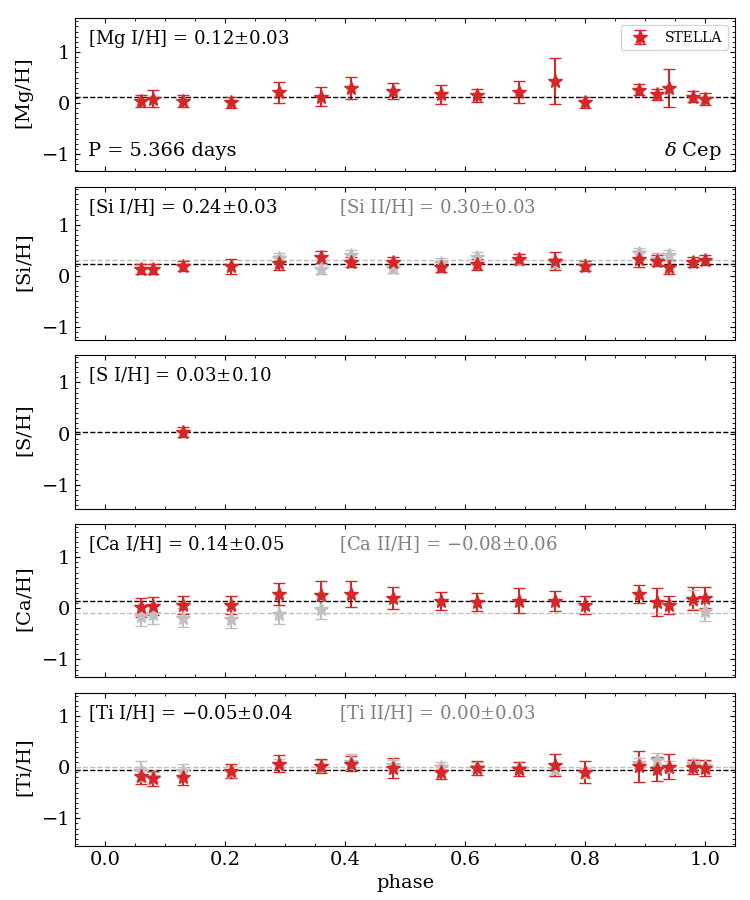}}
\end{minipage}
\begin{minipage}[t]{0.49\textwidth}
\centering
\resizebox{\hsize}{!}{\includegraphics{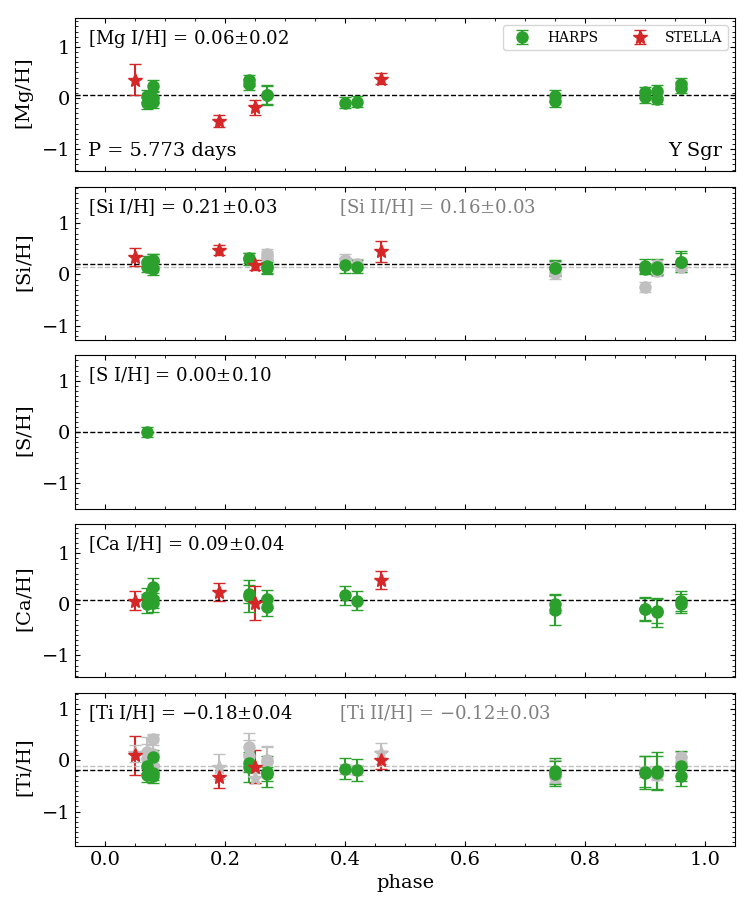}}
\end{minipage} \\
\begin{minipage}[t]{0.49\textwidth}
\centering
\resizebox{\hsize}{!}{\includegraphics{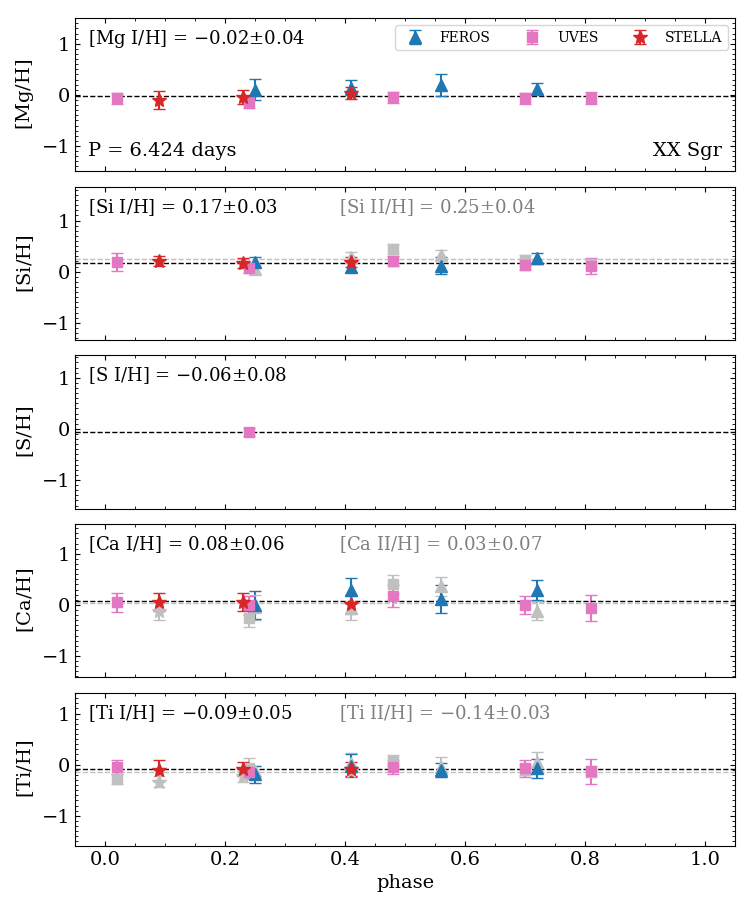}}
\end{minipage}
\begin{minipage}[t]{0.49\textwidth}
\centering
\resizebox{\hsize}{!}{\includegraphics{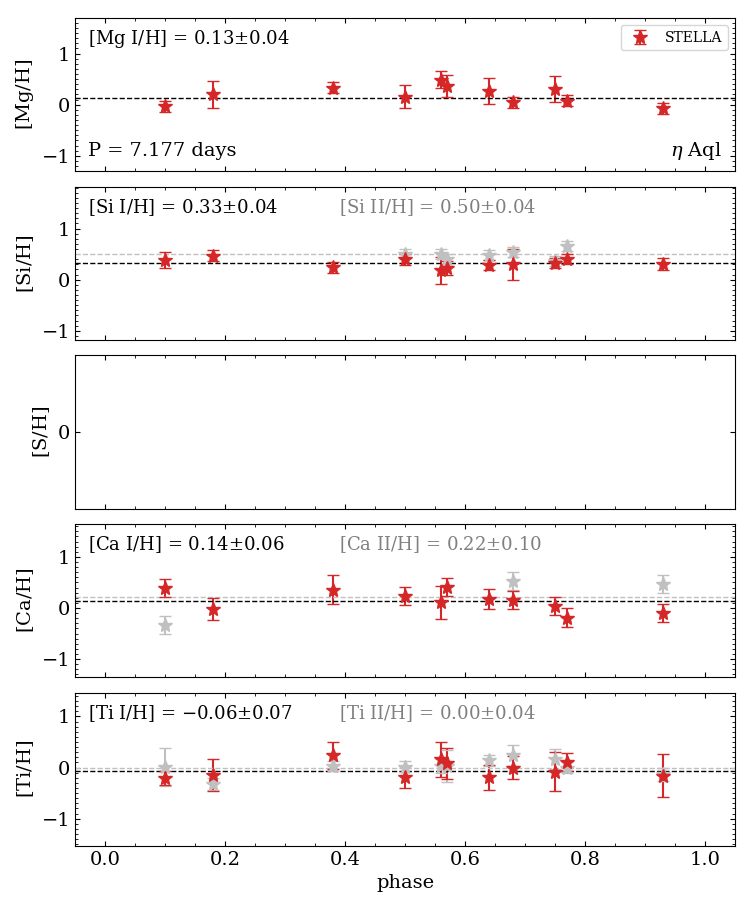}}
\end{minipage}
\caption[]{continued.}
\end{figure*}
\addtocounter{figure}{-1}
\begin{figure*}
\centering
\begin{minipage}[t]{0.49\textwidth}
\centering
\resizebox{\hsize}{!}{\includegraphics{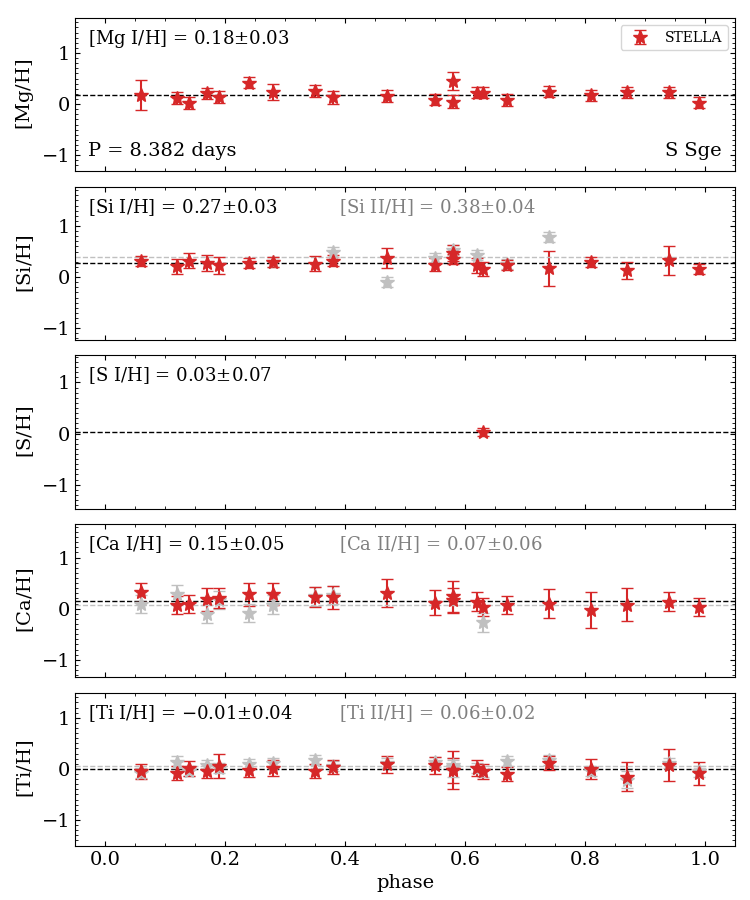}}
\end{minipage}
\begin{minipage}[t]{0.49\textwidth}
\centering
\resizebox{\hsize}{!}{\includegraphics{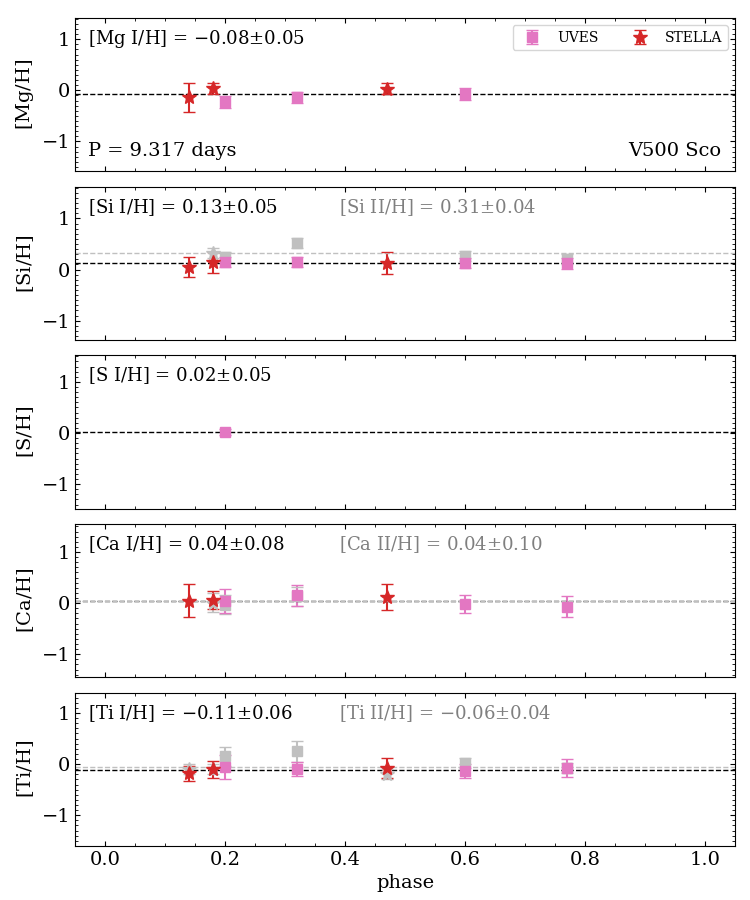}}
\end{minipage} \\
\begin{minipage}[t]{0.49\textwidth}
\centering
\resizebox{\hsize}{!}{\includegraphics{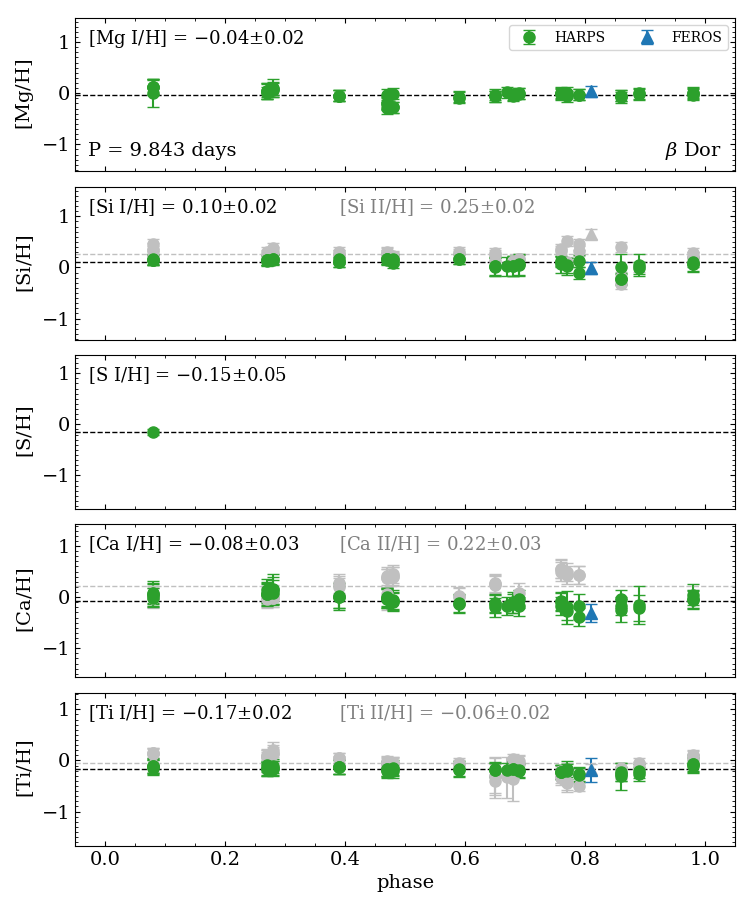}}
\end{minipage}
\begin{minipage}[t]{0.49\textwidth}
\centering
\resizebox{\hsize}{!}{\includegraphics{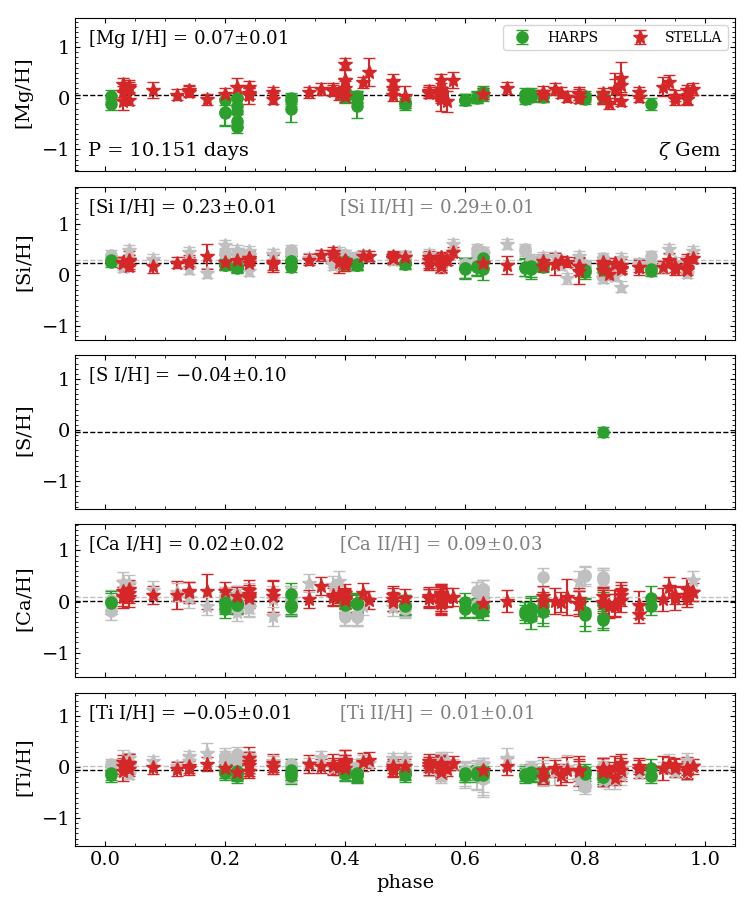}}
\end{minipage}
\caption[]{continued.}
\end{figure*}
\addtocounter{figure}{-1}
\begin{figure*}
\centering
\begin{minipage}[t]{0.49\textwidth}
\centering
\resizebox{\hsize}{!}{\includegraphics{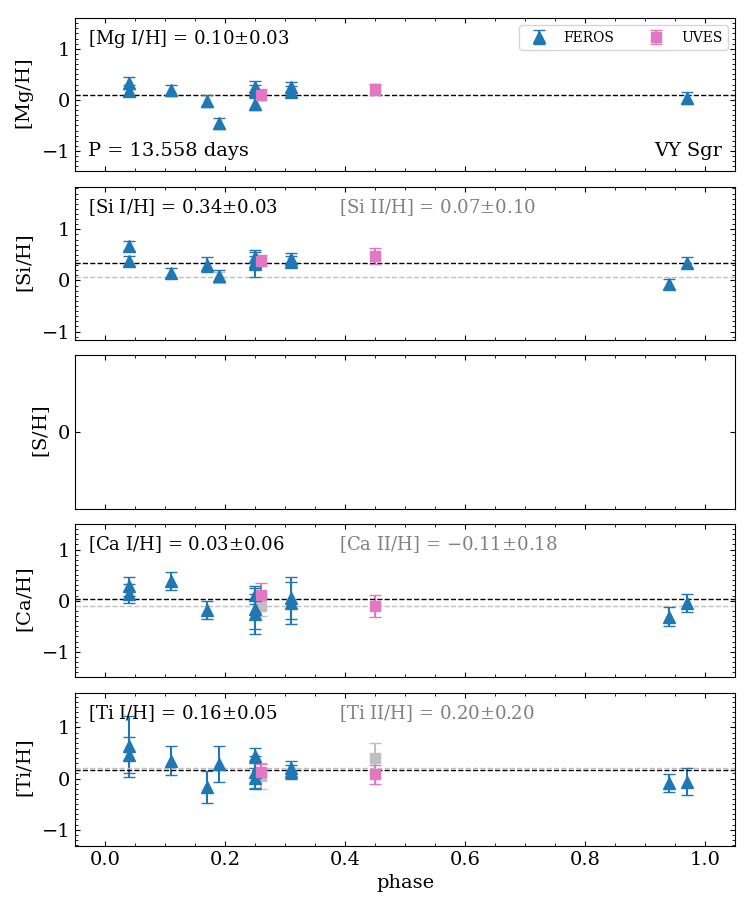}}
\end{minipage}
\begin{minipage}[t]{0.49\textwidth}
\centering
\resizebox{\hsize}{!}{\includegraphics{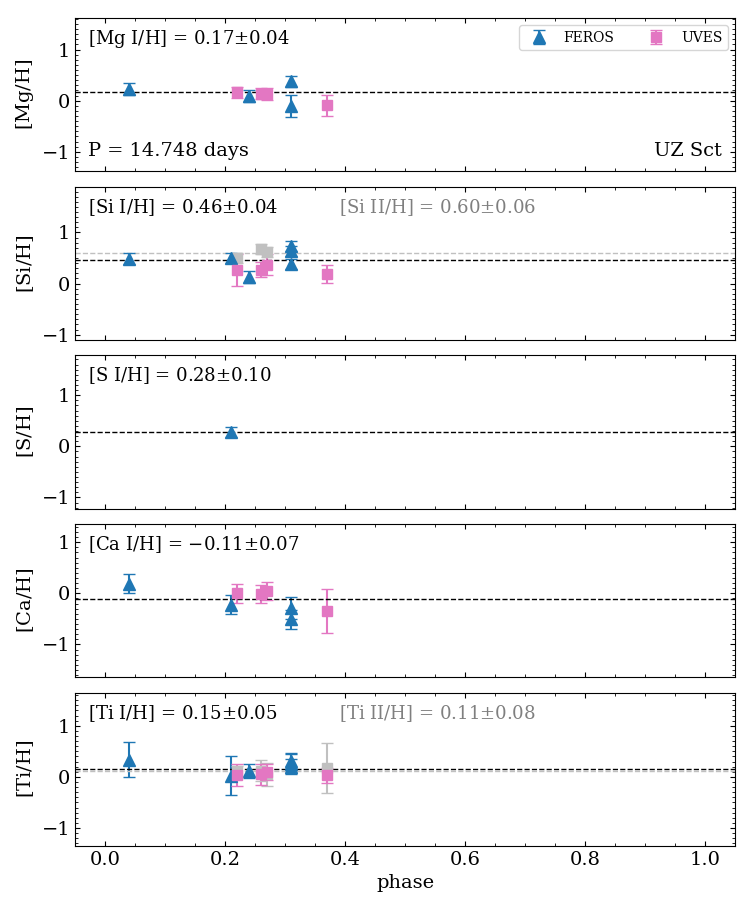}}
\end{minipage} \\
\begin{minipage}[t]{0.49\textwidth}
\centering
\resizebox{\hsize}{!}{\includegraphics{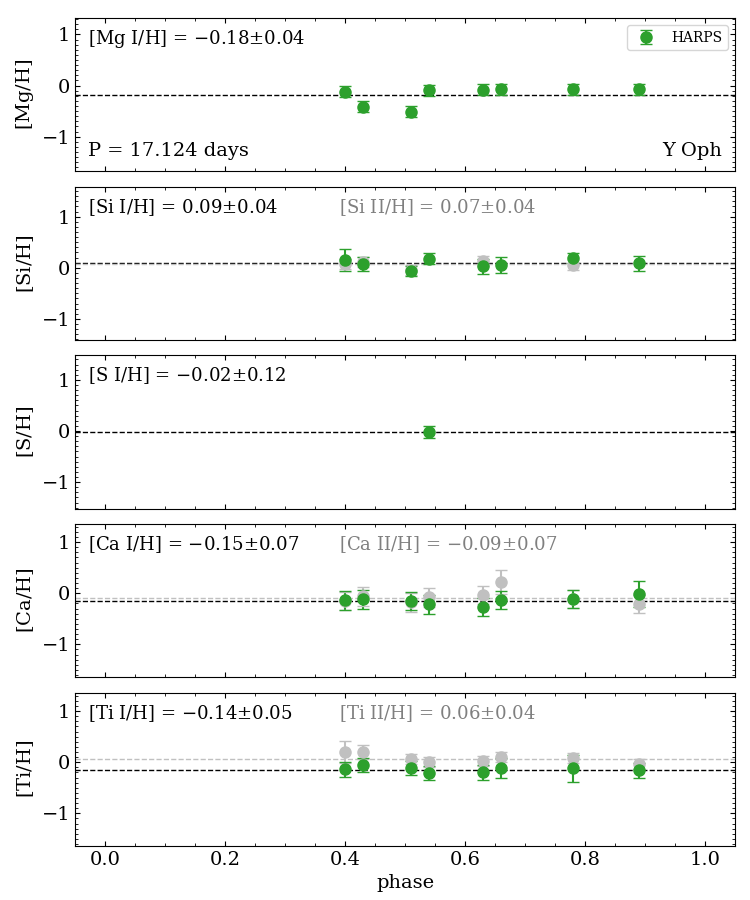}}
\end{minipage}
\begin{minipage}[t]{0.49\textwidth}
\centering
\resizebox{\hsize}{!}{\includegraphics{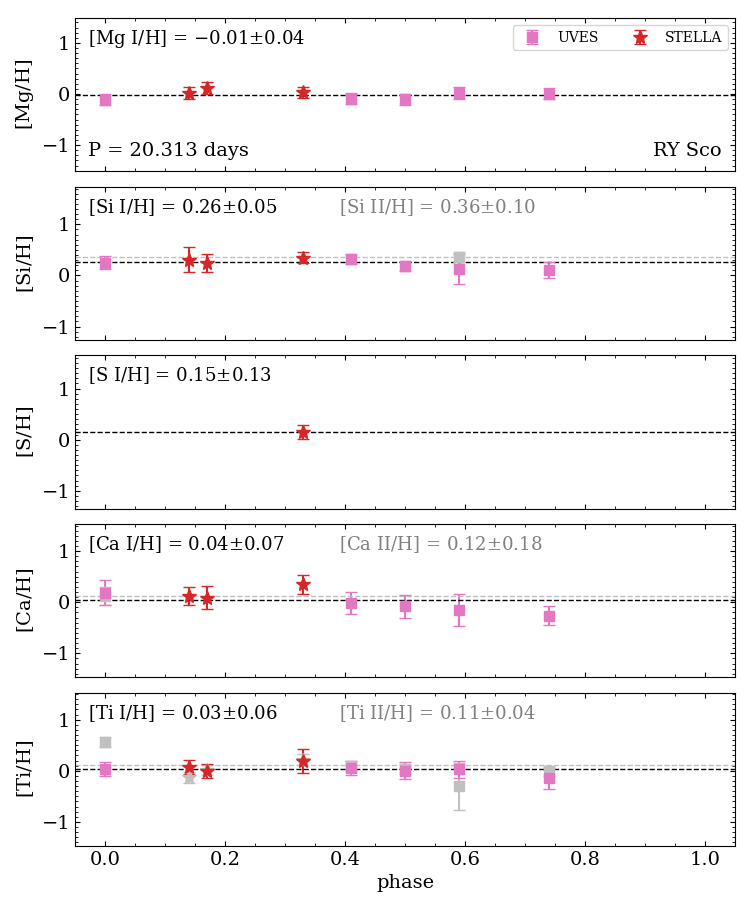}}
\end{minipage}
\caption[]{continued.}
\end{figure*}
\addtocounter{figure}{-1}
\begin{figure*}
\centering
\begin{minipage}[t]{0.49\textwidth}
\centering
\resizebox{\hsize}{!}{\includegraphics{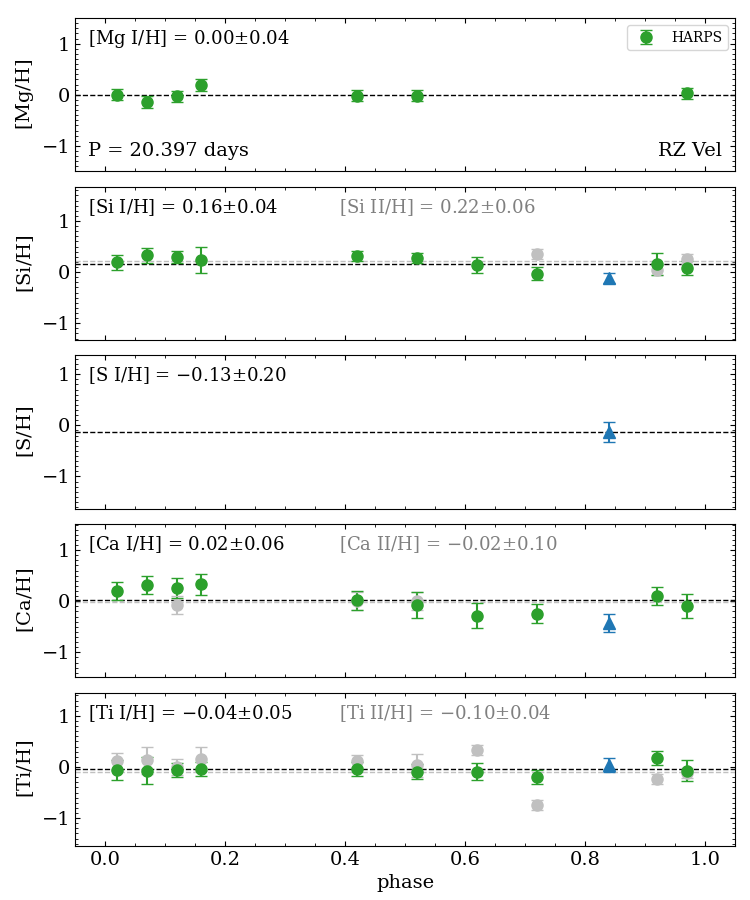}}
\end{minipage}
\begin{minipage}[t]{0.49\textwidth}
\centering
\resizebox{\hsize}{!}{\includegraphics{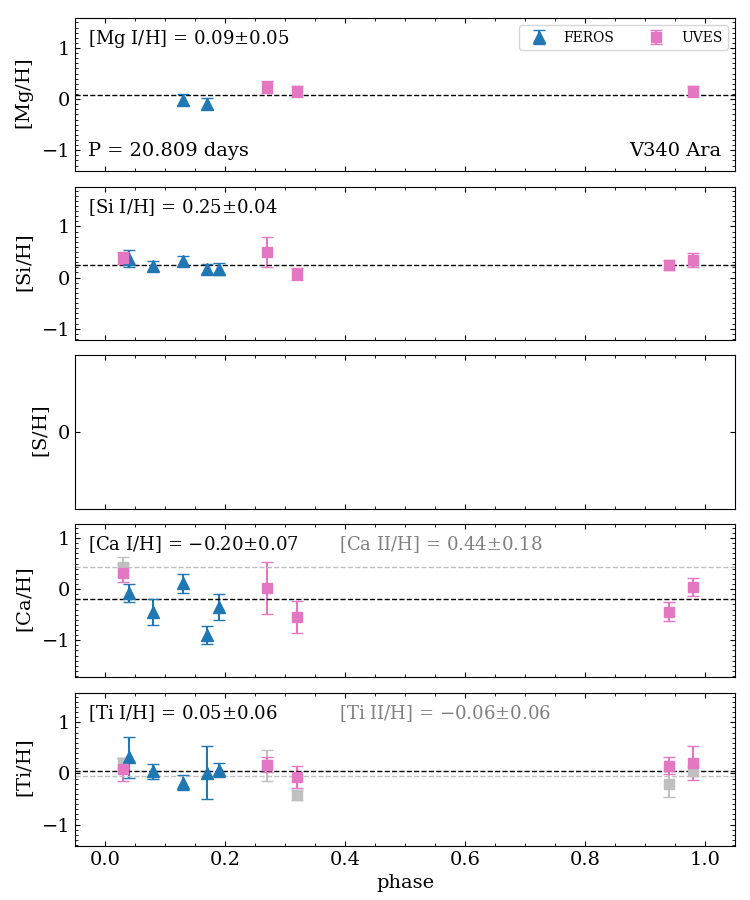}}
\end{minipage} \\
\begin{minipage}[t]{0.49\textwidth}
\centering
\resizebox{\hsize}{!}{\includegraphics{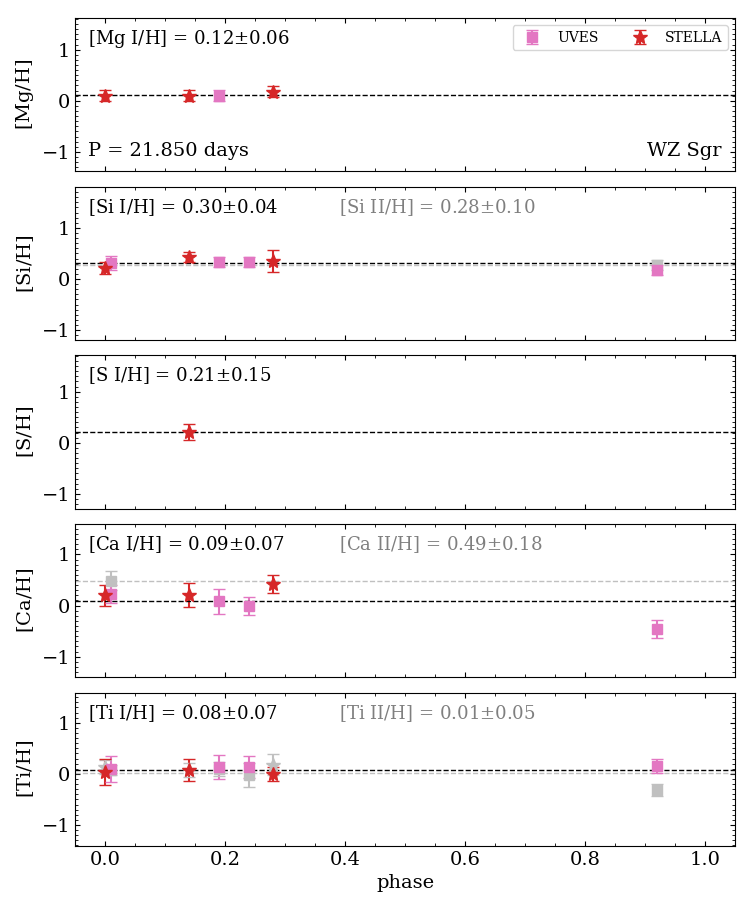}}
\end{minipage}
\begin{minipage}[t]{0.49\textwidth}
\centering
\resizebox{\hsize}{!}{\includegraphics{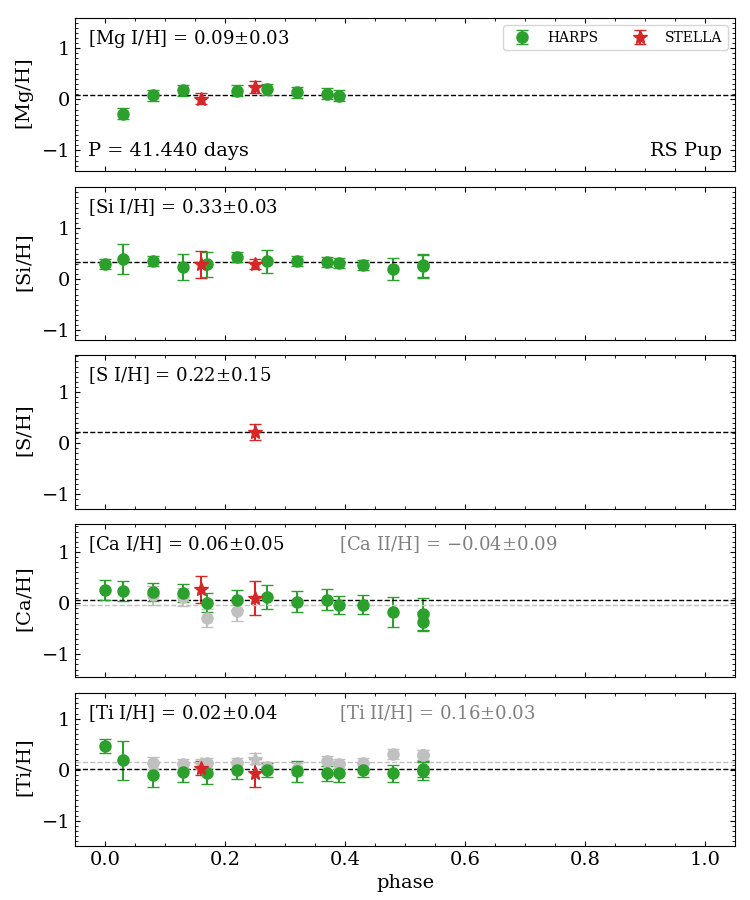}}
\end{minipage}
\caption[]{continued.}
\end{figure*}

\end{appendix}

\end{document}